\documentclass[fleqn,usenatbib]{mnras}

\usepackage{newtxtext,newtxmath}
\usepackage[T1]{fontenc}
\usepackage{ae,aecompl}
\usepackage{blindtext}
\usepackage{adjustbox}
\usepackage{makecell}
\usepackage{enumerate}
\usepackage[switch,displaymath,mathlines]{lineno}
\usepackage{graphicx}   % Including figure files
\usepackage{amsmath}    % Advanced maths commands
\usepackage{amssymb}    % Extra maths symbols

\title[Multiple populations mass loss]{Mass loss along the red giant branch in 46 Globular Clusters and their multiple populations} 

\author[Tailo, M. et al.]{
M. Tailo$^{1}$, \thanks{E-mail:mrctailo@gmail.com, marco.tailo@unipd.it}
A. P. Milone$^{1,2}$,
E. P. Lagioia$^{1,2}$, 
F. D'Antona$^{3}$,
A. F. Marino$^{1,2,4}$,
\newauthor
E. Vesperini$^{5}$,
V. Caloi$^{6}$,
P. Ventura$^{3}$,
E. Dondoglio$^{1}$,
G. Cordoni$^{1}$
\\
$^{1}$Dipartimento di Fisica e Astronomia ``Galileo Galilei'', Univ. di Padova, Vicolo dell'Osservatorio 3, Padova, IT-35122\\
$^{2}$Istituto Nazionale di Astrofisica - Osservatorio Astronomico di Padova, Vicolo dell'Osservatorio 5, Padova, IT-35122\\
$^{3}$Istituto Nazionale di Astrofisica - Osservatorio Astronomico di Roma, Via Frascati 33, I-00040 Monteporzio Catone, Roma, Italy\\
$^{4}$Centro di Ateneo di Studi e Attivita' Spaziali ``Giuseppe Colombo'' - CISAS, Via Venezia 15, Padova, IT-35131 \\
$^{5}$Department of Astronomy, Indiana University, Bloomington, IN 47405, USA\\
$^{6}$INAF -- IASF Roma, Via Fosso del Cavaliere, Roma, Italy, IT-00133  \\
}

\date{Accepted 2020 August 24. Received 2020 August 22; in original form 2020 June 1}

\pubyear{2020}

\begin{document}
\label{firstpage}
\pagerange{\pageref{firstpage}--\pageref{lastpage}}
%\linenumbers
\maketitle
\begin{abstract}
The location of Galactic Globular Clusters' (GC) stars on the horizontal branch (HB) should mainly depend on GC metallicity, the ``first parameter'', but it is actually the result of complex interactions between the red giant branch (RGB) mass loss, the coexistence of multiple stellar populations with different helium content, and the presence of a ''second parameter'' which produces dramatic differences in HB morphology of  GCs of similar metallicity and ages (like the pair M3--M13).
In this work, we combine the entire dataset from the {\it Hubble Space Telescope} Treasury survey and stellar evolutionary models, to analyse the HBs of 46 GCs. For the first time in a large sample of GCs, we generate population synthesis models, where the helium abundances for the first and the ``extreme'' second generations are constrained using independent measurements based on RGB stars. The main results are:
1) the mass loss of first generation stars is tightly correlated to cluster metallicity. 2) the location of helium enriched stars on the HB is reproduced only by adopting a higher RGB mass loss than for the first generation. The {\it difference} in mass loss correlates with helium enhancement and cluster mass. 
3) A model of ``pre-main sequence disc early loss'', previously developed by the authors, explains such a mass loss increase and is  consistent with the findings of multiple-population formation models predicting that populations more enhanced in helium tend to form with higher stellar densities and concentrations.
4) Helium-enhancement and mass-loss both contribute to the second parameter.
\end{abstract}


\section{Introduction}
\label{sec:into}

The study of the horizontal branch (HB), the locus of the color-magnitude diagram (CMD) populated by stars burning helium in their core, is crucial to understand stellar evolution and characterize the stellar populations in Globular Clusters (GCs).

The HB stars are the direct off-springs of the red giant branch ones (RGB) and reach the helium burning stage after the ignition of their degenerate helium core in a process dubbed core--helium flash.
The initial mass of the evolving stars (depending on the cluster age and metallicity ---iron and light elements, mainly CNO, content) and the mass loss on the RGB (subject to some cosmic spread) determine the final masses which populate the HB. Fixed the chemical composition, 
a larger RGB mass loss is needed to reach a larger effective temperature ($\rm T_{eff}$) on the HB, as the HB mass decreases for increasing $\rm T_{eff}$. Increasing the metallicity, each mass moves to lower $\rm T_{eff}$ values, so the metallicity constitutes the `first parameter' of the HB morphology \citep[e.g.][]{arp_1952}.

Assuming that all the cluster stars have the same helium content, probably the abundance emerging from the Big Bang nucleosynthesis, it was soon clear that the morphology of the HB could be widely different even in GCs with similar age and metallicity, and that a ''second parameter'' was at play \citep[see e.g.][]{sandage_1967, fusipecci_1993}. A classical example is the pair NGC\,5272 (M\,3) and NGC\,6205 (M\,13), showing radically different HBs. A different mass loss on the RGB was considered the main reason for the different HB morphology, but no clear association of this systemic mass loss difference with other cluster physical parameters was conclusively found. 

A major obstacle to understand the observed HB of GCs is that the parameters at play are often degenerate, so that a number of different combination of age, metallicity, helium and mass loss provide similar HBs.
In part, the parameter degeneracy is limited by adopting ages inferred from the main-sequence turn off and metallicities obtained from spectroscopy. 

The evidence that nearly all GCs host multiple stellar populations with different helium abundances has provided an additional challenge to explain their HBs. Indeed, helium enhanced stars evolve more rapidly than stars with Y$\sim$0.25, thus, for fixed metallicity, age and mass loss in the RGB stage, they produce less massive HB stars, which exhibit bluer colours \citep[e.g.][]{iben_1984,dantona_2002,dantona_2004}. Thus, in single age, monometallic GCs, the  second generation stars (2G), usually helium enhanced, will exhibit bluer colours  than the first generation ones (1G). 
 Nevertheless, it is difficult to disentangle between helium and mass loss from the observed HBs, because increasing the helium content  works in the same direction of  increasing mass loss \citep[e.g.][]{dantona_2002}. 
 Without external constraints, the approach adopted in most HB studies, used in both classical and more recent works \citep[such as][]{dantona_2002,dantona_2004,dantona_2005,caloi_2008,gratton2010,diCriscienzo_2010,dalessandro_2011,dantona_2013,dalessandro_2013,vandenberg_2016,vandenberg_2018} is to estimate both helium abundance and mass loss from the HB itself. Hence the complete set of parameters for each group of stars on the HB is still degenerate. 
 
A way to estimate the helium mass fraction in the 2G stars and break the parameters degeneracy on the HB is to use a theoretical scenario that describes the formation of  multiple populations and predicts their helium contents.
Recent examples are \cite{tailo_2016b,tailo_2017} and \cite{jang_2019}, who use the asymptotic giant branch (AGB) scenario \citep[][and references therein]{dercole_2008,ventura_2010,dercole_2012} and the scenario from \citet[][and references therein]{kim_2018}, respectively. The theoretical uncertainties on the nature of the polluter stars and the dynamical evolution of the populations, however, make this approach still uncertain \citep[see][for a review]{renzini_2015}. 

Recent work proposed a new approach to disentangle between the effect of helium and mass loss along the HB of GCs. \citet{tailo_2019b} studied M\,4, which is one of the simplest GCs in the context of multiple populations.
Indeed, it hosts two distinct groups of 1G and 2G stars 
 that can be identified along the main evolutionary phases \citep{marino_2008, marino_2017}, including the MS and the HB. In particular, the red HB of M\,4 is composed of 1G stars, whereas blue-HB stars belong to the 2G as inferred from high-resolution spectroscopy \citep{marino_2011}. 
Based on multi-band {\it Hubble Space Telescope} ({\it HST}) photometry of MS stars, \citet{tailo_2019b} first obtained accurate determinations of the helium abundances of 1G and 2G stars. Then, they used their helium determinations to fix the helium content of stellar populations along the HB and constrain their mass losses.  Intriguingly, Tailo and collaborators find that 2G stars lose more mass than the 1G and similar conclusions come from a similar investigation on multiple populations in M\,3 \citep{tailo_2019a}. 

The fact that the helium abundances of multiple stellar populations are now available for more than 70 GCs \citep[e.g.][and references therein]{lagioia_2018, lagioia_2019, zennaro_2019, milone_2018, milone_2020} allows to infer the mass loss in a large sample of GCs.

In this work, we extend the method by \citet{tailo_2019b} to a large sample of 46 GCs to estimate, for the first time, the RGB mass loss of their stellar populations. 
The paper is organized as follows. In \S\,\ref{sec:data_general} we present the observations and the theoretical models. \S\,\ref{sec:method} describes the procedure to infer the mass loss of the distinct stellar populations in GCs. Results are presented in \S\,\ref{sec:results} and discussed in \S\,\ref{sec:general_interpret}. Summary and conclusions follow in  \S\,\ref{sec:conc}.

\section{Data and models}
\label{sec:data_general}
To derive the mass loss of GCs, we combine multi-band photometry from the \textit{HST}  UV legacy survey of GCs \citep{piotto_2015}, helium abundances of multiple populations from \citet{milone_2018} and stellar models suitable for GC stars with different helium content \citep[][and references therein]{tailo_2016a}. The following subsections~\ref{sub:data},~\ref{sub:elio} and~\ref{sub:models} describe the photometry, the helium abundances and the theoretical models, respectively.

\subsection{Photometric dataset}\label{sub:data}
To analyze the HBs of GCs we exploited the photometric and astrometric catalogs from the \textit{HST}  UV  legacy survey of GCs \citep{piotto_2015, nardiello_2018}, which include astrometry and photometry in the F275W, F336W, F438W photometric bands of the ultraviolet and visual channel of the wide field camera 3 and in the F606W and F814W bands of the Wide Field Channel of the Advanced Camera for Surveys. We refer to \citet{piotto_2015} and \citet{nardiello_2018} for details on the data set and on the data reduction. Photometry has been corrected for differential reddening as in  \cite{milone_2012c}. 

\subsection{Helium abundances of multiple populations}
\label{sub:elio}

The helium abundances adopted in this study are provided by \citet{milone_2018}. 
These authors used multi-band {\it HST} photometry to analyze the groups of 1G and 2G stars and the group of 2G stars with extreme chemical composition (2Ge) identified by \citet{milone_2017} based on the pseudo two-color diagram called chromosome map \citep{milone_2015}.  They provide the average helium contents of 2G and 2Ge stars, relative to the helium abundance of 1G stars ($\rm \delta Y_{2G,1G}$ and $\rm \delta Y_{max}$). 
We point out that while in our previous studies and in the literature the term 'extreme 2G population' often refers to those stars with very high helium abundances present only in some clusters, here the 2Ge is defined for each cluster as {\it the most extreme 2G population  within that cluster}.

\subsection{Stellar models}
\label{sub:models}

We adopted the stellar-evolution models and the isochrones used by \cite{tailo_2016b,tailo_2017,tailo_2019b,tailo_2019a}, which are obtained with the stellar-evolution program ATON 2.0 by \cite{ventura_1998} and \cite{mazzitelli_1999}.  
We calculated a grid of models with different ages, metallicities (Z), and helium mass fractions (Y). In particular, our models range from [Fe/H]=$-$2.44 to %and 
$-$0.45  and from Y=0.25 to 0.40, thus accounting for the metallicity and helium enhancement values of all studied GCs. The helium mass fraction of the HB models includes the small correction due to the effect of the first dredge up. The HB evolution is followed until the end of the helium burning phase using the recipes of \cite{dantona_2002}.

We compared the $m_{\rm F438W}$ vs.\,$m_{\rm F438W}-m_{F814W}$ CMD of the observed HB of each clusters with a grid of synthetic CMDs derived from the corresponding models following the recipes of \citet[][and references therein]{dantona_2005}.
 In a nutshell, we determine the mass of the each HB star ($\rm M^{HB}$) in the simulations as follows: $\rm M^{HB}=M^{Tip}(Z, Y, A) - \Delta M(\mu,\delta)$. Here $\rm M^{Tip}$ is the stellar mass at the RGB tip, and depends on age (A), metallicity (Z) and helium content (Y); $\rm \Delta M$ is the mass lost by the star, which is described by a Gaussian profile with central value $\mu$ and standard deviation $\delta$. 
 The values of $\rm M^{Tip}$ are obtained from the isochrones database. Once the mass of a HB star has been determined, the program locates it on the models grid via a series of interpolations. The HB age of a star is then extracted from a uniform random distribution ranging from the zero age HB locus (ZAHB) to the end of the core helium burning phase. The uneven distribution with time of the points along the tracks ensures that the evolution speed information is preserved. 
 We simulate the effect of radiative levitation of metals in stars with effective temperatures between 11,500 and 18,000 K by increasing their atmospheric metal content to super solar values (equivalent to [Fe/H]$=$0.2) as suggested by \citet[][and references therein]{brown_2016} and \citet[][and references therein]{tailo_2017}. This process reproduces the effects of the \citet{grundahl_1999} jump.

\section{Deriving the mass loss of multiple populations}
\label{sec:method}
To infer the mass loss of stellar populations, we exploit the procedure introduced by \citet{tailo_2019b,  tailo_2019a},
which is based on the hypothesis that 1G stars mostly populate the reddest part of the HB in CMDs made with optical filter, whereas the  2Ge stars are located on the hottest HB side. As discussed in the introduction, such hypothesis, true for the Type I clusters (as defined in \cite{milone_2017}), is supported both by theoretical arguments \citep[e.g.\,][]{dantona_2002, caloi_2008} and by direct spectroscopic studies of 1G and 2G stars along the HB \citep[e.g.\,][]{marino_2011, marino_2014}.

By limiting the analysis to the easily defined groups at the lowest and highest T$_{\rm eff}$'s of the HB, we avoid in this work the more ambiguous identification and analysis of the intermediate populations, which require an extensive and non homogeneous cluster-by-cluster consideration.

Metal-rich GCs, with an only--red HB provide remarkable exceptions. Indeed, their 1G and 2G HB stars are partially mixed in optical CMDs and appropriate two-color diagrams are needed to identify multiple populations along the HB \citep[][]{milone_2012b}.

In the following, we describe the procedures used to investigate multiple populations in GCs with an HB extending to the blue, and in metal-rich GCs with an only--red HB, by exploiting the recipes by \cite{tailo_2019b,tailo_2019a}. 
The main quantities characterizing HB stars and estimated in this work include mass loss of 1G stars ($\rm \mu_{1G}$), mass loss of 2G stars ($\rm \mu_{2G}$) and of 2G stars with extreme chemical composition (2Ge, $\rm \mu_{2Ge}$). We will also derive the average HB masses of 1G, 2G and 2Ge stars ($\rm \bar{M}^{HB}_{1G}$, $\rm \bar{M}^{HB}_{2G}$ and $\rm \bar{M}^{HB}_{2Ge}$) and the difference between the mass loss of 2Ge and 1G stars, $\rm \Delta \mu_{e}$.
To do this, we consider as test cases two clusters with very different HB morphologies: NGC6752, which exhibits an extended HB and is representative of the majority of the studied GCs, and NGC6637, which has an only--red HB. 

\subsection{Clusters with blue HB: NGC 6752}

\subsubsection{Mass loss of first-generation stars}
\label{sec:method_6752_1G}

\begin{figure*}
    \centering
    \includegraphics[width=1.8\columnwidth,trim={4cm 2cm 3.5cm 1.0cm}]{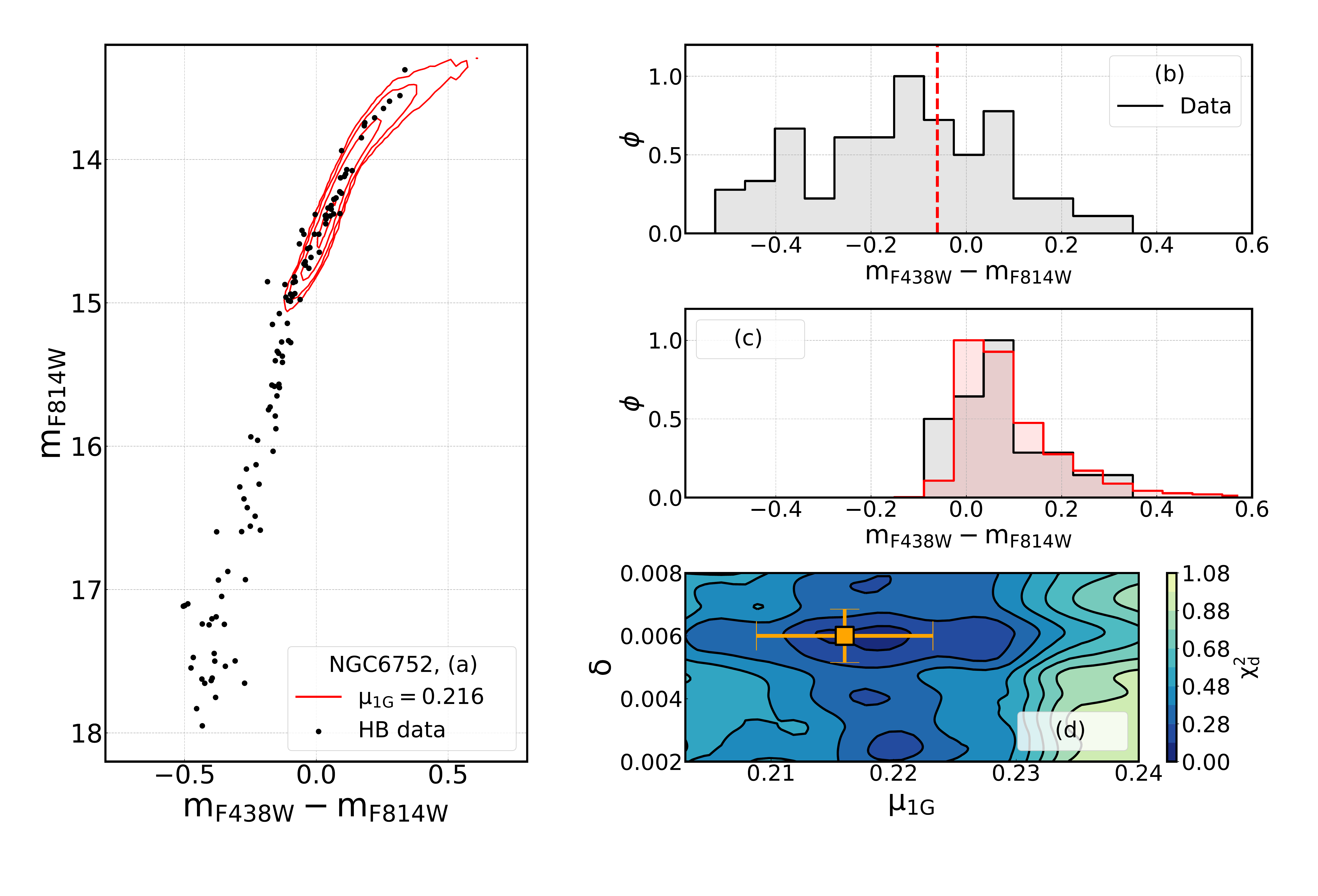}
    \caption{
    This figure illustrates the procedure to infer the mass loss of 1G stars in NGC\,6752. Panel a compares the observed CMD of HB stars (black dots) with the contours of the simulated 1G that provides the best match with the observation.
    The normalized histogram distribution of all HB stars is plotted in panel b, while panel c compares the histogram of simulated 1G stars (red histogram) with the histogram of the observed candidate 1G stars (gray histogram), which comprise HB stars with redder $m_{\rm F438W}-m_{\rm F814W}$ colors than the vertical dashed line plotted in panel b.
    Panel d shows the map of $\chi^2_d$ in the $\delta$ vs.\,$\mu_{\rm 1G}$ plane. The best determination of mass loss and mass-loss spread are indicated by the orange square, while the error bars indicate the uncertainties derived by means of bootstrapping (see text for details).
    }
    \label{pic:method_6752_1g}
\end{figure*}

To derive the mass loss of 1G stars in NGC\,6752 we generate simulated HB CMDs based on a grid of HB tracks for 1G. The tracks have the helium core mass at which the RGB evolution ignites the helium flash, and different masses in the hydrogen envelopes, standard helium abundance (Y=0.25) and the cluster metallicity. 
 
Each simulation adopts the parameters inferred from the best-fit isochrone\footnote{
The isochrone that provides the best fit with the observed CMD is derived as in previous papers from our group \citep[e.g.][]{tailo_2019a}. In a nutshell, we assumed Y=0.25, $\rm [\alpha/Fe]=+0.4$ and adopted the value of [Fe/H], distance modulus and reddening provided by the 2010 version of the \citet[][]{harris_1996} catalog. We produced a set of appropriate isochrones with ages between 10.0 and 14.5 Gyr in steps of 0.25\,Gyr. 
The best age determination is given by the isochrone that provides the best match with the region of the $\rm m_{F438W}-m_{F606W}\, vs. \,m_{F606W}$ CMD around the MS turn off ,  
the corresponding uncertainty corresponds to the range of ages that allow the isochrones to envelope 68.27\% of the turn off stars. We derive for NGC\,6752 an age of 13.00$\pm$0.50 Gyr.
} 
for the RGB mass, the mass lost during the RGB evolution 
 ($\rm \mu_{\rm 1G}$) 
 and a mass-loss spread ($\delta$).
A large grid of simulations is built, by varying $\rm \mu_{\rm 1G}$ from 0.100 to 0.310 $M_\odot$ in steps of 0.003 $M_\odot$ and $\delta$ from 0.002$M_\odot$ to 0.008$M_\odot$ in steps of 0.001$M_\odot$.

The normalized histogram color distribution of each simulation is compared with the corresponding  color distribution of candidate 1G stars of NGC\,6752, by means of the $\chi$-square distance \citep[hereinafter $\rm \chi^2_d$, see e.g.][]{chisq_cite}:  

$\rm \chi^2_d=0.5 \times\sum_{i}\dfrac{(p_i-q_i)^2}{(p_i+q_i)}$

Where $\rm p_i$ and $q_i$ are the values of each bin of the observed and simulated histogram, respectively.
Candidate 1G stars include the observed HB stars redder than $\rm \overline{col}_{sim} - 1.5\times\sigma_{\rm col,sim}$, where $\rm \overline{col}_{sim}$ is the average $m_{\rm F438W}-m_{F814W}$ color of simulated stars and $\sigma_{\rm col,sim}$ is the corresponding standard deviation.

To qualitatively discuss the effect of changing mass loss on the simulation, we note that, as the adopted value of $\rm \mu_{1G}$ increases (decreases), simulated HB stars with fixed mass-loss spread move towards bluer (redder) average colours. Hence, the blue boundary of the observed 1G candidates is also blue-(red-) shifted and the comparison between the distributions of simulated and observed histograms would involve progressively bluer (redder) observed stars. 
As a consequence, too-high values of $\rm \mu_{1G}$ would result into simulated HBs that are bluer than the bulk of observed 1G stars. On the other side,  too-small values of $\rm \mu_{1G}$ would provide HB stars that are redder than all observed HB stars. Both situations would provide high $\chi^{2}_{d}$ values.

In a similar way as $\rm \delta$ increases (decreases) the simulations span a wider (narrower) range of $\rm M^{HB}$ values thus covering a larger (smaller) portion of the theoretical HB. 
The value of $\sigma_{\rm col,sim}$ increases (diminishes) accordingly and so does the portion of observed HB that the simulation covers. 
If we are using a value of $\rm \delta$ too high (low), stars also belonging to the 2G might be included in the comparison (or stars belonging to the 1G might be excluded).

The best estimates of mass loss and mass-loss spread for 1G stars are given by the values of $\mu_{1G}$ and $\delta$ of the simulations that provide the minimum $\rm \chi^2_d$.  
Uncertainties are estimated by means of bootstrapping. Specifically, we  generated 5,000 realizations of the HB in NGC\,6752 and estimated $\rm \mu_{1G}$ and $\delta$ by using the procedure described above. The uncertainties correspond to the standard deviations of the 5,000 values of $\rm \mu_{1G}$ and $\delta$.
We obtain for NGC\,6752 $\rm \mu_{1G}=0.216\pm 0.007$ and  $\delta=0.006\pm 0.001 M_\odot$.

We derived the average mass of 1G stars along the HB ($\rm \bar{M}_{1G}^{HB}$) by subtracting from the mass of 1G stars at the RGB tip provided by the best-fit isochrone, $\rm M_{1G}^{Tip}=0.814 M_\odot$, the average mass loss of the 1G.  We find $\rm \bar{M}_{1G}^{HB}=0.598\pm 0.007M_\odot$ for NGC\,6752.

Results are listed in Table \ref{tab:param_dm} and illustrated in Figure~\ref{pic:method_6752_1g}, where we compare the observed $m_{\rm F814W}$ vs.\,$m_{\rm F438W}-m_{\rm F814W}$ CMD of HB stars in NGC\,6752 with the contours of the best-fit simulated 1G (panel a). For completeness, we also show the normalized histogram distribution of  the colors of HB stars (Figure~\ref{pic:method_6752_1g}b) and the comparison between the normalized histogram distributions of simulated stars and the candidate 1G stars (Figure~\ref{pic:method_6752_1g}c, which includes HB stars redder than the vertical dashed line plotted in panel b).
Finally, we show in Figure~\ref{pic:method_6752_1g}d the $\rm \chi^2_d$ map in the $\rm \mu_{1G}$ -- $\rm \delta$ plane, where the orange square marks the  best determinations of $\rm \mu_{1G}$  and $\rm \delta$.

\begin{figure}
    \centering
    %placeholders
    \includegraphics[width=0.9\columnwidth,trim={0cm 0.0cm 0cm 0cm}]{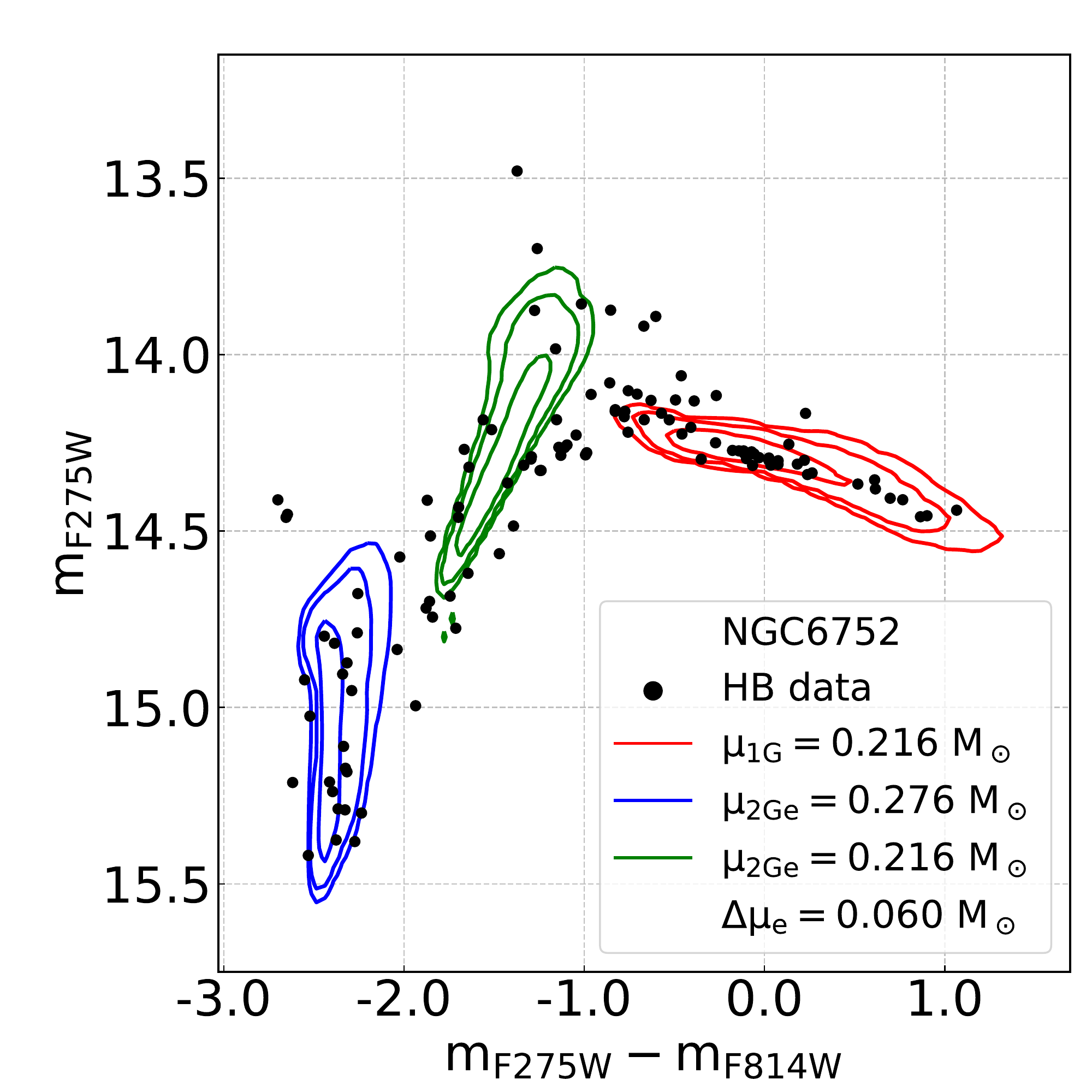}
    \caption{
     The $\rm m_{F275W}-m_{F814W}$ vs. $\rm m_{F275W}$ CMD for NGC\,6752 where we overplot our best fit simulations as contour plots, with red and blue respectively for the 1G and the 2Ge. The green contour plot represents the 2Ge simulation but with the same mass loss value of the 1G, as indicated in the label. 
    }
    \label{pic:checksim}
\end{figure}

As a sanity check, we have verified that the reddest group of stars has been correctly identified by also comparing the simulation with the data in the $\rm m_{F275W}-m_{F814W}$ vs. $\rm m_{F275W}$ CMD (see Figure \ref{pic:checksim}).

\subsubsection{Mass loss of extreme second-generation stars}
\label{sec:method_6752_2Ge}

\begin{figure*}
    \centering
    \includegraphics[width=1.8\columnwidth,trim={4cm 2cm 3.5cm 1.0cm}]{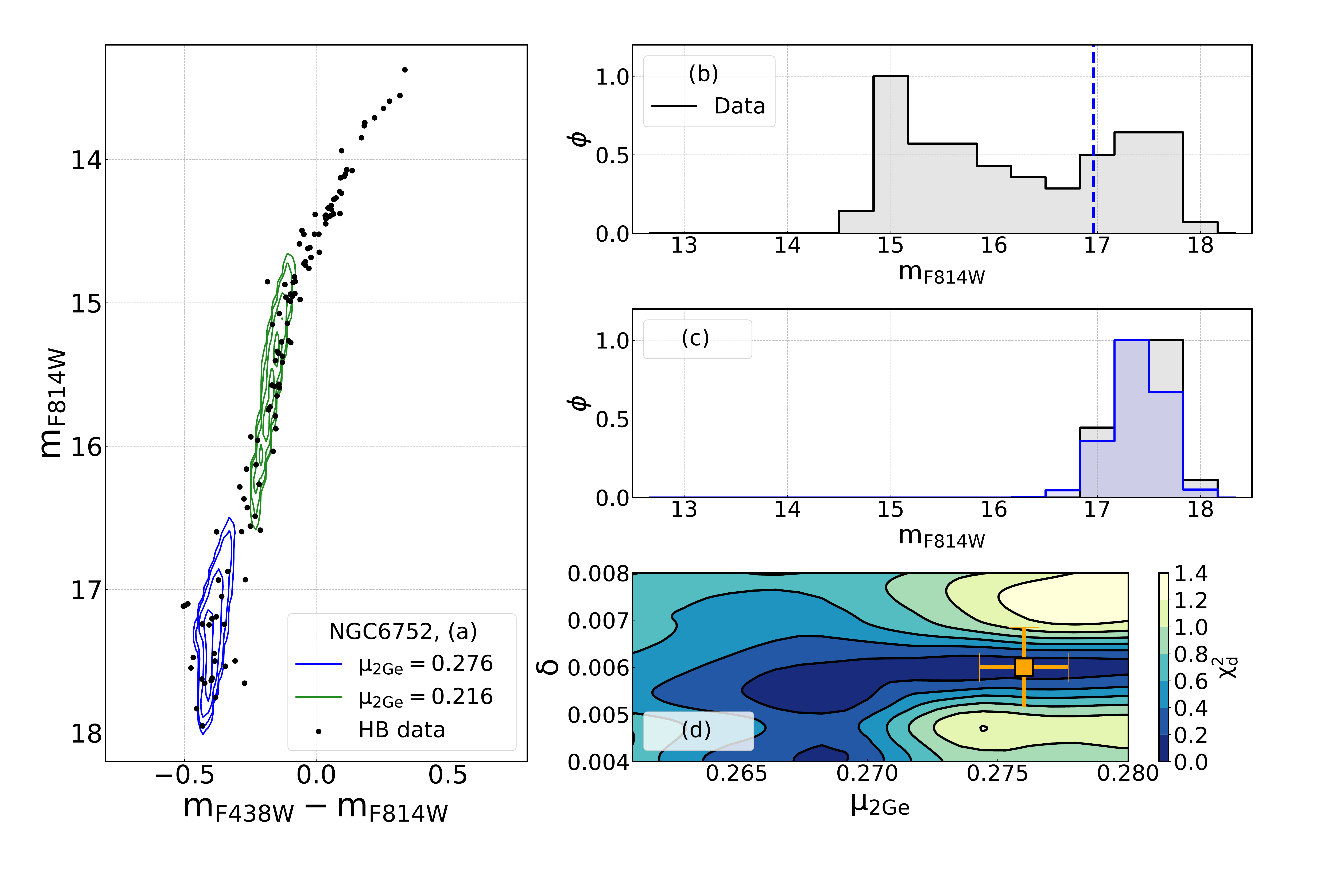}
    \caption{
    Procedure to derive the mass loss of 2Ge HB stars.  Panels a and b show the observed CMD of all HB stars (black points) and the histogram distributions of the F814W magnitude, respectively. The black histogram plotted in panel c shows the magnitude distributions of candidate 2Ge stars (i.e., stars fainter than the vertical dashed line plotted in panel b) and is compared with the distribution of the best-fit simulated HB 2Ge stars (blue histogram).  
    The blue contours shown in the panel-a CMD correspond the best-fit simulation of 2Ge stars, while the green contours correspond to a simulation with the same mass loss as 1G stars. 
     The map of $\rm \chi^2_d$ in the $\delta$ vs.\,$\mu_{\rm 1G}$ plane is plotted in panel d, where the orange square mark the best determination of mass loss and mass-loss spread. Error bars indicate the uncertainties obtained from  bootstrapping (see text for details).
    }
    \label{pic:method_6752_2ge}
\end{figure*}

The procedure to infer the mass loss of 2Ge stars is similar to the method described in the previous subsection for the 1G, but is based on the assumption that 2Ge stars populate the bluest, faintest tail of the HB.

We generate a grid of simulated HBs of 2Ge stars by assuming the same parameters (age, metallicity, mass loss and mass-loss spread) used for the 1G but different helium content. Specifically, the value of Y in the HB tracks is increased by the amount inferred by \citet{milone_2018} for the extreme population, which is Y=0.292 for NGC\,6752. 
 This same helium abundance is used to infer the 2Ge mass evolving on the RGB at the cluster age of 13\,Gyr.
 The normalized histogram distribution in $m_{\rm F814W}$\ of each simulation is compared with the corresponding magnitude distribution of the observed candidate 2Ge HB stars, by means of the $\chi$-square distance. 
Candidate 2Ge stars include HB stars fainter than $\rm \overline{mag}_{sim} - 1.5\times\sigma_{\rm mag,sim}$, where $\rm \overline{mag}_{sim}$ and $\sigma_{\rm mag,sim}$ are the mean and standard deviation, respectively, of the F814W magnitudes of observed HB stars.

As the value of $\rm \mu_{2Ge}$ increases (decreases), $\rm \overline{mag}_{sim}$\ and the whole simulation move toward higher (lower) magnitudes, describing a bunch of  progressively fainter (brighter) stars.  On the other hand if the value of $\rm \delta$ increases (decreases) the simulations overlap larger (smaller) sections of the HB.  
Considerations similar to those of the 1G case hold here, thus the value of $\rm \chi^2_d$ increases as the agreement between the two histograms worsens. The best estimates of the mass-loss of 2Ge stars $\rm \mu_{2Ge}$ and $\delta$ for 2Ge stars are derived by means of $\rm \chi^2_d$ minimization and the corresponding uncertainties are estimated by means of bootstrapping, in close analogy with what we did for the 1G.  
We find for 2Ge stars of NGC\,6752 $\rm \mu_{2Ge}=0.276\pm 0.002$ and $\rm \delta = 0.006\pm 0.001 M_\odot$. 

The comparison between the observed CMD of NGC\,6752 (black dots) and the simulated CMD of 2Ge stars that provides the best fit (blue contours) is shown in Figure~\ref{pic:method_6752_2ge}a.
The result thus requires a 2Ge mass loss larger by $\rm \mu_{2Ge}-\mu_{1G}=0.060M_\odot$\ than the 1G mass loss. In fact, if we simulate the CMD of 2Ge stars by assuming for both generations  the mass loss value inferred from 1G stars ($\rm \mu_{2Ge} =\mu_{1G}=0.216M_\odot$), the 2Ge group does not reproduce correctly the location of the extreme HB stars in NGC\,6752 (green contours in Figure~\ref{pic:method_6752_2ge}a). Consequently, we conclude that the 2Ge stars of NGC\,6752 lose more mass than the 1G.

The average mass of the 2Ge HB stars is the mass at the RGB tip $\rm M_{2Ge}^{Tip}=0.756 M_\odot$, derived by the best fit isochrone with Y$=$0.292, minus the best fit mass lost 
$\mu_{2Ge}=0.276M_\odot$, that is
$\rm \bar{M}_{2Ge}^{HB}=0.480 \pm 0.002 M_\odot$.

To illustrate the main steps towards determining the mass loss of 2Ge stars, we also show in Figure~\ref{pic:method_6752_2ge} the histogram distribution of $m_{\rm F814W}$ for HB stars (panel b), the comparison between the histograms of candidate 2Ge stars and the best-fit simulation (panel c) and the $\rm \chi^2_d$ map contour map in the $\delta$ vs.\,$\rm \mu_{2Ge}$ plane. As a sanity check, we verify that the algorithm has correctly identified the extreme HB stars by comparing the simulation with the data in the $\rm m_{F275W}-m_{F814W}$ vs. $\rm m_{F275W}$ CMD (see Figure \ref{pic:checksim}).
Finally, in Figure \ref{pic:checksim} and \ref{pic:method_6752_2ge}a we show, for completeness, the simulation of the 2Ge HB stars obtained with the same mass loss we found for the 1G (thus $\rm \mu_{2Ge} =\mu_{1G}=0.216\, M_\odot$). This is clearly a bad simulation of the 2Ge, as the green curves do not reach the faintest part of the locus. 

\subsubsection{Impact of age, metallicity and helium uncertainties on mass loss}
\label{sec:method_6752_obsunc}

To quantify the impact of the uncertainties of age, metallicity and helium abundances (hereafter $\rm \sigma_A$, $\rm \sigma_{Fe}$ and $\sigma_{Y}$) on mass loss determinations, we followed the recipe by \cite{tailo_2019b}. For simplicity, we assumed $\delta=0.006 M_\odot$.

To investigate the effect of age uncertainties, we repeated the procedures described in Section~\ref{sec:method_6752_1G} and ~\ref{sec:method_6752_2Ge} to derive the mass losses of 1G and 2Ge stars by adopting ages that differ from the best-fit age by $\pm \rm \sigma_A$ (i.e.\, 12.50\,Gyr and 13.50\,Gyr for NGC\,6752).  We find that a change in age by $\pm 0.50$ Gyr, corresponds to a variation of $\mp \rm 0.013 M_\odot$ in both $\rm \mu_{1G}$ and $\mu_{2Ge}$. Hence, age uncertainties provide negligible effect on the difference between the mass loss of 1G and 1G stars ($\Delta \mu_{e}$).
 
Similarly, to account for [Fe/H] uncertainties we estimated the mass losses of 1G and 2Ge stars by assuming iron abundances that differ from the value provided by \citet{harris_1996} by $\pm \rm \sigma_{Fe}$. A difference of $\pm 0.1$ dex, which is the typical error on [Fe/H] inferred from spectroscopy, affects $\rm \mu_{1G}$ and $\mu_{2Ge}$ by 0.017 and 0.010 $\rm M_\odot$, respectively. Hence, the adopted uncertainty on iron abundance have a small impact on $\Delta \mu_{e}$ by 0.007 $\rm M_\odot$.   
 
Finally, we considered the impact of helium uncertainties on mass loss by changing the helium abundance of 2Ge stars by $\pm \rm \sigma_{Y}$. We find that an helium variation of $\pm \rm 0.004$, as inferred by \citet{milone_2018} for 2Ge stars of NGC\,6752, corresponds to a mass-loss change of $\mp$0.007$\rm M_\odot$ on both $\mu_{2Ge}$ and $\Delta \mu_{e}$.     

We conclude that, by combining in quadrature the effects of age, iron abundance and helium content uncertainties, together with the one obtained from the bootstrapping procedure, our best estimate of the mass loss in 1G stars is $\mu_{1G}=0.216\pm0.022$, the estimate of mass loss in 2Ge stars $\mu_{2Ge}=0.276\pm0.023$ and, finally, the mass loss difference in NGC\,6752 is $\Delta \mu_e= 0.060\pm0.017 M_\odot$. 

\subsection{Clusters with no blue HB: the case of NGC 6637}
\label{sec:method_fehrich}

\begin{figure*}
    \centering
    \includegraphics[width=1.8\columnwidth,trim={2.5cm 2.5cm 1.8cm 1.0cm}]{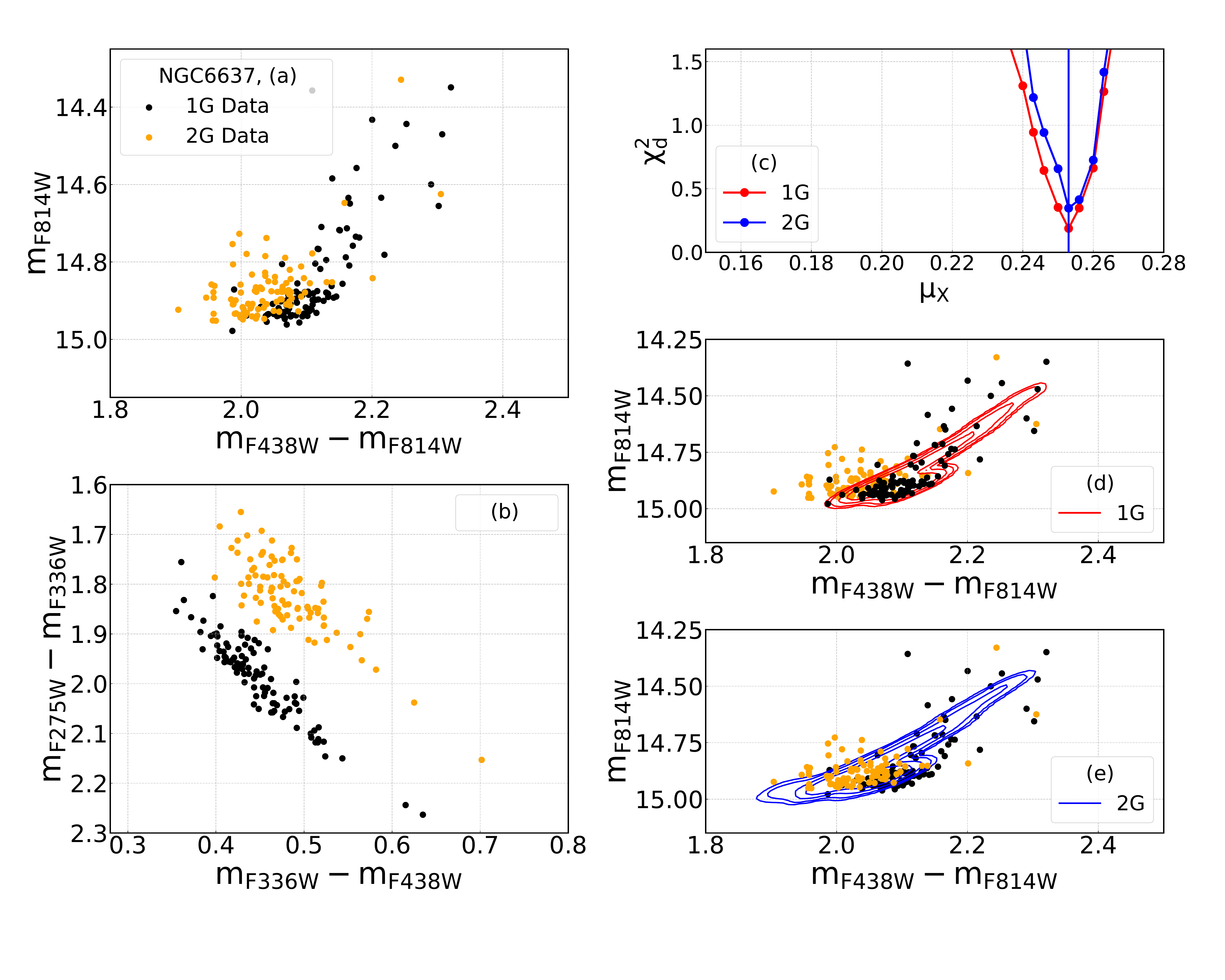}
    \caption{
 Procedure to derive the mass loss of 1G and 2G stars of NGC\,6637. Panels a and b show the $m_{\rm F814W}$ vs.\,$\rm m_{F438W}-m_{F814W}$ CMD and the $\rm m_{F275W}-m_{F336W}$ vs.\,$\rm m_{F336W}-m_{F438W}$ two-color diagram of HB stars.  Panel c shows the $\chi^2_d$ profiles for the mass loss of 1G (red) and 2G stars (blue), while in panels d and e we superimpose on the observed CMD the contours of 1G and 2G stars that correspond to the best-fit simulations.
Observed 1G and 2G stars selected in the two-color diagram of panel b are colored  black and yellow, respectively. 
    }
    \label{pic:method_6637}
\end{figure*}

1G and 2G stars of metal-rich GCs with only-red HB are mostly mixed in CMDs made with optical filters as illustrated in Figure~\ref{pic:method_6637}a for NGC\,6637. 
Hence, we exploited the $\rm  m_{F275W}-m_{F336W}$ vs.\,$\rm m_{F336W}-m_{F438W}$  diagram, where 1G and 2G stars define distinct sequences, to identify multiple populations along the red HB \citep[][see Figure ~\ref{pic:method_6637}b for NGC\,6637]{milone_2012b}. 
Since this two-color diagram does not provide clear separation among 2Ge stars and the remaining 2G stars, we limit the investigation to the entire sample of 2G stars. 

To estimate the mass loss of both 1G and 2G stars in GCs with the red HB alone, we used the procedure adopted for 1G stars of NGC\,6752 (see Section~\ref{sec:method_6752_1G}), thus analysing the $\rm m_{F438W}-m_{F814W}$ colour distribution of the HB stars. This is necessary because, when the HB stars are all on the red side, a large number of simulations occupy the same magnitude level. An additional remarkable difference is that we adopted the helium abundance of 2G stars inferred by \citet{milone_2018}. 

Results are summarized in the right panels of Figure~\ref{pic:method_6637}. We show  the $\chi$-square (see Equation 1) resulting from the simulations with  $\delta=0.003$, corresponding to the position of the minima,
against the mass loss of 1G and 2G stars (panel c), and compare the contours of the simulations of 1G and 2G stars that correspond to the minimum $\chi^{2}_{\rm d}$ with the observed CMDs (panels c and d).
Noticeably, both 1G and 2G stars of the HB in NGC\,6637 are described by assuming the same mass loss $\rm \mu_{1G}=\mu_{2G} \sim 0.253 M_\odot$. The uncertainties on these values are evaluated with the procedure in subsection\,\ref{sec:method_6752_obsunc} and are reported in Table\,\ref{tab:param_dm}.

\begin{table*}
    \centering
    \caption{ 
    Main GC parameters used in this work. For each cluster we provide: ID, average reddening, E(B-V), distance module ($\rm (m-M)_V$) and [Fe/H] \citep[from the 2010 version of the][catalogue]{harris_1996}, cluster ages (this work), helium difference between 2G and 1G stars ($\rm \Delta Y_{2G,1G}$) and maximum helium abundance variation \citep[from][]{milone_2018} and GC group \citep[adapted from][see Section \ref{sec:corr_1g}]{milone_2014}.
    }
	\begin{adjustbox}{width=1.95\columnwidth,center}
    \renewcommand{\arraystretch}{1.2} 
    \begin{tabular}{lcccccccccccc}
    \hline
    \hline
    Cluster Names& E(B-V) (mag)&$\rm (m-M)_V (mag)$& [Fe/H]& $\rm [\alpha/Fe]$&Age (Gyr) &  $\rm \delta Y_{2G,1G}$& $\rm \delta Y_{max}$&Group\\
    \hline
    NGC0104, 47 Tuc& 0.04& 13.37& -0.72& 0.4& 13.00$\pm 0.75$ &  0.011 $\pm$ 0.005 & 0.049 $\pm$ 0.005&M3-like\\
    NGC0288& 0.03& 14.84& -1.32& 0.4& 12.75$\pm 0.50$ &  0.015 $\pm$ 0.010 & 0.016 $\pm$ 0.012&M13-like\\
    NGC2298& 0.14& 15.60& -1.92& 0.4& 13.25$\pm 0.50$ & -0.003 $\pm$ 0.008 & 0.011 $\pm$ 0.012&M13-like\\
    NGC2808& 0.22& 15.59& -1.14& 0.4& 12.00$\pm 0.75$ &  0.048 $\pm$ 0.005 & 0.124 $\pm$ 0.007&M3-like\\
    NGC3201& 0.24& 14.20& -1.59& 0.4& 12.00$\pm 0.50$ & -0.001 $\pm$ 0.013 & 0.028 $\pm$ 0.032&M3-like\\
    NGC4590, M\,68& 0.05& 15.21& -2.23& 0.4& 12.75$\pm 0.75$ &  0.007 $\pm$ 0.009 & 0.012 $\pm$ 0.009&M3-like\\
    NGC4833& 0.32& 15.08& -1.85& 0.4& 13.25$\pm 0.50$ &  0.016 $\pm$ 0.008 & 0.051 $\pm$ 0.009&M3-like\\
    NGC5024, M\,53& 0.02& 16.32& -2.10& 0.4& 13.25$\pm 0.75$ &  0.013 $\pm$ 0.007 & 0.044 $\pm$ 0.008&M3-like\\
    NGC5053& 0.01& 16.23& -2.27& 0.4& 13.00$\pm 0.50$ & -0.002 $\pm$ 0.013 & 0.004 $\pm$ 0.025&M3-like\\
    NGC5466& 0.00& 16.02& -1.98& 0.4& 12.75$\pm 0.50$ &  0.002 $\pm$ 0.013 & 0.007 $\pm$ 0.024&M3-like\\ 
    NGC5904, M\,5& 0.03& 14.46& -1.29& 0.4& 12.25$\pm 0.75$ &  0.012 $\pm$ 0.004 & 0.037 $\pm$ 0.007&M3-like\\
    NGC5927& 0.45& 15.82& -0.49& 0.2& 12.25$\pm 0.50$ &  0.011 $\pm$ 0.004 & 0.055 $\pm$ 0.015&M3-like\\
    NGC5986& 0.28& 15.96& -1.59& 0.4& 13.25$\pm 0.50$ &  0.005 $\pm$ 0.006 & 0.031 $\pm$ 0.012&M13-like\\
    NGC6093, M\,80& 0.18& 15.56& -1.75& 0.4& 13.50$\pm 0.75$ &  0.011 $\pm$ 0.008 & 0.027 $\pm$ 0.012&M13-like\\
    NGC6101& 0.05& 16.10& -1.98& 0.4& 12.75$\pm 0.50$ &  0.005 $\pm$ 0.010 & 0.017 $\pm$ 0.011&M13-like\\
    NGC6144& 0.36& 15.86& -1.76& 0.4& 13.00$\pm 0.50$ &  0.009 $\pm$ 0.011 & 0.017 $\pm$ 0.013&M13-like\\
    NGC6171, M\,107& 0.33& 15.05& -1.00& 0.4& 12.50$\pm 0.25$ &  0.019 $\pm$ 0.011 & 0.024 $\pm$ 0.014&M3-like\\
    NGC6205, M\,13& 0.02& 14.33& -1.52& 0.4& 12.25$\pm 0.75$ &  0.020 $\pm$ 0.004 & 0.052 $\pm$ 0.004&M13-like\\
    NGC6218, M\,12& 0.19& 14.01& -1.37& 0.4& 13.00$\pm 0.50$ &  0.009 $\pm$ 0.007 & 0.011 $\pm$ 0.011&M13-like\\
    NGC6254, M\,10& 0.28& 14.08& -1.56& 0.4& 12.75$\pm 0.50$ &  0.006 $\pm$ 0.008 & 0.029 $\pm$ 0.011&M13-like\\
    NGC6304& 0.54& 15.52& -0.45& 0.2& 12.50$\pm 0.50$ &  0.008 $\pm$ 0.005 & 0.025 $\pm$ 0.006&M3-like\\
    NGC6341, M\,92& 0.02& 14.65& -2.31& 0.4& 13.50$\pm 0.75$ &  0.022 $\pm$ 0.004 & 0.039 $\pm$ 0.006&M3-like\\
    NGC6352& 0.22& 14.43& -0.64& 0.2& 12.75$\pm 0.25$ &  0.019 $\pm$ 0.014 & 0.027 $\pm$ 0.006&M3-like\\
    NGC6362& 0.09& 14.68& -1.00& 0.4& 12.25$\pm 0.50$ &  0.004 $\pm$ 0.011 & 0.004 $\pm$ 0.011&M3-like\\
    NGC6366& 0.71& 14.94& -0.59& 0.2& 12.25$\pm 0.50$ &  0.011 $\pm$ 0.010 & 0.011 $\pm$ 0.015&M3-like\\
    NGC6397& 0.18& 12.37& -2.00& 0.4& 13.00$\pm 0.50$ &  0.006 $\pm$ 0.009 & 0.008 $\pm$ 0.011&M13-like\\
    NGC6441& 0.47& 16.78& -0.46& 0.2& 12.25$\pm 0.75$ &  0.029 $\pm$ 0.006 & 0.081 $\pm$ 0.022&M3-like\\
    NGC6496& 0.15& 15.74& -0.46& 0.2& 12.00$\pm 0.50$ &  0.009 $\pm$ 0.011 & 0.021 $\pm$ 0.006&M3-like\\
    NGC6535& 0.34& 15.34& -1.79& 0.4& 13.25$\pm 0.50$ &  0.003 $\pm$ 0.021 & 0.003 $\pm$ 0.022&M13-like\\
    NGC6541& 0.14& 14.82& -1.81& 0.4& 13.00$\pm 0.50$ &  0.024 $\pm$ 0.005 & 0.045 $\pm$ 0.006&M13-like\\
    NGC6584& 0.10& 15.96& -1.50& 0.4& 12.25$\pm 0.75$ &  0.000 $\pm$ 0.007 & 0.015 $\pm$ 0.011&M3-like\\
    NGC6624& 0.28& 15.36& -0.44& 0.2& 12.50$\pm 0.50$ &  0.010 $\pm$ 0.004 & 0.022 $\pm$ 0.003&M3-like\\
    NGC6637& 0.18& 15.28& -0.64& 0.2& 12.25$\pm 0.50$ &  0.004 $\pm$ 0.006 & 0.011 $\pm$ 0.005&M3-like\\
    NGC6652& 0.09& 15.28& -0.81& 0.4& 13.25$\pm 0.25$ &  0.008 $\pm$ 0.007 & 0.017 $\pm$ 0.011&M3-like\\
    NGC6681, M\,70& 0.07& 14.99& -1.62& 0.4& 13.25$\pm 0.75$ &  0.009 $\pm$ 0.008 & 0.029 $\pm$ 0.015&M13-like\\
    NGC6717, Pal 9& 0.22& 14.94& -1.28& 0.4& 13.00$\pm 0.25$ &  0.003 $\pm$ 0.006 & 0.003 $\pm$ 0.009&M13-like\\
    NGC6723& 0.05& 14.84& -1.10& 0.4& 13.00$\pm 0.75$ &  0.005 $\pm$ 0.006 & 0.024 $\pm$ 0.007&M3-like\\
    NGC6752& 0.04& 13.13& -1.54& 0.4& 13.00$\pm 0.50$ &  0.015 $\pm$ 0.005 & 0.042 $\pm$ 0.004&M13-like\\
    NGC6779, M\,56& 0.26& 15.68& -1.99& 0.4& 13.50$\pm 0.50$ &  0.011 $\pm$ 0.007 & 0.031 $\pm$ 0.008&M13-like\\
    NGC6809, M\,55& 0.08& 13.89& -1.94& 0.4& 13.25$\pm 0.50$ &  0.014 $\pm$ 0.008 & 0.026 $\pm$ 0.015&M13-like\\
    NGC6838, M\,71& 0.25& 13.80& -0.78& 0.4& 12.75$\pm 0.50$ &  0.005 $\pm$ 0.009 & 0.024 $\pm$ 0.010&M3-like\\
    NGC6981, M\,72& 0.05& 16.31& -1.42& 0.4& 12.25$\pm 0.75$ &  0.011 $\pm$ 0.006 & 0.017 $\pm$ 0.006&M3-like\\
    NGC7078, M\,15& 0.10& 15.39& -2.37& 0.4& 13.00$\pm 0.75$ &  0.021 $\pm$ 0.009 & 0.069 $\pm$ 0.006&M3-like\\
    NGC7099, M\,30& 0.03& 14.64& -2.27& 0.4& 13.50$\pm 0.50$ &  0.015 $\pm$ 0.010 & 0.022 $\pm$ 0.010&M13-like\\
    \hline     \hline
    \end{tabular}
    \end{adjustbox}
    \label{tab:param_sim}
\end{table*}

\begin{table*}
    \centering
    \caption{
    This table lists the main quantities derived in this paper, including  
    stellar mass at the RGB tip ($\rm M^{Tip}$), mass loss ($\mu$), average mass on HB stars ($\rm M^{HB}$) for 1G, 2Ge and 2G stars.  We also provide the values of the Reimers' parameter $\rm \eta_R$ inferred for each cluster, the mass-loss difference between 2Ge and 1G stars ($\Delta \mu_e$) and the internal mass loss spread ($\delta$). We point out that, although evaluated independently, we obtain equal values of $\delta$ for both the 1G and the 2Ge in all the examined GCs. We therefore report a single value for this quantity in the table. The error values of these quantities, where appropriate, are obtained from the combination in quadrature of all different sources (see text). (a) From \citet{tailo_2019a},(b) From \citet{tailo_2019b}.
    }
    \begin{adjustbox}{width=2\columnwidth,center}
    \renewcommand{\arraystretch}{1.20} 
    \begin{tabular}{lccccccccc}
    \hline
    \hline
    ID  &$\rm M^{Tip}_{1G}/M_\odot$&$\rm \mu_{1G}/M_\odot$ &$\rm \bar{M}^{HB}_{1G}/M_\odot$& $\rm \eta_{R,1G}$&$\rm M^{Tip}_{2Ge}/M_\odot$&$\rm \mu_{2Ge}/M_\odot$ &$\rm \bar{M}^{HB}_{2Ge}/M_\odot$&$\rm \Delta \mu_{e}/M_\odot$&$\delta/M_\odot$\\
    \hline
    NGC0288         & 0.817& 0.213$\pm$0.021& 0.604$\pm$0.021   &0.516$\pm$0.021& 0.795& 0.246$\pm$0.025& 0.546$\pm$0.025& 0.033$\pm$0.020   & 0.004$\pm0.001$\\
    NGC2298         & 0.789& 0.116$\pm$0.015& 0.676$\pm$0.015   &0.278$\pm$0.020& 0.775& 0.156$\pm$0.023& 0.619$\pm$0.023& 0.040$\pm$0.023   & 0.004$\pm0.001$\\
    NGC2808         & 0.828& 0.113$\pm$0.024& 0.715$\pm$0.024   &0.242$\pm$0.032& 0.661& 0.216$\pm$0.025& 0.445$\pm$0.025& 0.103$\pm$0.012   & 0.005$\pm0.001$\\
    NGC3201         & 0.821& 0.116$\pm$0.024& 0.705$\pm$0.024   &0.270$\pm$0.033& 0.782& 0.139$\pm$0.044& 0.642$\pm$0.044& 0.023$\pm$0.042   & 0.005$\pm0.001$\\
    NGC4590, M\,68	& 0.794& 0.043$\pm$0.020& 0.751$\pm$0.020   &0.120$\pm$0.027& 0.778& 0.063$\pm$0.020& 0.715$\pm$0.020& 0.020$\pm$0.026   & 0.006$\pm0.001$\\
    NGC4833         & 0.791& 0.123$\pm$0.016& 0.668$\pm$0.016   &0.306$\pm$0.022& 0.724& 0.193$\pm$0.022& 0.531$\pm$0.022& 0.070$\pm$0.025   & 0.003$\pm0.001$\\
    NGC5024, M\,53	& 0.790& 0.100$\pm$0.017& 0.690$\pm$0.017   &0.263$\pm$0.023& 0.737& 0.120$\pm$0.019& 0.617$\pm$0.019& 0.020$\pm$0.013   & 0.006$\pm0.001$\\
    NGC5053			& 0.790& 0.116$\pm$0.014& 0.674$\pm$0.014   &0.320$\pm$0.019& ---& ---          & ---          & ---             & 0.004$\pm0.001$\\
    NGC5272, M\,3$^{(a)}$  & 0.847& 0.188$\pm$0.017& 0.659$\pm$0.017   &0.459$\pm$0.023& 0.789& 0.240$\pm$0.022& 0.550$\pm$0.022& 0.052$\pm$0.014   & 0.005$\pm$0.001	 \\
    NGC5466			& 0.798& 0.103$\pm$0.017& 0.695$\pm$0.017   &0.264$\pm$0.023& 0.788& 0.119$\pm$0.023& 0.669$\pm$0.023& 0.016$\pm$0.022   & 0.002$\pm0.001$\\
    NGC5904, M\,5	& 0.833& 0.176$\pm$0.021& 0.657$\pm$0.021   &0.404$\pm$0.029& 0.782& 0.216$\pm$0.022& 0.566$\pm$0.022& 0.040$\pm$0.015   & 0.006$\pm0.001$\\
    NGC5986         & 0.798& 0.170$\pm$0.019& 0.628$\pm$0.019   &0.419$\pm$0.026& 0.756& 0.263$\pm$0.024& 0.493$\pm$0.024& 0.093$\pm$0.025   & 0.003$\pm0.001$\\
    NGC6093, M\,80  & 0.791& 0.156$\pm$0.021& 0.635$\pm$0.021   &0.397$\pm$0.029& 0.755& 0.266$\pm$0.027& 0.489$\pm$0.027& 0.110$\pm$0.020   & 0.003$\pm0.001$\\
    NGC6101         & 0.798& 0.110$\pm$0.015& 0.688$\pm$0.015   &0.282$\pm$0.020& 0.775& 0.120$\pm$0.019& 0.655$\pm$0.019& 0.010$\pm$0.015   & 0.006$\pm0.001$\\
    NGC6121, M\,4$^{(b)}$ & 0.850& 0.209$\pm$0.024& 0.624$\pm$0.024   &0.481$\pm$0.023& 0.833& 0.236$\pm$0.027& 0.597$\pm$0.027& 0.027$\pm$0.007   & 0.006$\pm$0.001	 \\
    NGC6144         & 0.797& 0.166$\pm$0.019& 0.631$\pm$0.019   &0.429$\pm$0.026& 0.775& 0.176$\pm$0.026& 0.599$\pm$0.026& 0.010$\pm$0.027   & 0.002$\pm0.001$\\
    NGC6171, M\,107 & 0.859& 0.230$\pm$0.025& 0.629$\pm$0.025   &0.528$\pm$0.034& 0.823& 0.243$\pm$0.028& 0.580$\pm$0.028& 0.013$\pm$0.017   & 0.006$\pm0.001$\\
    NGC6205, M\,13  & 0.821& 0.210$\pm$0.020& 0.611$\pm$0.020   &0.526$\pm$0.027& 0.750& 0.273$\pm$0.021& 0.477$\pm$0.021& 0.063$\pm$0.015   & 0.003$\pm0.001$\\
    NGC6218, M\,12  & 0.815& 0.223$\pm$0.023& 0.592$\pm$0.023   &0.540$\pm$0.032& 0.799& 0.270$\pm$0.029& 0.559$\pm$0.029& 0.047$\pm$0.022   & 0.003$\pm0.001$\\
    NGC6254, M\,10  & 0.815& 0.206$\pm$0.019& 0.609$\pm$0.019   &0.519$\pm$0.026& 0.775& 0.233$\pm$0.026& 0.542$\pm$0.027& 0.026$\pm$0.021   & 0.004$\pm0.001$\\
    NGC6341, M\,92  & 0.781& 0.053$\pm$0.020& 0.728$\pm$0.020   &0.149$\pm$0.027& 0.730& 0.120$\pm$0.022& 0.610$\pm$0.022& 0.067$\pm$0.018   & 0.004$\pm0.001$\\
    NGC6362         & 0.851& 0.213$\pm$0.025& 0.638$\pm$0.025   &0.482$\pm$0.034& 0.845& 0.250$\pm$0.028& 0.595$\pm$0.028& 0.037$\pm$0.015   & 0.003$\pm0.001$\\
    NGC6397         & 0.792& 0.136$\pm$0.015& 0.659$\pm$0.015   &0.365$\pm$0.020& 0.781& 0.146$\pm$0.020& 0.635$\pm$0.020& 0.010$\pm$0.016   & 0.005$\pm0.001$\\
    NGC6441         & 0.918& 0.223$\pm$0.016& 0.695$\pm$0.016   &0.499$\pm$0.025& 0.795& 0.296$\pm$0.047& 0.499$\pm$0.047& 0.073$\pm$0.044   & 0.006$\pm0.001$\\
    NGC6535         & 0.794& 0.180$\pm$0.021& 0.614$\pm$0.021   &0.476$\pm$0.029& 0.790& 0.203$\pm$0.025& 0.587$\pm$0.025& 0.023$\pm$0.015   & 0.002$\pm0.001$\\
    NGC6541         & 0.796& 0.170$\pm$0.016& 0.626$\pm$0.016   &0.448$\pm$0.022& 0.736& 0.220$\pm$0.026& 0.516$\pm$0.026& 0.050$\pm$0.022   & 0.006$\pm0.001$\\
    NGC6584         & 0.821& 0.150$\pm$0.019& 0.671$\pm$0.019   &0.351$\pm$0.026& 0.800& 0.166$\pm$0.024& 0.633$\pm$0.024& 0.016$\pm$0.020   & 0.005$\pm0.001$\\
    NGC6681, M\,70  & 0.800& 0.183$\pm$0.018& 0.617$\pm$0.018   &0.461$\pm$0.025& 0.761& 0.256$\pm$0.029& 0.505$\pm$0.029& 0.073$\pm$0.029   & 0.004$\pm0.001$\\
    NGC6717, Pal\,9 & 0.822& 0.220$\pm$0.022& 0.602$\pm$0.022   &0.520$\pm$0.030& 0.818& 0.256$\pm$0.024& 0.562$\pm$0.024& 0.036$\pm$0.015   & 0.005$\pm0.001$\\
    NGC6723         & 0.827& 0.180$\pm$0.023& 0.647$\pm$0.023   &0.401$\pm$0.032& 0.793& 0.233$\pm$0.024& 0.560$\pm$0.024& 0.053$\pm$0.013   & 0.004$\pm0.001$\\
    NGC6752         & 0.814& 0.216$\pm$0.022& 0.598$\pm$0.022   &0.544$\pm$0.030& 0.756& 0.276$\pm$0.023& 0.480$\pm$0.023& 0.060$\pm$0.017   & 0.006$\pm0.001$\\
    NGC6779, M\,56  & 0.789& 0.140$\pm$0.015& 0.649$\pm$0.015   &0.374$\pm$0.020& 0.747& 0.173$\pm$0.020& 0.574$\pm$0.020& 0.033$\pm$0.016   & 0.005$\pm0.001$\\
    NGC6809, M\,55  & 0.790& 0.140$\pm$0.017& 0.650$\pm$0.017   &0.370$\pm$0.023& 0.756& 0.173$\pm$0.025& 0.583$\pm$0.025& 0.033$\pm$0.017   & 0.005$\pm0.001$\\
    NGC6981, M\,72  & 0.825& 0.160$\pm$0.022& 0.665$\pm$0.022   &0.371$\pm$0.030& 0.801& 0.183$\pm$0.024& 0.618$\pm$0.024& 0.023$\pm$0.015   & 0.004$\pm0.001$\\
    NGC7078, M\,15  & 0.791& 0.073$\pm$0.021& 0.718$\pm$0.021   &0.205$\pm$0.029& 0.700& 0.210$\pm$0.024& 0.490$\pm$0.024& 0.137$\pm$0.019   & 0.008$\pm0.001$\\
    NGC7099, M\,30         & 0.781& 0.066$\pm$0.014& 0.715$\pm$0.014   &0.193$\pm$0.019& 0.752& 0.083$\pm$0.019& 0.669$\pm$0.019& 0.017$\pm$0.019   & 0.006$\pm0.001$\\
    \hline  
    ID$^*$  &$\rm M^{Tip}_{1G}/M_\odot$&$\rm \mu_{1G}/M_\odot$ &$\rm \bar{M}^{HB}_{1G}/M_\odot$&$\rm \eta_{R,1G}$&$\rm M^{Tip}_{2G}/M_\odot$&$\rm \mu_{2G}/M_\odot$ &$\rm \bar{M}^{HB}_{2G}/M_\odot$&$\rm \Delta \mu_{e}/M_\odot$&$\delta/M_\odot$\\
    \hline
    NGC0104, 47\,Tuc & 0.871& 0.233$\pm$0.045& 0.638$\pm$0.045&0.516$\pm$0.059 & 0.859& 0.233$\pm$0.044& 0.626$\pm$0.044& ---           & 0.004$\pm0.001$\\
    NGC5927         & 0.919& 0.263$\pm$0.031& 0.656$\pm$0.031&0.601$\pm$0.042 & 0.902& 0.263$\pm$0.033& 0.632$\pm$0.033& ---           & 0.006$\pm0.001$\\
    NGC6304         & 0.914& 0.270$\pm$0.031& 0.644$\pm$0.031&0.618$\pm$0.042 & 0.900& 0.270$\pm$0.032& 0.630$\pm$0.032& ---           & 0.004$\pm0.001$\\
    NGC6352         & 0.909& 0.256$\pm$0.039& 0.653$\pm$0.039&0.583$\pm$0.053 & 0.880& 0.256$\pm$0.041& 0.624$\pm$0.041& ---           & 0.004$\pm0.001$\\ 
    NGC6366         & 0.895& 0.273$\pm$0.041& 0.622$\pm$0.041&0.626$\pm$0.056 & ---& ---          & ---          & ---           & 0.003$\pm0.002$\\
    NGC6496         & 0.924& 0.280$\pm$0.020& 0.644$\pm$0.020&0.644$\pm$0.026 & 0.910& 0.280$\pm$0.023& 0.634$\pm$0.023& ---           & 0.004$\pm0.001$\\
    NGC6624         & 0.914& 0.276$\pm$0.014& 0.638$\pm$0.014&0.633$\pm$0.018 & 0.898& 0.276$\pm$0.015& 0.608$\pm$0.015& ---           & 0.005$\pm0.001$\\
    NGC6637         & 0.881& 0.253$\pm$0.039& 0.648$\pm$0.039&0.575$\pm$0.056 & 0.875& 0.253$\pm$0.042& 0.621$\pm$0.042& ---           & 0.003$\pm0.001$\\
    NGC6652         & 0.842& 0.180$\pm$0.031& 0.662$\pm$0.031&0.390$\pm$0.042 & 0.832& 0.180$\pm$0.032& 0.654$\pm$0.032& ---           & 0.004$\pm0.001$\\
    NGC6838, M\,71  & 0.858& 0.210$\pm$0.018& 0.648$\pm$0.018&0.466$\pm$0.024 & 0.850& 0.210$\pm$0.020& 0.640$\pm$0.020& ---           & 0.005$\pm0.001$\\
    \hline
    \hline
    \end{tabular}
    \end{adjustbox}
    \label{tab:param_dm}
\end{table*}

\section{Results}
\label{sec:results}

The procedure described in the previous section for NGC\,6752 has been extended to 34 GCs with blue HB. The parameters used as input for the simulations are listed in Table~\ref{tab:param_sim}, while the resulting RGB-tip masses, mass losses, average HB masses of 1G and 2Ge stars are listed in upper part of Table~\ref{tab:param_dm}, where we also provide the average mass loss difference between 1G and 2Ge stars. For completeness, we included the results for NGC\,5272 and NGC\,6121 from \citet{tailo_2019a} and \citet{tailo_2019b}.  

The input parameters adopted for 10 GCs with red--only HB,  and the resulting RGB-tip masses, mass losses, average masses of 1G and 2G stars, and mass loss difference between 1G and 2G stars are listed in Table~\ref{tab:param_sim} and in the lower part of Table~\ref{tab:param_dm}, respectively. 

We conducted a series of tests to ensure that our results are not affected by the most common biases of this kind of analysis. A) We verified that the results for our GC sample are not dependent on the exact binning choices. We tested a few sample cases halving the binning step and verified that the values we get are compatible B) We verified that our results are not biased by the exact metric normalization by re-examining a few cases changing the normalization criterion of the simulation histograms; specifically, we repeated the comparison normalizing the simulation histogram to the same number of stars in the target populations and verified that the results obtained with the two method are compatible.  C) We verified that the identifications provided by our searching algorithm are compatible with the ones obtained from independent sources (e.g. in the case of 6441 by comparing our identification with the one provided via a colour-colour diagram similar to the one in Figure \ref{pic:method_6637}. D) We verified that our algorithm is capable of recovering the parameters of the target populations from a toy HB model, where we know exactly the input parameters of the 1G and 2Ge stars.  E) We look at our results in the $\rm m_{F275W}-m_{F814W}$ vs. $\rm m_{F275W}$ CMD, similar to the one in Figure \ref{pic:checksim}, as a sanity check for the population identification.

The complete showcase of the HBs analysed in this work is provided in Appendix\,\ref{sec:app_showcase}, where we compare the CMDs our GCs with the contours of the simulated CMDs that provide the best match with the observed 1G and 2Ge HB stars.

In the next subsection we further investigate the mass of HB 1G stars and the mass loss of 1G stars, while Section~\ref{sec:pattern_2g} is dedicated to HB mass and mass loss of 2Ge stars.

\subsection{Mass loss of first-generation stars}
\label{sec:corr_1g}

\begin{figure*}
    \centering
    \includegraphics[width=2.00\columnwidth,trim={1cm 1cm 1cm 0cm}]{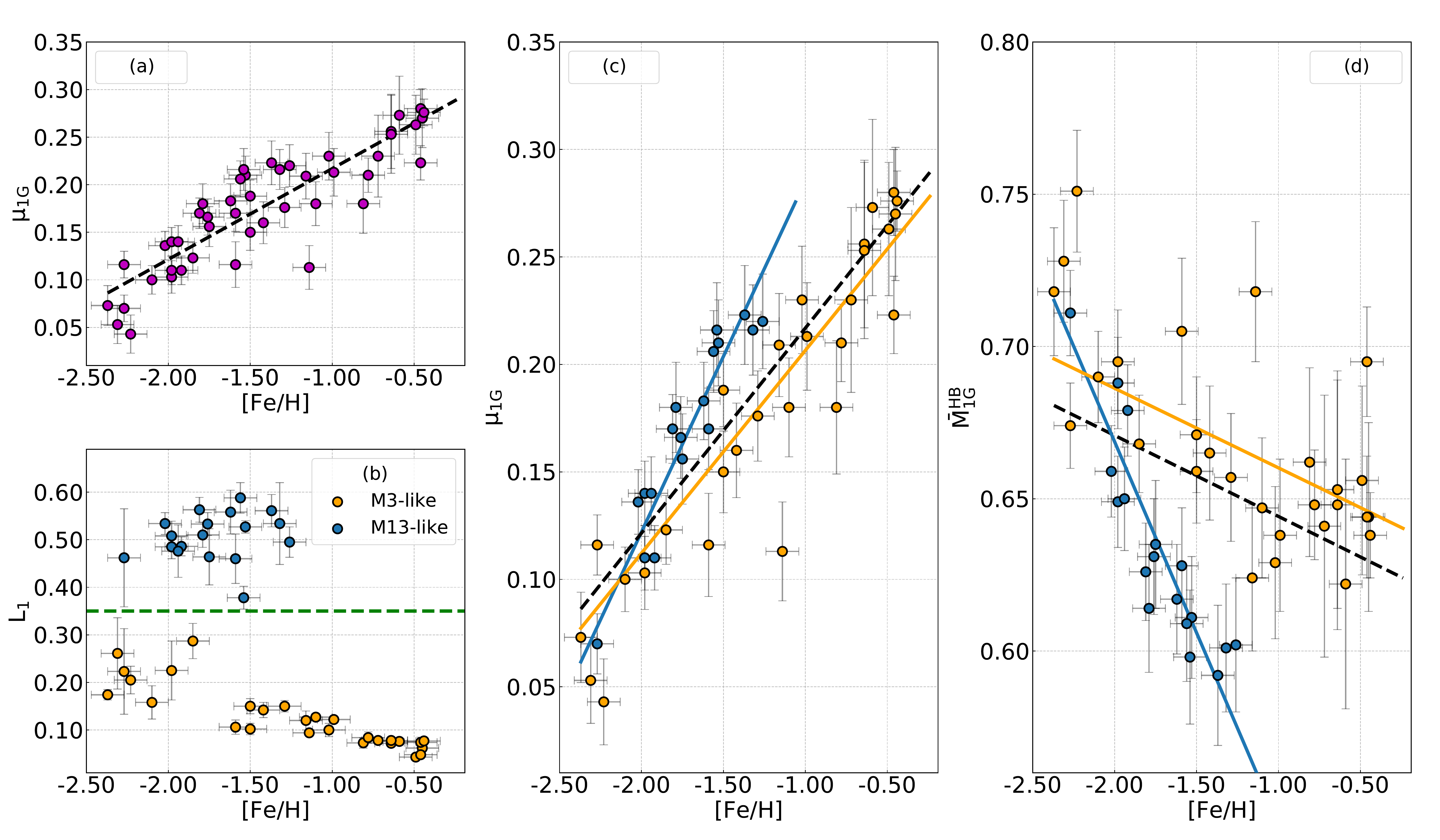}
    \caption{
    \textit{Panel a:} Mass loss of 1G stars, $\rm \mu_{1G}$, as a function of [Fe/H].
    \textit{Panel b:} Color distance between the reddest part of the HB, $\rm L_1$ \citep[][]{milone_2014}, against [Fe/H]. 
    The dashed line at $\rm L_1=0.35$ mag separates the groups of M3- (orange) and M13-like (blue) GCs selected in this paper. 
    \textit{Panel c:} Same as panel \textit{a} but with the two groups of GCs identified in panel \textit{b} marked with the corresponding colors. 
    \textit{Panel d:} Average HB mass of 1G stars, $\rm \bar{M}_{1G}^{HB}$, as a function of [Fe/H] for M3- (orange) and M13-like GCs (blue).
    The black dashed lines mark are the best-fit straight lines derived from all the studied clusters, while blue and orange lines represent the the best-fit straight lines derived for M3- and M13-like clusters, respectively. 
    The equations of these lines are provided in Table~\ref{Tab:rette}.
    }
    \label{pic:mu1g_feh_scheme}
\end{figure*}

The results listed in Table~\ref{tab:param_dm}, where we provide mass losses for 1G stars of 46 GCs, show that the mass loss of 1G stars clearly depends on the cluster metallicity.
This fact is illustrated in Figure~\ref{pic:mu1g_feh_scheme}a,  which shows a strong correlation between 1G mass loss and the iron abundance.
%, with a Spearman's rank correlation coefficient, $\rm R_S=0.88$.

The values of $\rm \mu_{1G}$ and [Fe/H] of Figure~\ref{pic:mu1g_feh_scheme}a   follow  the linear relation:
\begin{equation}
\rm \mu_{1G}=(0.095\pm0.007)\times [Fe/H]+(0.313\pm0.012).
\label{eq:mu1_ge_rel}
\end{equation}

 obtained by means of least squares and represented with a black line in Figure~\ref{pic:mu1g_feh_scheme}a. The  Pearson rank coefficient  of this linear fit has the high value $\rm R_p$=0.88, showing that the points are well reproduced by a straight line. Nevertheless, the 1G mass-loss spread among GCs with the same iron abundance is larger than that expected based on the  observational errors.

To investigate the reasons of this scatter, we analyze the two distinct groups of GCs selected in Figure~\ref{pic:mu1g_feh_scheme}b, where we plot  the color distance between the RGB and the reddest part of HB \citep[$\rm L_1$,][]{milone_2014} against [Fe/H]. 
By adopting the names of the GC prototypes, M3 and M13, we define M3-like GCs with $\rm L_1 \leq 0.35$, which mostly include GCs having HB stars on the red side of the RR\,Lyr region, and M13-like GCs with $\rm L_1 > 0.35$, in which stars redder than the RR\,Lyr are missing\footnote{M3-like objects include both the G1 and G2 groups defined by \citet{milone_2014}, while the M13-like objects correspond to their G3 sample.}. 
The evidence that M13-like GCs provide different $L_1$ values than the remaining GCs with the same [Fe/H] reflects the fact that GCs with the same metallicity exhibit different HB morphologies, and that at least one second parameter, in addition to metallicity, is needed to explain the HB of GCs.     

\begin{table*}
    \centering
	\caption{Linear fits in the form $\rm \alpha\times [Fe/H] + \beta$ derived in this paper. We also provide the  Pearson rank coefficient, $\rm R_P$, and the r.m.s of the residuals respect the best-fit line.}
    \begin{adjustbox}{width=2.00\columnwidth,center}
     \renewcommand{\arraystretch}{1.30} 
    \begin{tabular}{l|cccl|cccl|cccl}
    \hline
    & & ALL & & & & M3-like & & & & M13-like & & \\
    \hline
    Variable &$\rm \alpha$&$\rm \beta$&$\rm R_P$& scatter &$\rm \alpha$&$\rm \beta$&$\rm R_P$&scatter&$\rm \alpha$&$\rm \beta$&$\rm R_P$& scatter\\
    \hline
	$\rm \mu_{1G}$&$0.095\pm0.007$&$0.313\pm0.012$&0.88&0.030&$0.094\pm0.007$&$0.302\pm0.011$&0.93&0.027&$0.164\pm0.015$&$0.450\pm0.029$&0.94&0.016\\
	$\rm M^{HB}_{1G}$&$-0.027\pm0.008$&$0.616\pm0.014$&-0.36&0.034&$-0.027\pm0.007$&$0.641\pm0.010$&-0.89&0.025&$-0.126\pm0.015$&$0.416\pm0.029$&-0.90&0.015\\
    $\rm \eta_{R}$&$0.183\pm0.021$&$0.682\pm0.032$&0.79&0.082&$0.189\pm0.018$&$0.672\pm0.023$&0.90&0.070&$0.329\pm0.044$&$0.994\pm0.077$&0.88&0.047\\     
    $\rm \mu_{2Ge}$&$0.135\pm0.018$&$0.421\pm0.034$&0.72&0.044&$0.101\pm0.021$&$0.345\pm0.037$&0.80&0.037&$0.216\pm0.023$&$0.581\pm0.043$&0.80&0.030\\  
	$\rm M^{HB}_{2Ge}$&$-0.097\pm0.027$&$0.411\pm0.049$&-0.35&0.061&$-0.059\pm0.034$&$0.493\pm0.062$&-0.66&0.063&$-0.190\pm0.037$&$0.233\pm0.068$&-0.67&0.047\\      
    \hline
    \hline
	\end{tabular}
	\end{adjustbox}
	\label{Tab:rette}
\end{table*}

In Figure \ref{pic:mu1g_feh_scheme}c we show again the $\rm \mu_{1G}$ vs.\,[Fe/H] plot of panel a, highlighting the M3- (yellow dots) and M13-like (blue dots) clusters. Clearly, M13-like GCs lie on a steeper relation than M3-like GCs. The plot clearly shows the evidence coming out from the different HB morphology of the two groups: the 1G stars of M13-like GCs loose more mass than the 1G stars of M3-like GCs with the same metallicity, if no other parameter is at play.

\subsubsection{HB and RGB-tip masses of 1G stars}

 To give a different look at the correlation between mass loss of 1G stars and GC metallicity shown in  Figure \ref{pic:mu1g_feh_scheme}c, we analyse the dependence between [Fe/H], $\rm \bar{M}_{1G}^{HB}$ and $\rm M^{Tip}_{1G}$.
Obviously, the mass loss has been simply obtained as the difference between the mass at the RGB tip and the average HB mass.
 
Figure~\ref{pic:mu1g_feh_scheme}d shows that the average HB masses of 1G stars vary in the range from $\sim 0.60 M_{\odot}$ to $\sim 0.75 M_{\odot}$ 
\footnote{
We compared our results with other in the literature. For M\,13 (NGC\,6205) we obtain average mass loss value in agreement with \citet{dalessandro_2013}. Our average mass loss value for NGC\,0104 (47\,Tuc) is in agreement within 1$\sigma$ with the ones in \citet{diCriscienzo_2010} and \citet{salaris_2016}. The average HB masses of 1G stars in 47\,Tuc and M\,13 are consistent with those derived by \citet{denissenkov_2017}, while results on NGC\,6809 and NGC\,6362 are consistent with those by \citet{vandenberg_2018}.
Our results are not directly comparable with those by \cite{jang_2019}, who assumed  lower helium content (Y=0.23) for 1G stars. Anyway, we notice that the values of $\rm M^{HB}_{1G}$ and $\rm \mu_{1G}$  that they obtained for the GCs M4, M5, M15, and M80 are consistent with the conclusion that average HB masses and the mass losses correlate with the cluster metallicity. 
}

We also note a mild anticorrelation between  $\rm \bar{M}_{1G}^{HB}$  and [Fe/H], although clusters with similar metallicities span a wide range of HB masses, with M3- and M13-like GCs following distinct patterns in the $\rm \bar{M}_{1G}^{HB}$ vs.\,[Fe/H] plane. 
Specifically, 1G stars of M13-like clusters have smaller HB masses when compared with their counterparts of M3-like GCs. This latter group exhibits a broad distribution around the best-fit least-squares line (yellow line in Figure \ref{pic:mu1g_feh_scheme}d) corresponding to a residual HB-mass spread of 0.025 $M_{\odot}$. In contrast, M13-like clusters show a small mass scatter of 0.015 $M_{\odot}$ around the corresponding best-fit straight line (blue line). 

\begin{figure}
    \centering
    \includegraphics[width=0.90\columnwidth,trim={1cm 0.0cm 0.7cm 0cm}]{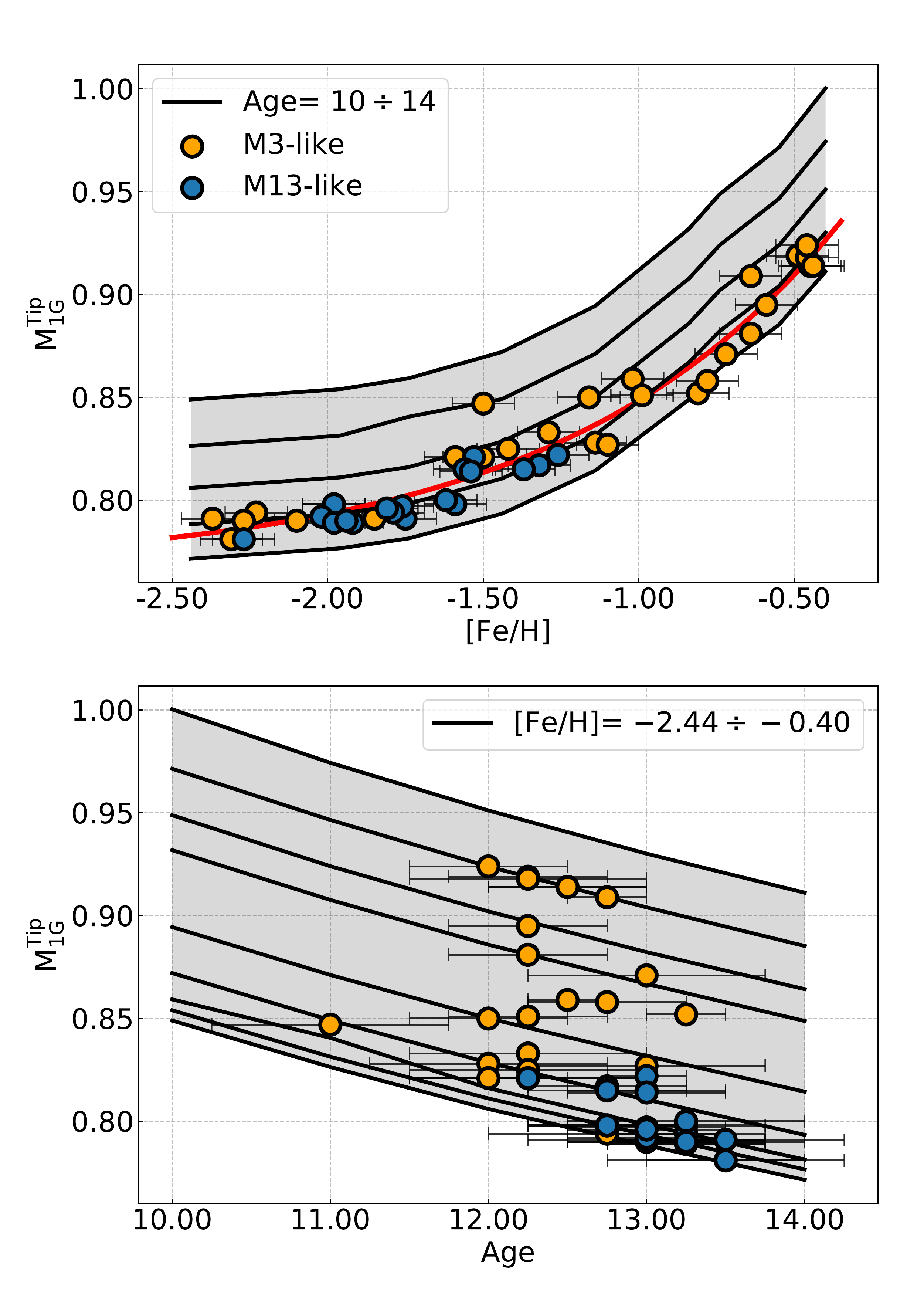}
    \caption{\textit{Upper panel:} $\rm M_{1G}^{Tip}$ as function of iron abundance. The black solid lines correspond to ages of 10,11,12,13,14 Gyr with the oldest one located at the bottom. The red solid line is the least-square exponential fit of the data.
     \textit{Lower panel:} Stellar mass of 1G stars at the tip of the RGB , $\rm M_{1G}^{Tip}$ provided by the fest-fit isochrones \citep[from][]{tailo_2016b, tailo_2017, tailo_2019b, tailo_2019a}, as function of GC age. Here the black solid lines correspond to [Fe/H] $-$2.44,$-$1.96,$-$1.74,$-$1.44,$-$1.14,$-$0.84,$-$0.74,$-$0.55,$-$0.40 the lowest at the bottom. }
    \label{pic:mtip_1g_params}
\end{figure}

On the contrary, the stellar masses of 1G stars at the RGB tip exhibit a strong correlation with [Fe/H] as illustrated in the top panel of Figure~\ref{pic:mtip_1g_params}, with $\rm M_{1G}^{Tip}$ ranging from less than 0.8$M_{\rm \odot}$ at [Fe/H]$\sim -2.4$ to more than 0.9$M_{\rm \odot}$ in the most metal-rich studied GCs. Observations are reproduced by the red line, which is the best-fit exponential function ($\rm M^{Tip}_{1G}=(0.250\pm0.011)\times 10^{(0.483\pm 0.068)\times [Fe/H]}+(0.766\pm0.009)$) obtained by means of least squares
 \footnote{The scatter of the observed points around the red line is mostly due to the fact that, for a fixed metallicity, $\rm M_{1G}^{Tip}$ also depends on cluster age. Clearly, age dependence is small when compared to dependence  from metallicity for the studied old GCs.}.
Thus, the correlation between $\rm \mu_{1G}$ and [Fe/H] comes out from the combination of the mild decrease of $\rm M^{HB}_{1G}$ (Figure \ref{pic:mu1g_feh_scheme}e) and the increase of $\rm M^{Tip}_{1G}$ with metallicity (Figure \ref{pic:mtip_1g_params}). 
 
\begin{figure}
    \centering
    %placeholders
    \includegraphics[width=0.9\columnwidth,trim={1cm 1.0cm 0cm -1cm}]{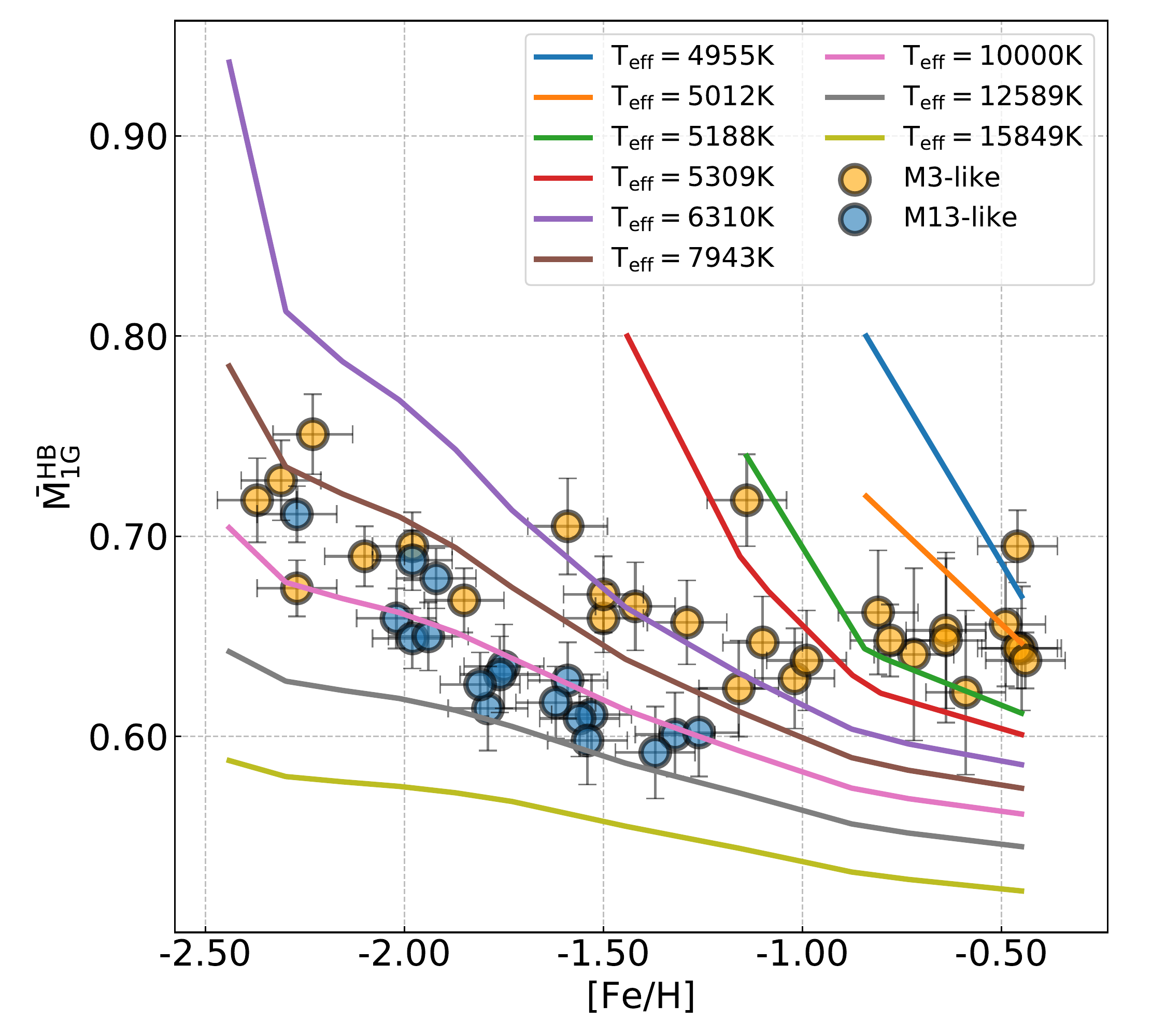}
    \caption{
    Reproduction of Figure~\ref{pic:mu1g_feh_scheme}, where we show the average HB mass of 1G stars $\rm \bar{M}^{HB}_{1G}$ as function of [Fe/H]. Here, we superimposed the curves from ZAHB models for different values of $\rm T_{eff}$, as quoted in the inset. 
    }
    \label{pic:mhb1g_teff}
\end{figure}

To further explore HB 1G stars, we show in Figure~\ref{pic:mhb1g_teff} the $\rm \bar{M}_{1G}^{HB}$ vs.\,[Fe/H] relations derived from the zero age HB models used in this paper for various effective temperatures. As expected, 1G stars of metal-rich clusters ([Fe/H]$\gtrsim -1.2$) are, on average, cooler than those of metal-poor GCs. 
The 1G stars of M13-like GCs share similar effective temperatures ($\sim$9,000-12,000 K) and are typically hotter than the 1G stars if the M3-like GCs with similar  [Fe/H]. This further reflects the effect of the still-debated second parameter on the HB of GCs. 

\subsection{Mass loss in extreme second generation stars}
\label{sec:pattern_2g}

\begin{figure*}
    \centering
    %placeholders
    \includegraphics[width=1.8\columnwidth,trim={0cm 0cm 0cm 0cm}]{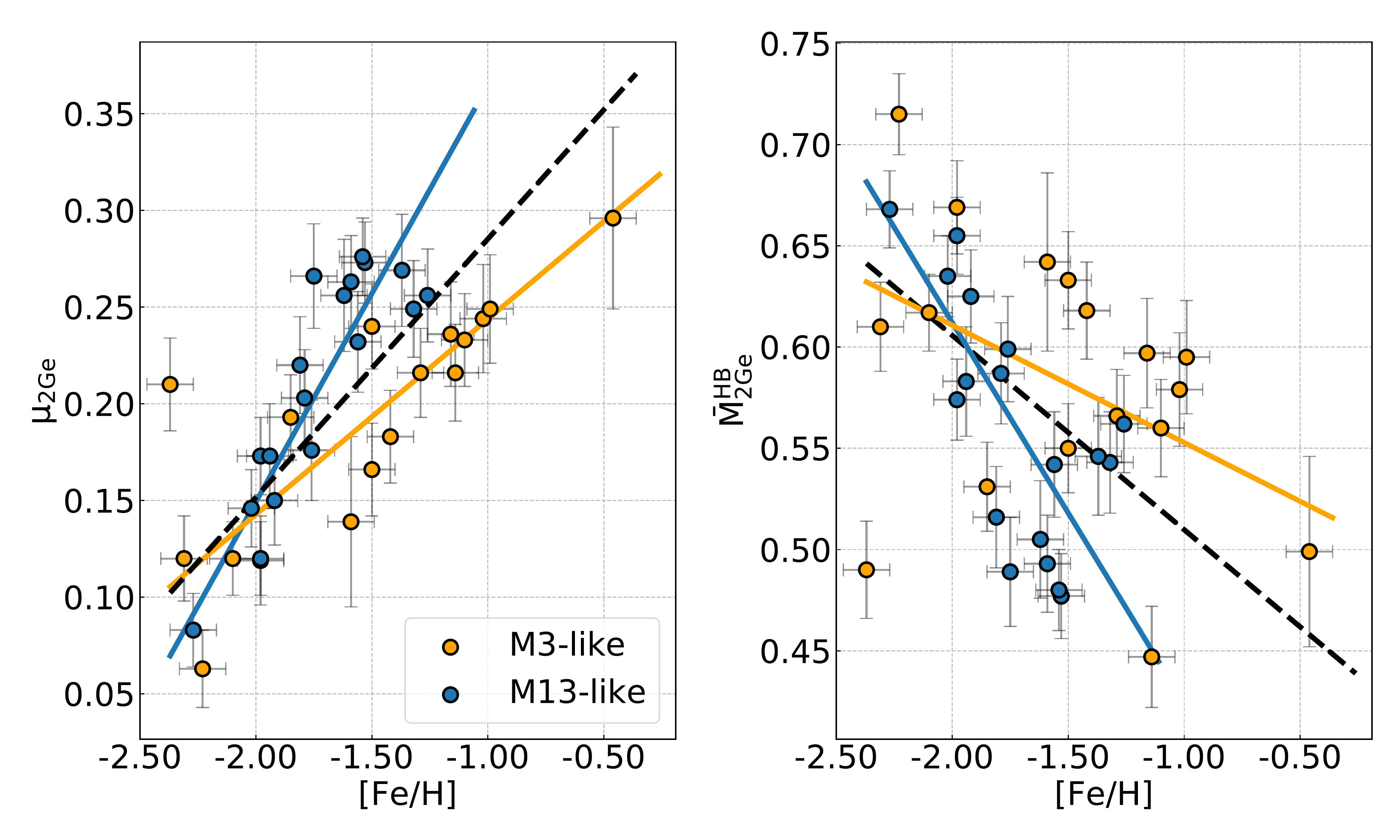}
    \caption{Mass loss, ($\rm \mu_{2Ge}$, left panel), and HB mass of 2Ge stars, ($\rm \bar{M}^{HB}_{2Ge}$, right panel)  against [Fe/H]. M3- and M13-like GCs identified in Figure \ref{pic:mu1g_feh_scheme}, are colored orange and blue, respectively. The best fit straight lines derived from all GCs, and for M3- and M13 like clusters alone are colored black, orange and blue, respectively. Their Equations are provided in Table~\ref{Tab:rette}.}
    \label{pic:mu2gevsfe}
\end{figure*}

RGB mass loss and average HB masses of 2Ge stars significant vary from one cluster to another and both span an interval of about 0.25 $M_{\odot}$.  
Figure~\ref{pic:mu2gevsfe} shows that these two quantities  exhibit significant correlation ($\rm \mu_{2Ge}$) and anticorrelation ($\rm \bar{M}^{HB}_{2Ge}$) with [Fe/H], with metal-rich clusters having, on average smaller HB masses and higher mass losses than metal-poor GCs.
M3- and M13-like clusters are described by distinct linear correlations in both $\rm \mu_{2Ge}$ vs.\,[Fe/H] and $\rm \bar{M}^{HB}_{2Ge}$ vs.\,[Fe/H] planes, with M13-like GCs defining steeper slopes and larger scatters (Figure~\ref{pic:mu2gevsfe}). The comparison with Fig.\ref{pic:mu1g_feh_scheme}  shows that mass loss and HB masses of 1G and 2Ge stars share similar qualitative behaviours as function of [Fe/H].

\begin{figure}
    \centering
    \includegraphics[width=0.9\columnwidth,trim={0.0cm 0.0cm 0.0cm 0.0cm}]{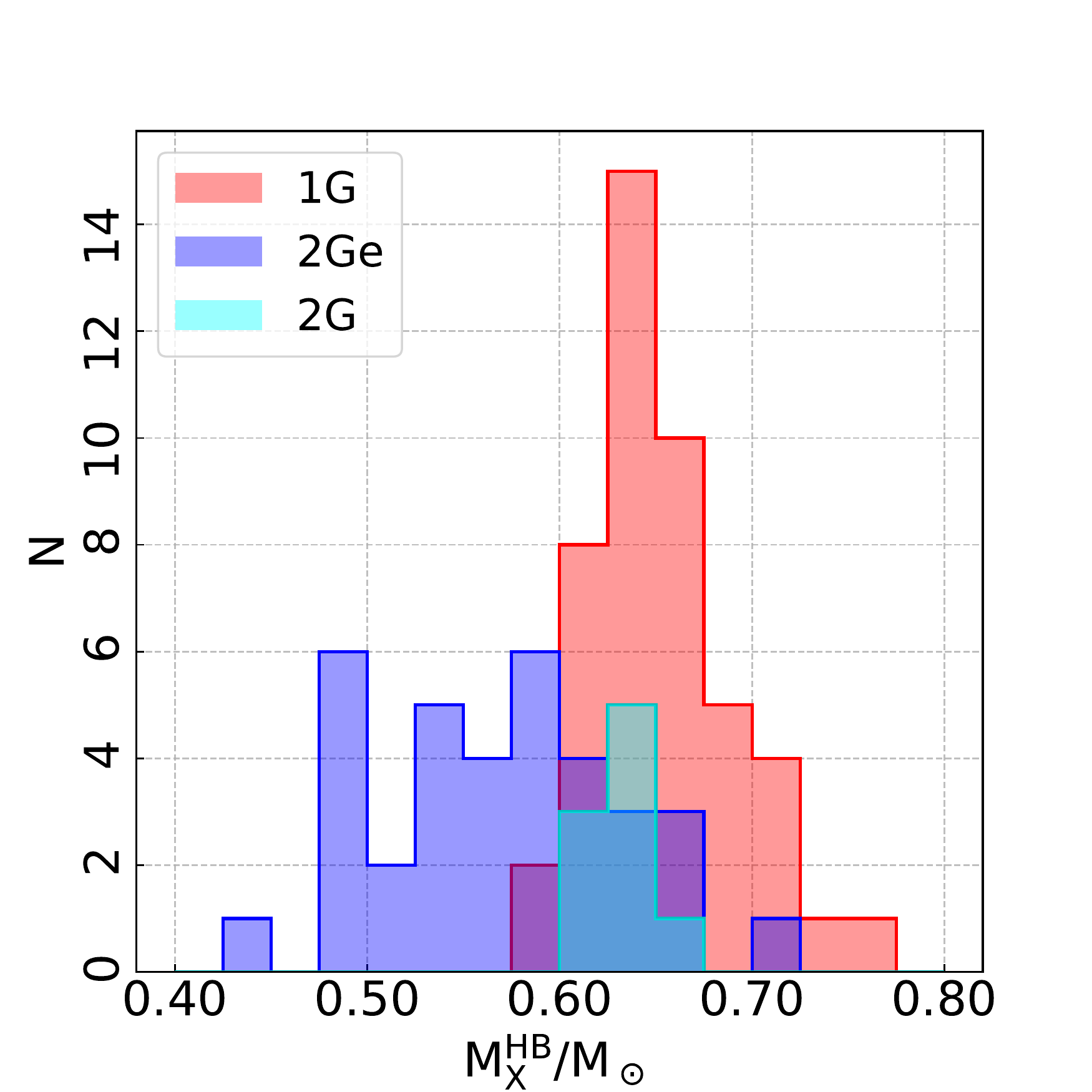}
    \caption{
    Histogram distributions of the average HB mass, $\rm M^{HB}_{X}$, for all studied clusters. The histograms corresponding to X=1G, X=2Ge and X=2G are colored, blue, purple, and cyan, respectively.}
    \label{pic:hbparam_distribution_M}
\end{figure}

As expected, 2Ge exhibit, on average, smaller HB masses and span a wider range of masses than the 2G and the 1Gs,  as shown in Figure~\ref{pic:hbparam_distribution_M}, where we show the histogram distributions of $\rm \bar{M}^{HB}_{x}$ (x=1G, 2G, 2Ge). 

\begin{figure*}
    \centering
    \includegraphics[width=0.9\columnwidth,trim={0cm 0.0cm 0.0cm 0.0cm}]{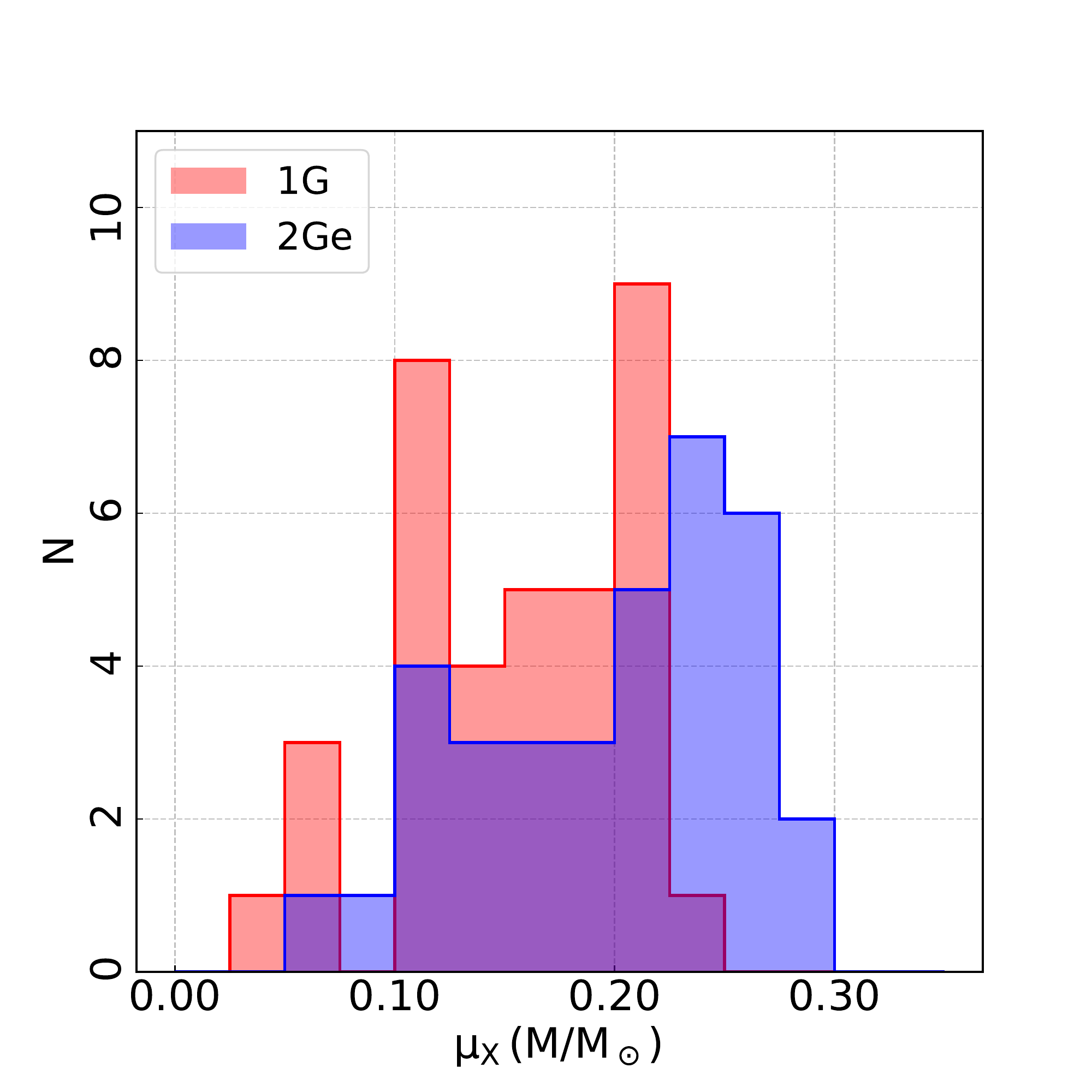}
    \includegraphics[width=0.9\columnwidth,trim={0cm 0.0cm 0.0cm 0.0cm}]{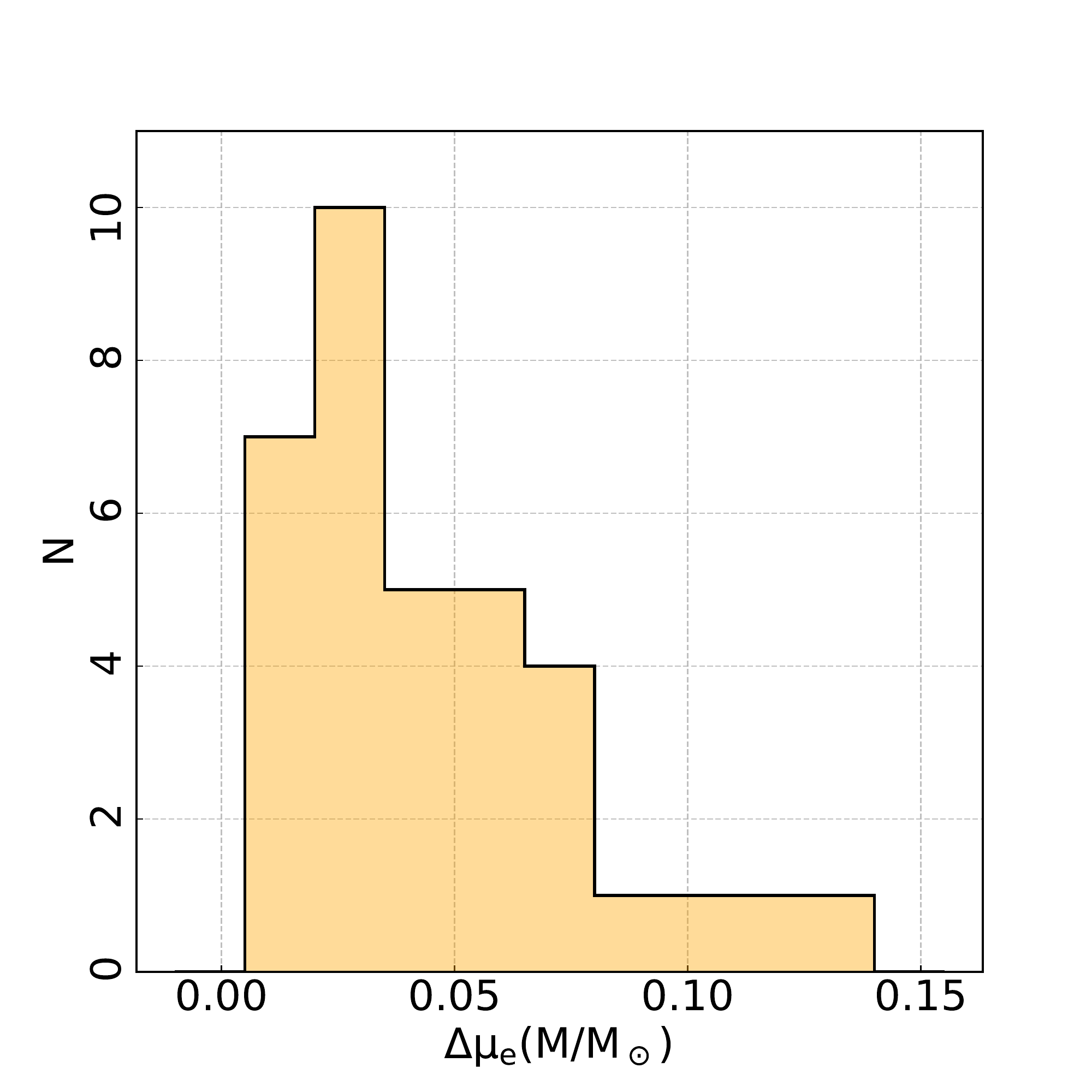}
    \caption{
    Histogram distributions of the mass loss of 1G (red) and 2Ge stars (blue) in all clusters with blue HB (left) and histogram distribution of the differences between the mass losses of 2Ge and 1G stars (right). }
    \label{pic:hbparam_distribution_mu}
\end{figure*}

2Ge stars also cover a wider mass loss interval than the 1G, ranging from  less than 0.05$M_\odot$ to $\sim$0.30$M_\odot$. This is illustrated in the left panel of Figure~\ref{pic:hbparam_distribution_mu} where we compare the histogram distributions of $\rm \mu_{1G}$ and $\rm \mu_{2Ge}$ for GCs with the blue HB.

\begin{figure}
    \centering
    \includegraphics[width=0.9\columnwidth,trim={0.0cm 0.0cm 0.0cm 0.0cm}]{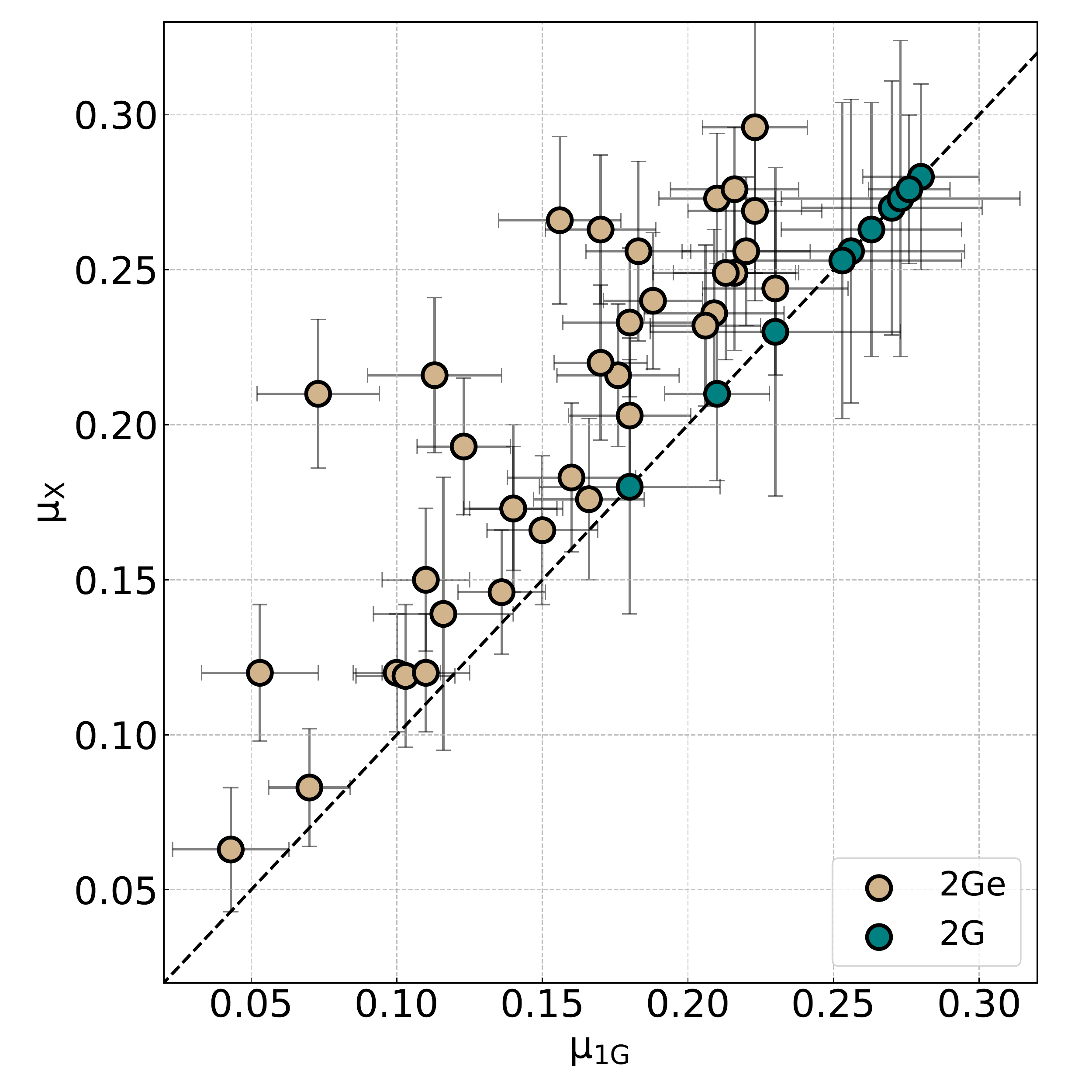}
    \caption{
    Average mass loss of 2Ge stars (tan) or 2G stars (teal) as a function of the 1G average mass loss. The black-dashed line represents the locus where $\rm \mu_{X}= \mu_{1G}$.
    }
    \label{pic:hbparam_mu1vsmu2}
\end{figure}

The histogram distribution of the mass-loss difference between 2Ge and 1G stars (right panel of  Figure~\ref{pic:hbparam_distribution_mu}) shows that 2Ge stars of all GCs with blue HB lose more mass than 1G stars (i.e.$\rm \Delta \mu_{e}>0$). Hence, \textit{additional mass loss is required for the 2Ge to reproduce their location on the HB}.
This result is emphasized in Figure~\ref{pic:hbparam_mu1vsmu2}, where we plot the mass loss of 2Ge stars as a function of the $\rm \mu_{1G}$ (tan dots). Clearly, all GCs with blue HB lie above the dashed line, which is the locus of points with $\rm \mu_{1G} = \mu_{X}$ (where X=2G or 2Ge).  
On the contrary, we find that in clusters with only--red HB the same mass loss properly reproduces the HB of both 1G and 2G stars.

We notice that several authors adopted enhanced mass losses for their second-generation stars to reproduce the HB morphology, although this finding is not always explicitly highlighted in their papers. This fact includes work on both Galactic GCs \citep[e.g.][]{dalessandro_2011,dalessandro_2013,dicriscienzo_2015,tailo_2017, denissenkov_2017, vandenberg_2018} and extragalactic GCs \citep[e.g.][]{dantona_2013}. 

\begin{figure}
    \centering
    %placeholders
    \includegraphics[width=0.9\columnwidth,trim={0cm 0cm 0cm 0cm}]{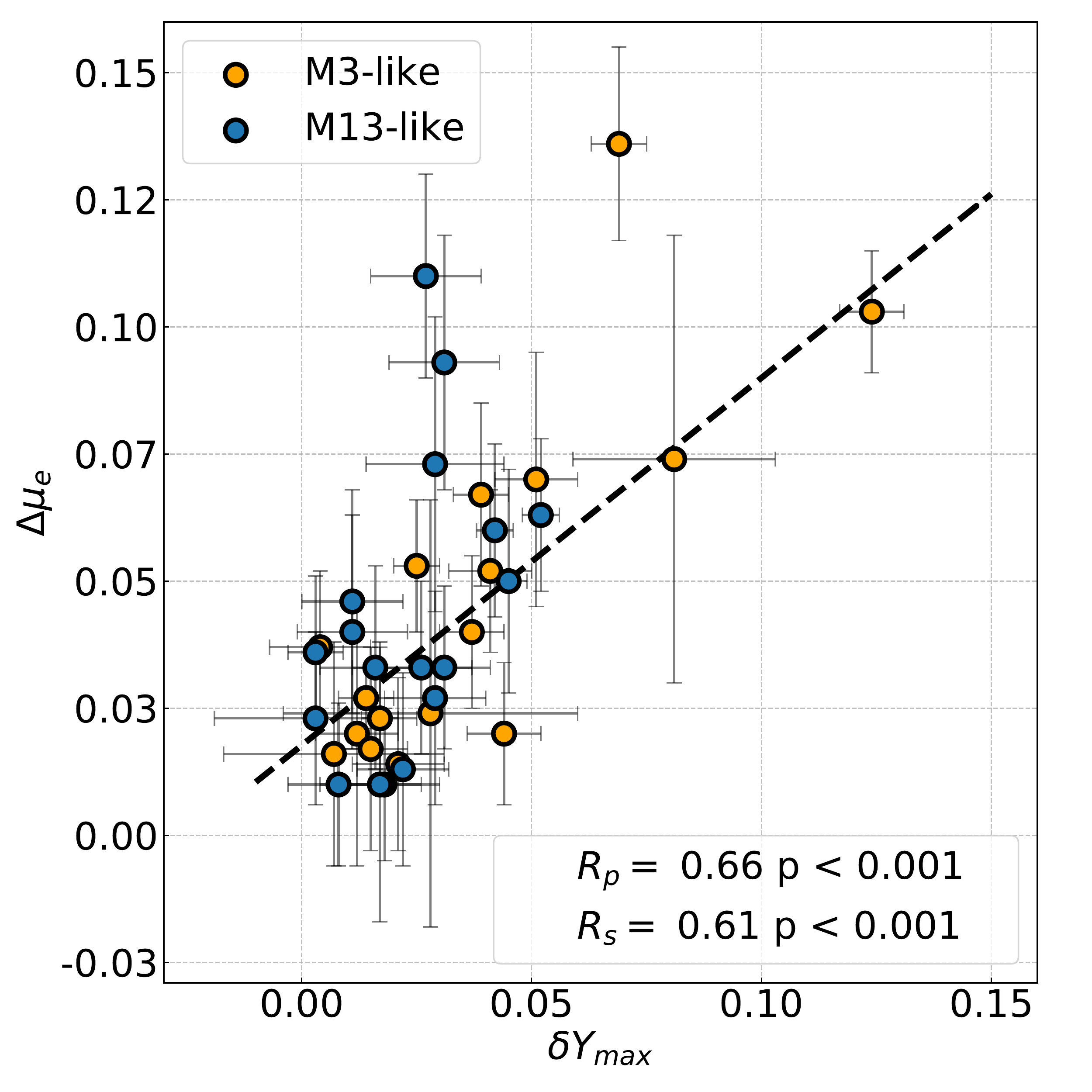}
    \caption{ 
     Difference between the mass loss of 2Ge and 1G stars, $\rm \Delta \mu_e$, as a function of the maximum helium variation of the host GC. M3- and M13-like GCs are the orange and blue dots, respectively. The black, dashed line is the best lest-squares fit: $\rm \Delta \mu_e/M_\odot=(0.708\pm0.018)\times \delta Y_{max}+(0.018\pm0.003)$.
    }
    \label{pic:dmue_dymax}
\end{figure}

\begin{figure*}
    \centering
    %placeholders
    \includegraphics[width=0.9\columnwidth,trim={0cm 0cm 0cm 0cm}]{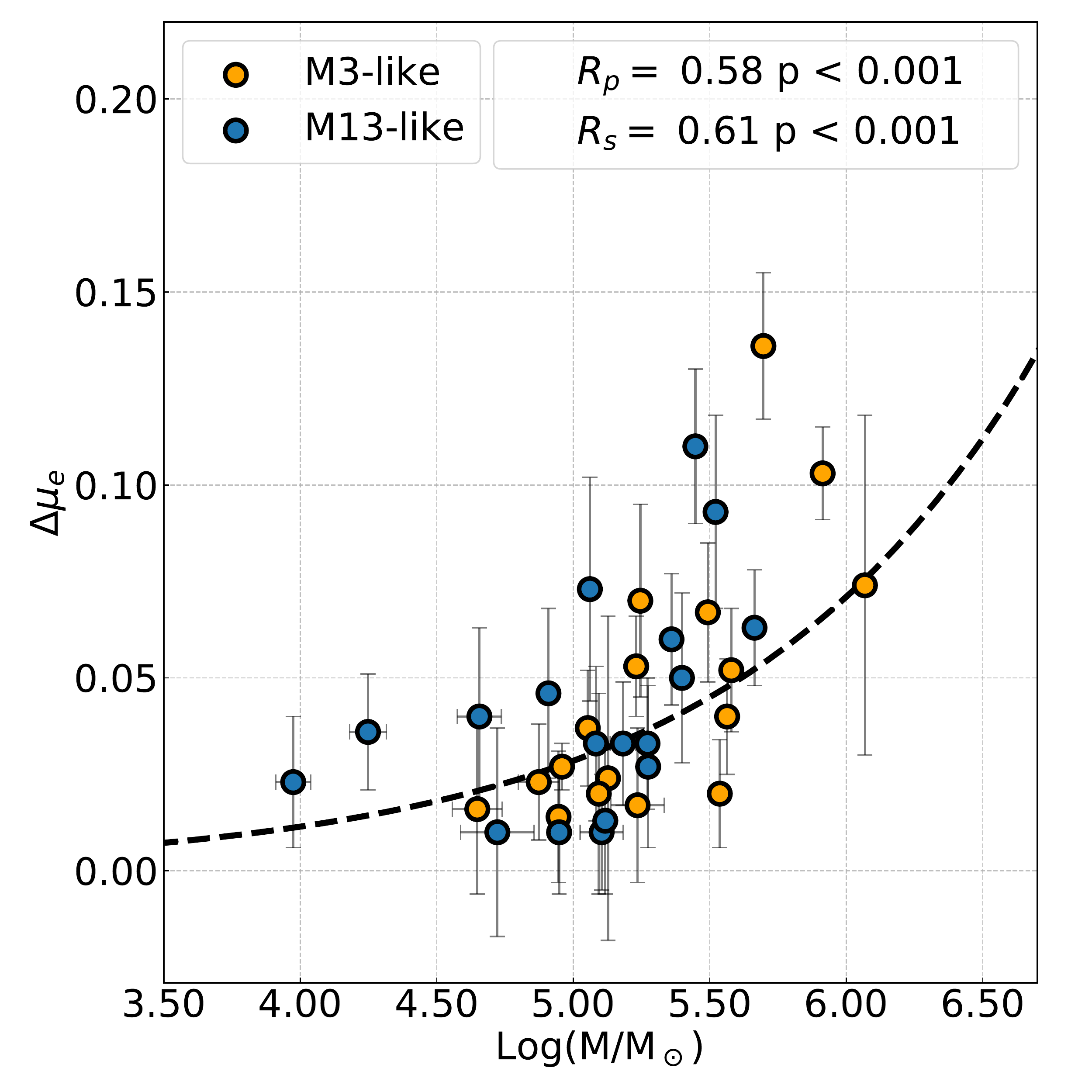}    
    \includegraphics[width=0.9\columnwidth,trim={0cm 0cm 0cm 0cm}]{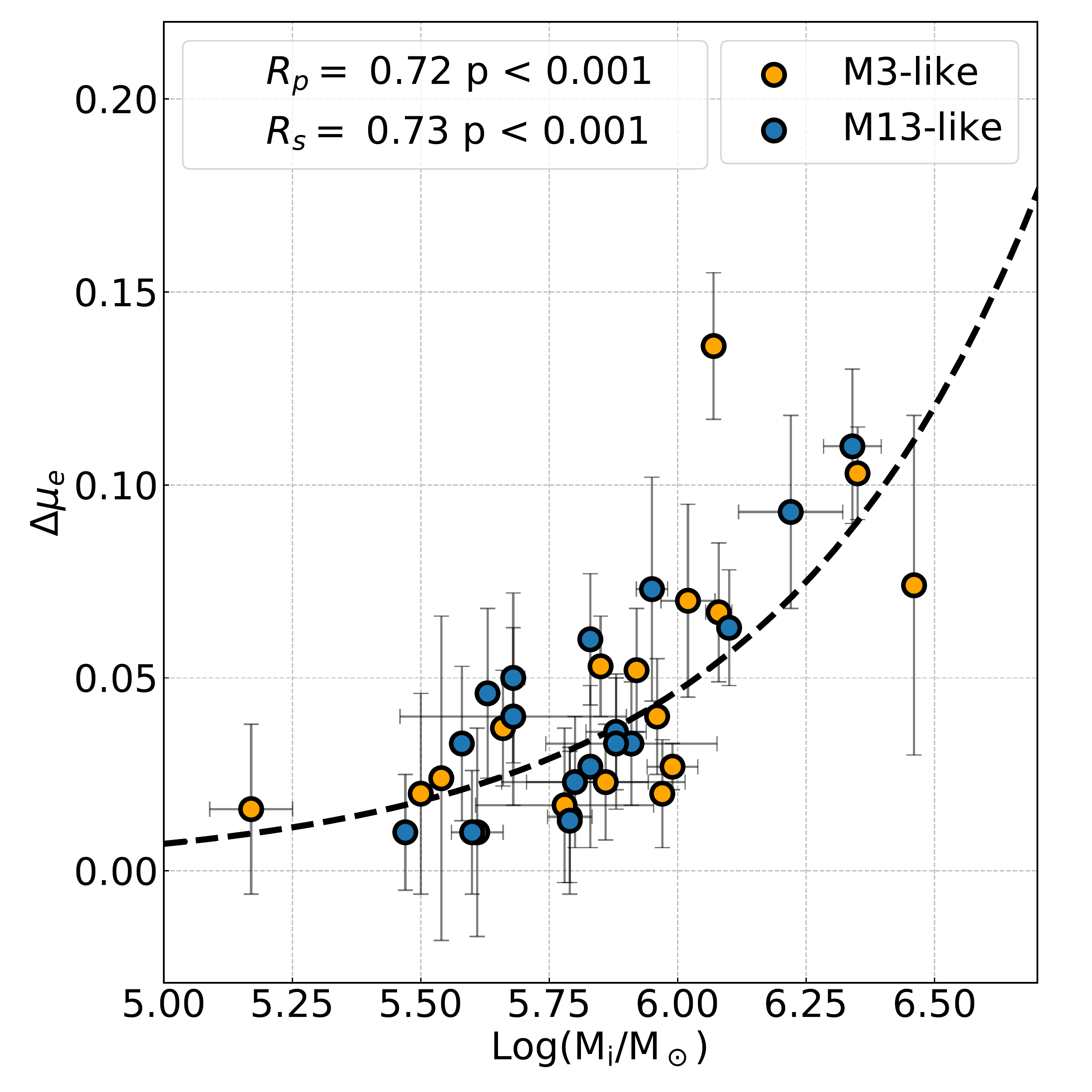}
    \caption{ 
     Difference between the mass loss of 2Ge and 1G stars, $\rm \Delta \mu_e$, as a function the present-day (left) and the initial GC mass (right).
    M3- and M13-like GCs are represented with orange and blue dots, respectively.
Black dashed lines are derived by means of lest-squares, and correspond to the relations that provide the best fit with the data: 
$
\rm log(\Delta \mu_e/M_\odot)=(0.385\pm0.087)\times log(M/M_\odot)-(3.469\pm0.445);
\rm log( \Delta \mu_e/M_\odot)=(0.823\pm0.143)\times log(M_i/M_\odot)-(6.268\pm0.836).
$
    }
    \label{pic:dmue_mcluster}
\end{figure*}

\begin{figure}
    \centering
    %placeholders
    \includegraphics[width=0.9\columnwidth,trim={0cm 0cm 0cm 0cm}]{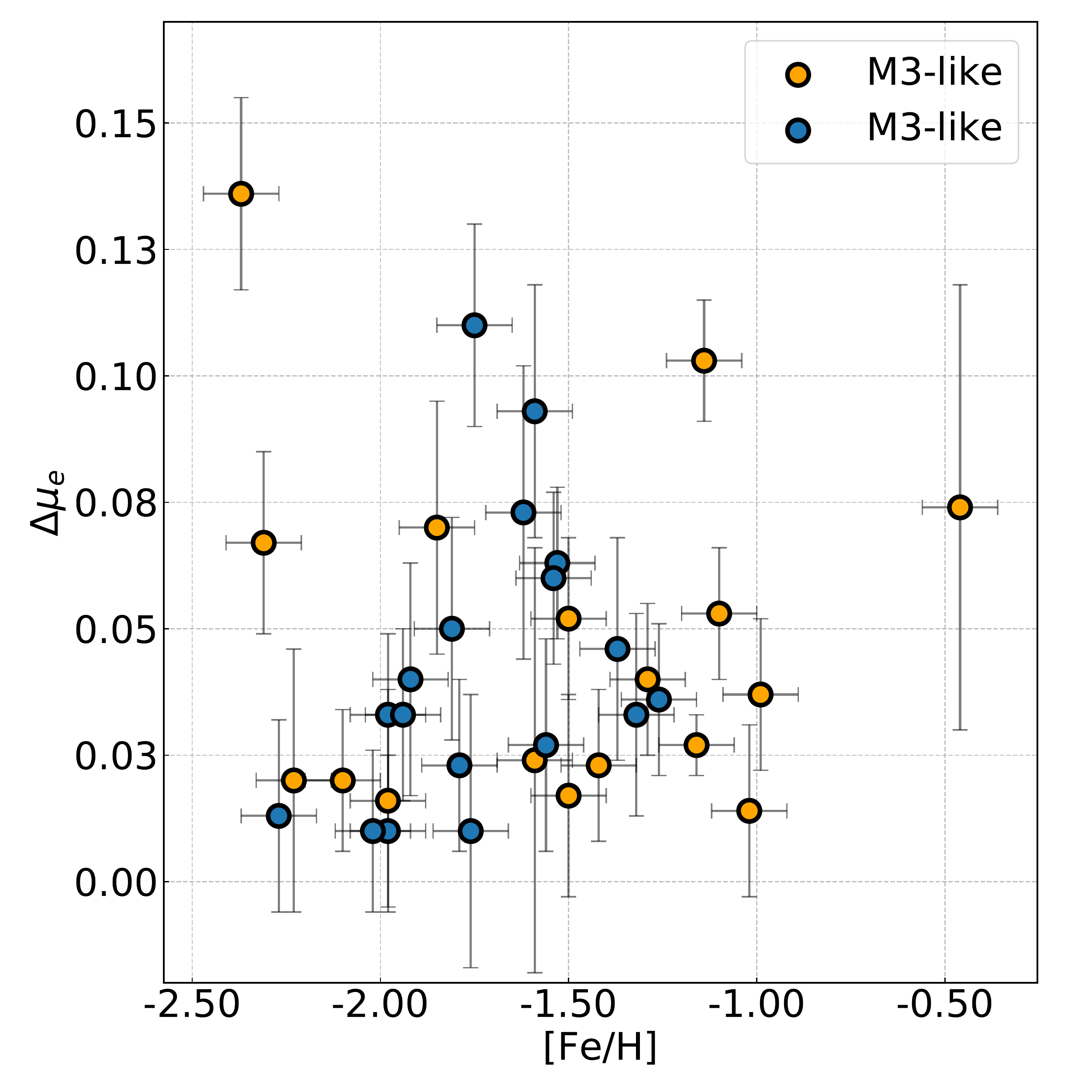}
    \caption{ 
    Difference between the mass loss of 2Ge and 1G stars, $\rm \Delta \mu_e$, against [Fe/H]. We color the M3- and M13-like clusters identified in Figure~\ref{pic:mu1g_feh_scheme}b, orange and blue, respectively.
    }
    \label{pic:dmue_feh}
\end{figure}

The mass-loss difference between 2Ge and 1G stars exhibits significant correlations with the maximum internal helium variation $\delta Y_{\rm max}$ (Figure \ref{pic:dmue_dymax}), the present-day and initial masses\footnote{From \cite{baumgardt_2018,baumgardt_2019}} of the host GCs (Figure~\ref{pic:dmue_mcluster}). 
M3- and M13-like GCs share similar patterns in all the panels of Figure~\ref{pic:dmue_dymax}. 
On the other side, there is no evidence for significant correlation between $\rm \Delta \mu_e$ and [Fe/H], as shown in Figure~\ref{pic:dmue_feh}.

\section{Discussion}
\label{sec:general_interpret}
The two main results of the analysis shown in Section~\ref{sec:results} are:
\begin{enumerate}
    \item There is a tight correlation between the average mass loss of 1G stars and the GC metallicity; if we separate the `second parameter' groups, the M13-like clusters follow a steeper relation than M3-like clusters.
    \item The mass loss of 2Ge stars is larger than the mass loss of 1G stars, and this mass loss difference is correlated with the helium abundance of the 2Ge stars and the mass of the host cluster.
\end{enumerate}
In this Section we discuss these results in the context of their respective fields of stellar astrophysics. In  subsection~\ref{sub:mass_loss} we compare the relation RGB 1G mass loss versus [Fe/H] with the mass loss resulting in models employing the Reimers' mass loss expression \citep{reimers_1975, reimers_1977}.  Subsection~\ref{sub:2par} is focused on the second-parameter problem of the HB morphology. Subsection~\ref{sub:multipop} discusses the impact of the result of a larger mass loss in the 2Ge on the formation scenarios of multiple stellar populations. 

\subsection{The mass--loss law}
\label{sub:mass_loss}
The problem of RGB mass loss is certainly as old as the first attempts to interpret the features of GC CMDs.
Figure \ref{pic:mu1g_feh_scheme}a shows a clear linear relation between $\rm \mu_{1G}$\ and [Fe/H].
This result depends on two main novelties of the present analysis: 1) we examined a large sample of GCs, \textit{by means of an homogeneous set of data and models}; 2) in the multiple population framework, \textit{we anchor the HB 1G stars to the coolest HB location in each cluster} which are then directly compared with the simulations. In this sense our relation is a direct measurement of the mass loss needed to describe the 1G stars.

Previous evidence that the mass loss depend on the cluster has been provided in several papers.
\citet{gratton2010} show a strong relation between mass loss and metallicity based on optical CMDs of a large sample of Galactic GCs. 
However, their results are based on the median mass of all HB stars and do not take the presence of MPs into account. 
As a consequence, the `median' HB masses derived by Gratton and collaborators typically correspond to 2G stars, which are helium enhanced and hotter than the 1G and are higher than those derived in our paper from 1G stars.  
 The difference is mostly due to the fact that the approach by \citet{gratton2010} does not disentangle between the effect of helium abundance and mass loss on the median HB masses thus overestimating the amount of mass loss. 
 \citet[][]{origlia_2014} inferred a similar relation between mass loss and cluster metallicity by using Spitzer Infra Red Array Camera photometry obtained in the 3.6-8 $\mu$m of 16 GCs. 
 Their results are based on mid-infrared excess of light in RGB stars, which was
interpreted as the results of dust formation around these stars \citep[but see][for alternative interpretation in the case of 47\,Tuc]{momany2012a}.
Recently, \citet{salaris_2013} and \citet{savino_2019} have proposed a relation between mass loss and metallicity from the study of the HB data of Sculptor and Tucana dwarf spheroidal galaxies, respectively. Both works involve direct comparisons of the observed HBs with synthetic HB stellar models.

\begin{figure}
    \centering
    %placeholders
    \includegraphics[width=0.9\columnwidth,trim={0cm 0cm 0cm 0cm}]{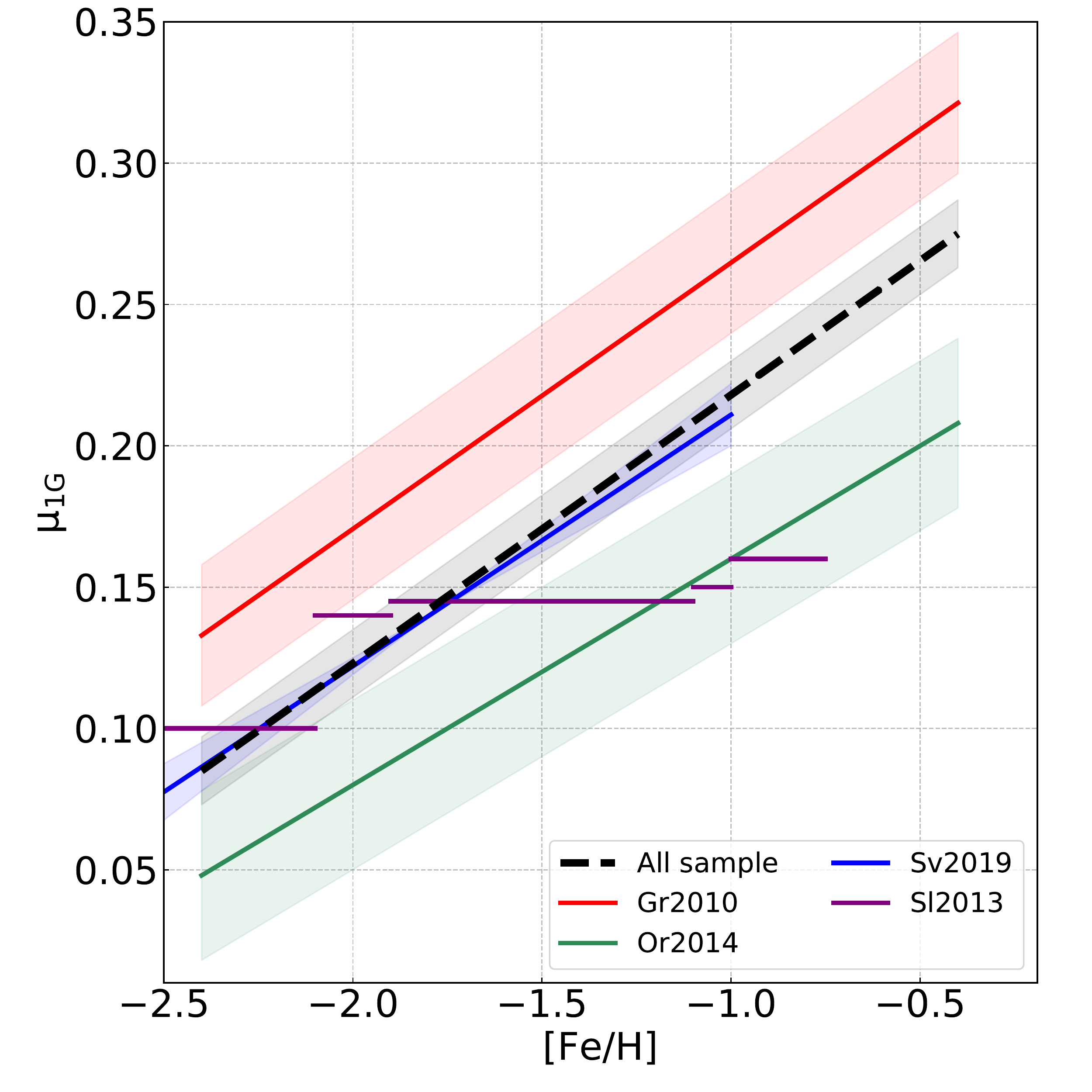}
    \caption{ Comparison of $\rm \mu_{1G}$ vs. [Fe/H] relations found in this work with other relations in literature. The shaded areas are the $1\sigma$ interval of each relation, when available.  Sources are \citet[][GR2010]{gratton2010}, \citet[][Or2014]{origlia_2014} ,\citet[][Sv2019]{savino_2019} and \citet[][Sl2013]{salaris_2013}.
    }
    \label{pic:mu_lit}
\end{figure}

For the sake of comparison we report in Figure \ref{pic:mu_lit} our Eq.\,1 and the different relations listed above, with their 1$\sigma$ intervals. Our relation is compatible with the one from \citet{savino_2019}.

\begin{figure*}
    \centering
    %placeholders
    \includegraphics[width=0.9\columnwidth,trim={0cm 0.5cm 0.2cm 0cm}]{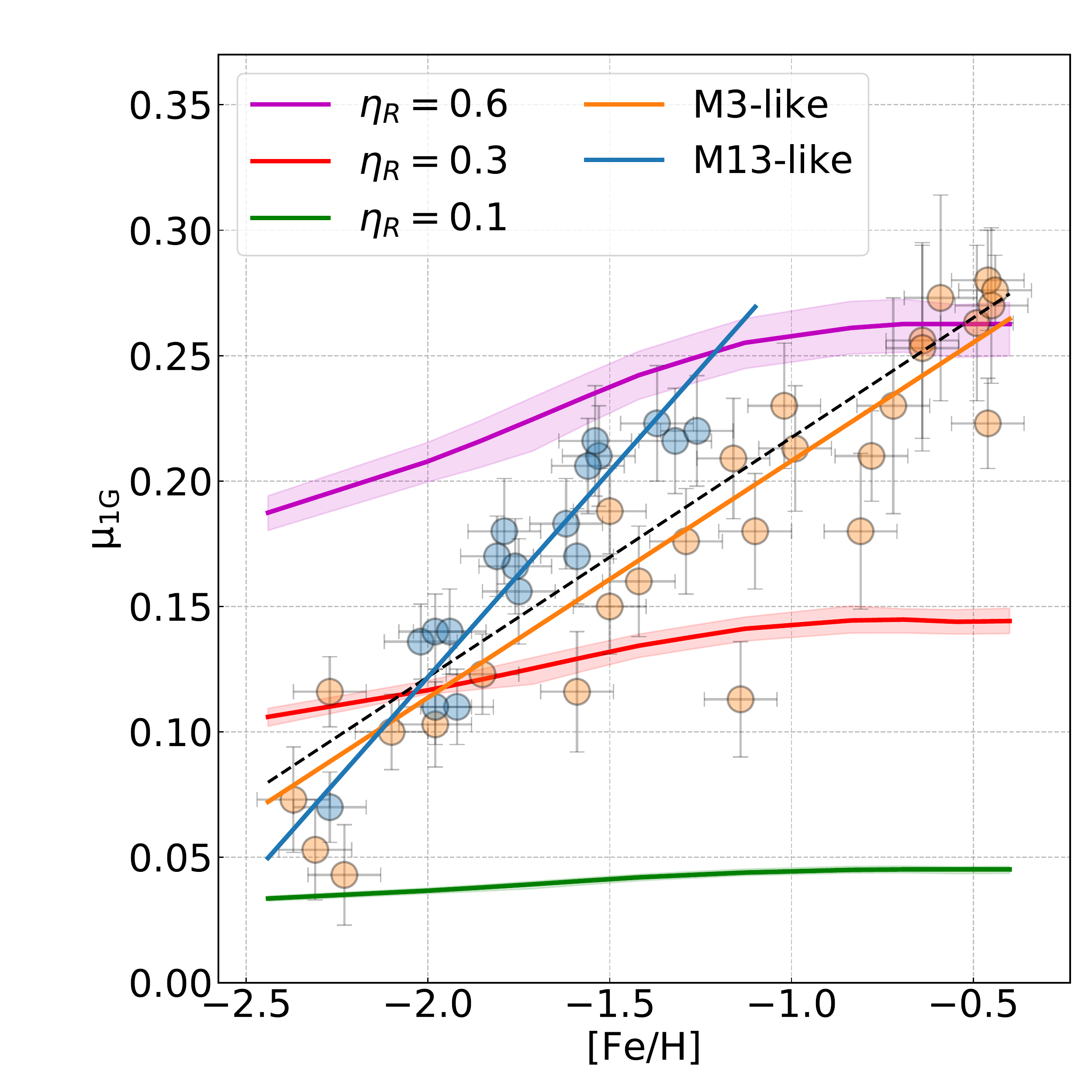}
    \includegraphics[width=0.9\columnwidth,trim={0cm 0.5cm 0.2cm 0cm}]{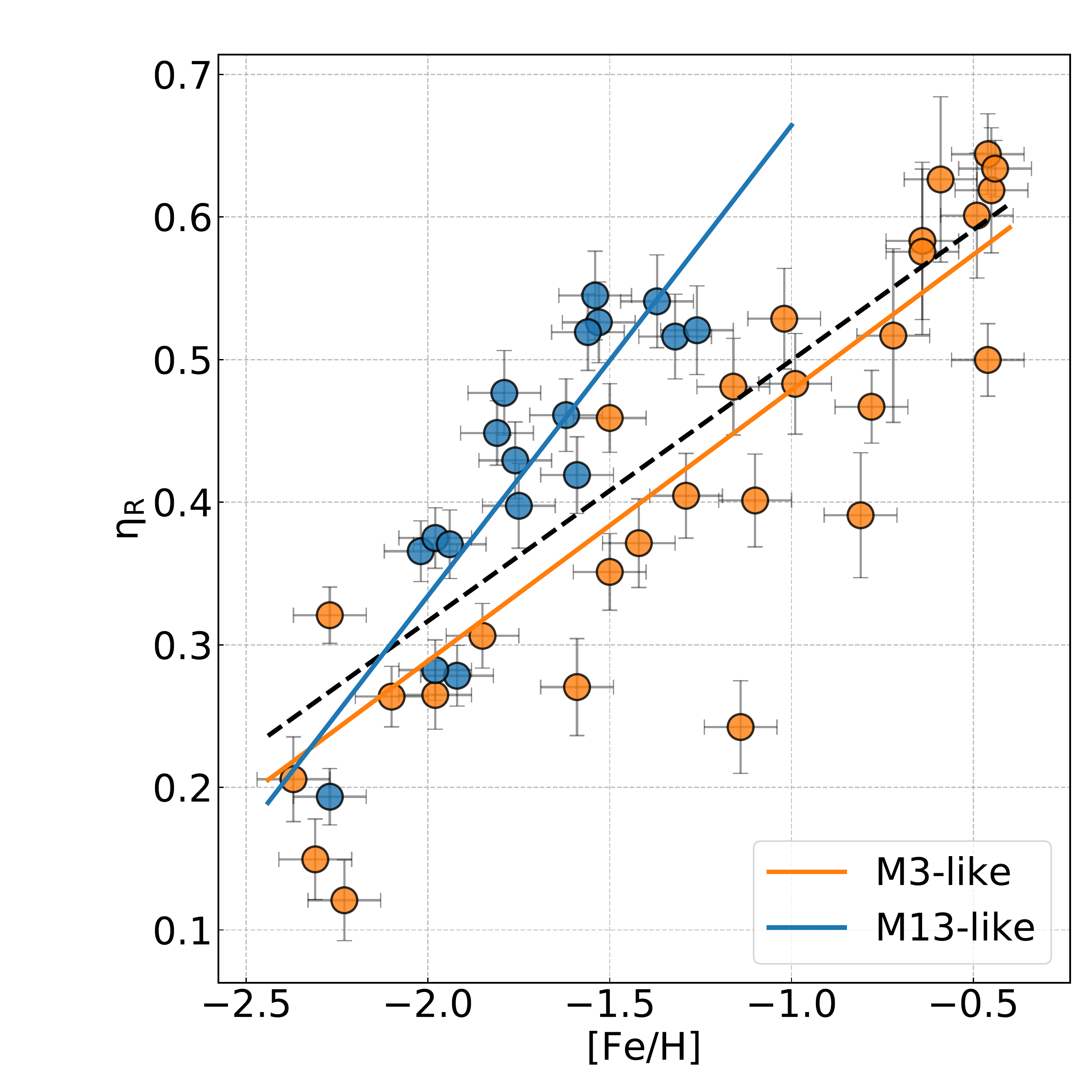}
    \caption{
    \textit{Left panel.} Reproduction of Figure~\ref{pic:mu1g_feh_scheme}a where we show the 1G mass loss against iron abundance for the studied GCs. The green, red and purple lines are the mass losses predicted by the Reimers'law for ages of 12 Gyr and $\eta_{\rm R}=0.1, 0.3$ and 0.6 respectively. The shaded areas include mass losses for ages between 13 and 11 Gyr. See text for details.  
    \textit{Right panel.} The value of $\eta_{\rm R}$ inferred for the 1G of each GC is plotted against cluster iron abundance. 
    The black-dashed, blue, and orange lines of both panels represent the best-fit straight lines derived for all GCs, and for M3-like and M13-like GCs, respectively. Their equations are provided in Table~\ref{Tab:rette}.
    }
    \label{pic:mloss_models_comparison}
\end{figure*}

Historically, the mass loss rate ($\dot{M_*}$, in $M_\odot/yr$) of late type giants has been described by \citet{reimers_1975, reimers_1977} formulation, given as 
 $\dot{M_*}\rm =4 \times 10^{-13}\eta_R  L_*~R_*/M_*$ with L$_*$, R$_*$, M$_*$ as stellar luminosity, radius and mass, respectively, given in solar units, and $\eta_R$ is a fitting parameter.
In the left panel of Figure~\ref{pic:mloss_models_comparison}, we show the relations between mass loss and [Fe/H] derived from the Reimers' law for $\rm \eta_R$ = 0.1, 0.3 and 0.6 (green, red and purple colors).  The continuous lines correspond to ages of 12 Gyr and the shaded areas enclose the mass losses derived from models between 11 Gyr (upper boundaries) and 13 Gyr (lower boundaries). 
Clearly, the relations between 1G mass loss and [Fe/H] derived from the observations are steeper than those expected from the Reimers' law for all the choices of $\rm \eta_R$. 
We conclude that a constant value of  $\rm \eta_R$ does not properly describe the mass loss of 1G stars in GCs.
We stress again that the relation we find refers to the 1G only, and not to the whole HB. Also recently, \citet{mcdonald2015} find  $\eta_R=0.477 \pm 0.007$, a value independent of metallicity, because they select a `median' HB mass, extracted from \citet{gratton2010}, to compute the RGB mass loss.

A match with the observed $\rm \mu_{1G}$\ versus  [Fe/H] relation can then be obtained by assuming a dependence between the fitting parameter $\rm \eta_R$ and the [Fe/H].  To do this, we estimate the value of $\rm \eta_R$ that reproduces mass loss and metallicity of each GC, and plot in the right panel of Figure~\ref{pic:mloss_models_comparison} $\rm \eta_R$ against [Fe/H]. 
$\rm \eta_R$ must increase with the cluster metallicity and we can describe mass-loss of 1G stars by the Reimers' formula with: 
 \begin{equation}
\begin{split}
\rm \eta_R=(0.183\pm0.021)\times [Fe/H]+(0.682\pm0.032);\\
%\rm R_P=0.79.
\end{split}
\label{eqn:etar_1}
\end{equation}
The expression above is the least-squares best-fit straight line derived from the points plotted in the right panel of Figure~\ref{pic:mloss_models_comparison}. Noticeably, M3- and M13-like GCs define different relations in the $\rm \eta_R vs.\,$[Fe/H] plane (see equations in Table~\ref{Tab:rette}).  

\subsection{Second parameter problem}
 \label{sub:2par}
 In the following, we discuss the results of this paper in the context of the second-parameter of the HB morphology. Specifically, in Section~\ref{sub:2pA} we discuss the impact of mass loss of 1G stars in shaping the HB, while  Section~\ref{sub:2pB} is focused on the mass-loss difference between 2Ge and 1G stars.
  
 \subsubsection{1G mass loss and HB morphology}
 \label{sub:2pA}
 
 The evidence that the mass loss of 1G stars exhibits a strong correlation with the iron abundance of the host cluster confirms that  metallicity is the first parameter in shaping the HB of GCs.
 The different HB morphology of the M3- and M13-like GCs, highlighted by the higher values of L1 \citep{milone_2014} in the M13-like class, is the evidence that at least a second parameter is at play.
 Our analysis nicely shows (Figure \ref{pic:mu1g_feh_scheme}c) the straightforward result that M13-like GCs exhibit higher $\rm \mu_{1G}$ than the other GCs with similar metallicity {\it if no other input is playing a role}. From the similar age determination for the clusters in our sample, we know that this second parameter can not be the age.

 If we apply to the M3- M13-like clusters the scenario which will be discussed in subsection~\ref{sub:multipop}, the larger mass loss of 1G stars in M13-like clusters may indicate that they formed in high-density environment compared to the 1G of M3-like GCs with the same metallicity.

It is worth to discuss two other possible scenarios:

\begin{enumerate}
\item The 1G stars of the M13-like clusters may be born with a helium content larger than in the M3-like clusters. For example in the case of M\,13, by combining the results in Section~\ref{sec:corr_1g} with the variation of $\rm M^{Tip}$ with helium, $\rm \Delta M^{Tip}=-1.33\times \delta Y$, obtained from our isochrones database, we can reproduce the 1G HB stars by assuming an helium content Y$\sim$0.27 and the same mass loss of M3. Similar conclusion can be extended to all M13-like clusters. This scenario would imply that the 1G stars in this group of GCs originated from a primordial cloud that was already helium enhanced, but it would be challenging to explain the origin of such a high helium abundance in the gas at high redshift.
\item An intriguing explanation has been put forward by \cite{dantona_2008} in their study of the second-parameter GC pair M\,3-M\,13. As predicted by various scenarios for the formation of multiple populations, GCs would lose a large amount of 1G stars as they evolve \citep[see e.g.][and references therein]{renzini_2015}. Hence, M13-like GCs could be extreme cases of clusters that have entirely lost their 1G. In this scenario, the stars that we used as 1G in M13-like GCs are actually fake-1G \citep[see][and references therein for details]{dantona_2008, dantona_2016} and, as such, we can apply to them the scenario discussed below in subsection\,\ref{sub:multipop}, and suggest that have a small helium enhancement, to which we can associate and extra-mass loss. Taking into account the `2G-like mass loss', the helium needed to fit the cooler side of the HB in M\,13 is a quite modest Y$\lesssim$0.26, a value very close to the standard 1G value. 
 These fake 1G stars, with a very modest helium enhancement, and correspondingly small or negligible abundance anomalies, are identified, for instance, in NGC\,2808 as the stars of the `population C'\citep{dantona_2016} or with the population G1 in \citet{kim_2018}, and should be in fact the 2G populations closest in abundances to the 1G. 
 \end{enumerate}
 
 \subsubsection{Enhanced mass loss of 2Ge stars and HB morphology}\label{sub:2pB}
When the presence of multiple populations in GCs was not yet clear, and researchers did not expect that the helium content of GC stars differ, even significantly, it was obvious that the more extended in color was the HB, the larger had to be the mass loss spread during the red giant evolution. 

Introducing the helium content as a parameter of the morphology resulted to be a possible way to reduce this mass loss spread, because, for fixed age and metallicity, helium rich stellar populations exhibit bluer HBs than  stellar populations with pristine helium content. It appeared possible to attribute to the enhanced helium the hotter locations along the HB \citep[e.g.][]{dantona_2002, dantona_2005, gratton2010} and maybe get rid of the differences in mass loss. This fact was also consistent with the strong correlation between the maximum helium variation derived from multi-band photometry of GCs, and the color extension of the HB \citep[L2,][]{milone_2014, milone_2018}. 

\begin{figure*}
    \centering
    \includegraphics[width=0.9\columnwidth,trim={0.0cm 0.0cm 0.0cm 0.0cm}]{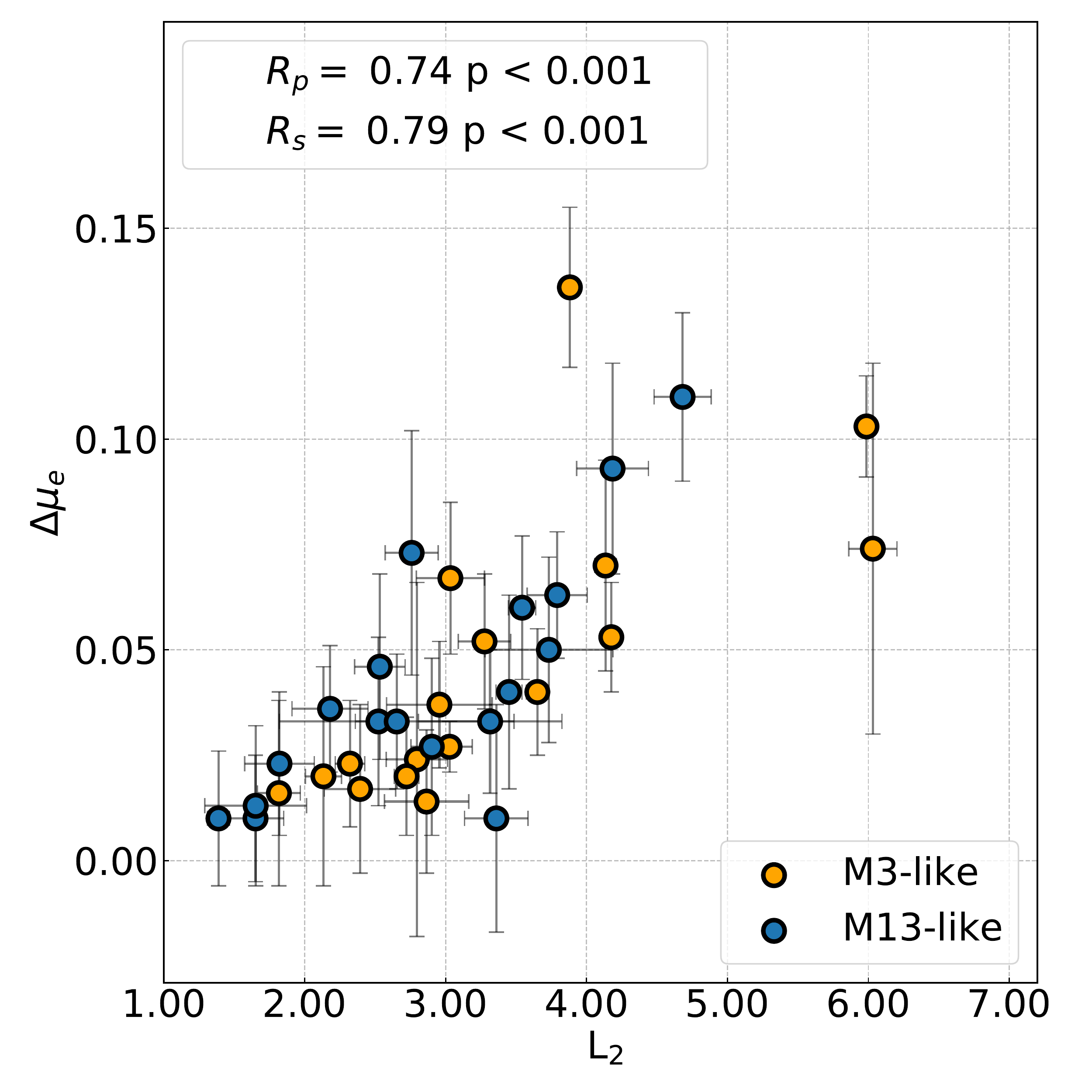}
    \includegraphics[width=0.9\columnwidth,trim={0.0cm 0.0cm 0.0cm 0.0cm}]{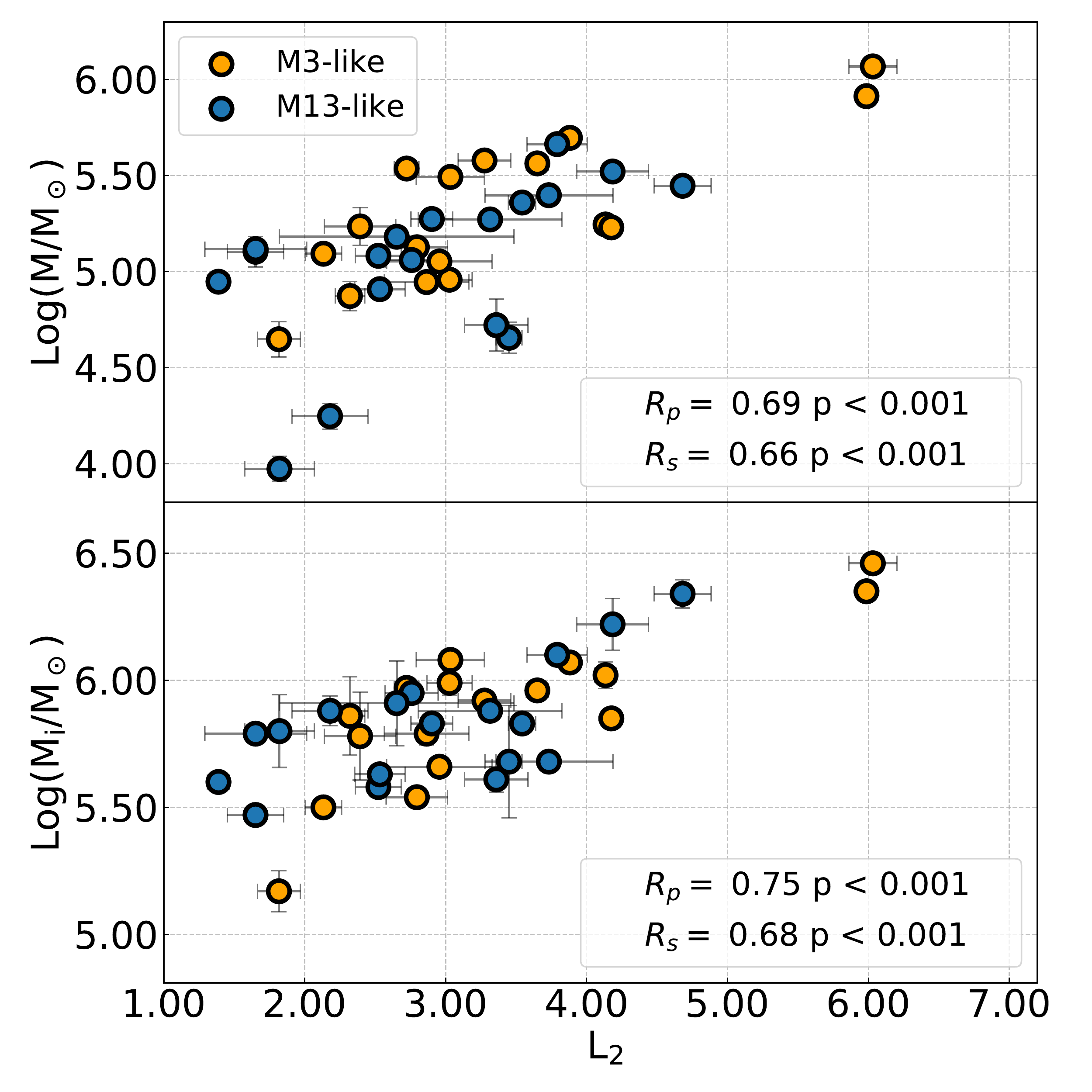}
    \caption{ \textit{Lef panel:} Difference between the mass loss of 2Ge and 1G stars, $\rm \Delta \mu_e$, against the F275W - F814W colour extension of the HB. The groups of M3- and M13-like clusters are represented as orange and blue dots, as labelled. \textit{Right panel:} Present day and initial cluster mass against the colour extension of the HB (top and bottom panels respectively). }
    \label{pic:figure_l2_m18}
\end{figure*} 

Nevertheless, here we found conclusively that helium variations alone are not able to entirely reproduce the observed HBs \citep[see also][]{tailo_2019b, tailo_2019a}, and that extra-mass loss is necessary to fit the observations. The left panel of Figure \ref{pic:figure_l2_m18} shows that $\Delta \mu_e$ correlates with the color extension of the HB (the parameter L2). 
The right panel shows the correlation between L2 and the total present-day mass of clusters.
Previous suggestions for the presence of such a type of correlation can be found in the literature, e.g. in \citet{recioblanco2006, gratton2010}.  A similar correlation is found by adopting the initial masses \citep[from][]{baumgardt_2019}. 
We conclude here that {\it the enhanced mass loss of second-generation stars,  together with helium variations are the parameters which determine the HB extension in GCs}, and both are related to the present-day and initial mass. In addition, we are driven to conclude that helium variations and extra-mass loss may be two concomitant but different aspects of the formation of multiple populations, as we are going to outline in the next.

\subsection{Extra-mass loss in the 2Ge as a tracer of the multiple populations formation }\label{sub:multipop}
The discovery that 2Ge stars of all GCs lose more mass than the 1G \citep[Figure~\ref{pic:hbparam_mu1vsmu2}, see also][]{tailo_2015, tailo_2019a, tailo_2019b}  
 and, in particular, the correlations between the extra-mass loss $\Delta \mu_e$\ and the extra-helium content of the 2Ge population $\delta Y_{max}$\ (Figure\,\ref{pic:dmue_dymax}) and the total (present-day and initial) mass of the clusters (Figure\,\ref{pic:dmue_mcluster}) represent powerful constraints for the mechanisms of formation of multiple stellar populations. Since evolutionary effects alone can not provide such increase (see Appendix \ref{app:standard_flash}) the need of an additional mechanism arises, allowing us to gain information on the environment where the distinct generations were born. 

\begin{figure*}
    \centering
    \includegraphics[width=1.8\columnwidth,trim={0.0cm 0.0cm 0.0cm 0.0cm}]{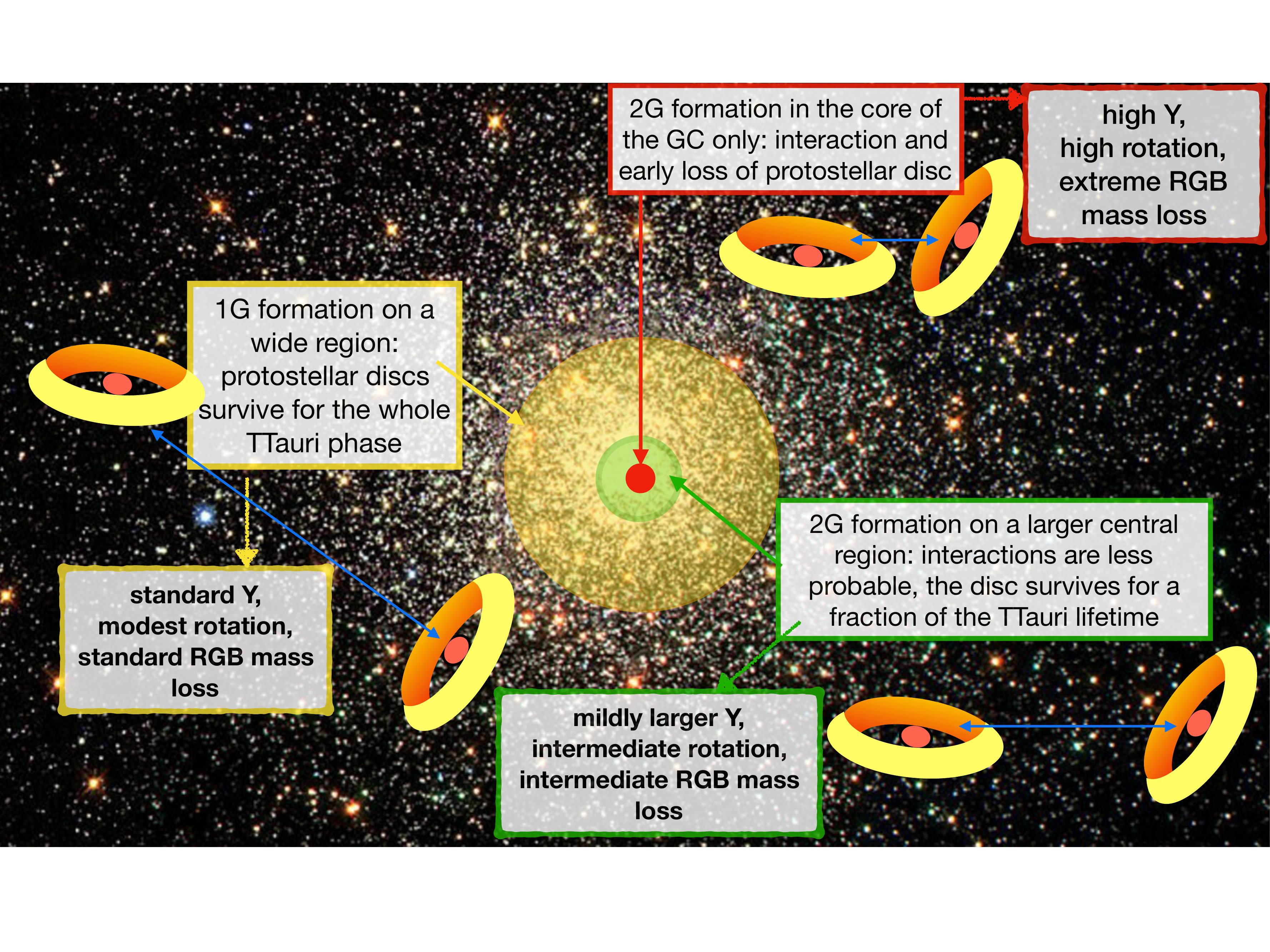}
    \caption{Schematic interpretation of the different rotation of 1G and 2G stars. During the pre-main sequence evolution, the rotation of the central contracting star is magnetically locked to the rotation of its residual accretion disk, having a diameter of 50--100\,AU. While the 1G formation occurs in an ambient of moderate density (yellow region), where the disk-disk interactions occur on a timescale longer than the typical stellar contraction time (few $10^6$yr), the 2G formation occurs in much smaller central regions (green and red regions, representing less and more extreme 2G populations; these regions represent either  star formation in the same cluster at successive times, or star formation in clusters of different initial mass), where collisions and an early loss of the accretion disk are much more probable. Free stellar contraction and increase of the rotation rate follow the loss of the disk. Rapidly rotating cores delay ignition of the core-helium flash and allow a larger RGB mass loss.  
    This scheme, supported by hydrodynamical simulations, is at the basis of the direct correlation between the extra-mass loss, the helium over-abundance of the 2Ge populations, and the total initial mass of the cluster.}
    \label{pic:schema}
\end{figure*}

\citet{tailo_2015} discuss, in the specific context of the blue hook morphology of $\omega$\,Cen, how the formation environment of the different multiple populations can affect the stellar evolution during the RGB phase, mainly by affecting the initial stellar rotation.
Here we apply the same scenario to the less extreme environments encountered by the 2Ge of most clusters (see the schematic illustration in Figure \ref{pic:schema}).  
Very young low mass stars ($M/M_\odot\sim 0.1-1.0$) behave as the T\,-Tauri stars in the Galactic field, pre-main sequence stars in the convective phase of contraction towards the main sequence.
T\,-Tauri's typically rotate with a period of 1--12\,d, and their rotation rate is maintained constant in time by the magnetic coupling of the star with the accretion disk \citep[magnetic disk-locking, see ][]{armitage_1996}. \citet{bouvier_1997a,bouvier_1997b} developed a model to explain why main sequence stars in young clusters (such as $\alpha$\,Persei) show a wide distribution of rotation velocities. They showed that, the earlier would pre-main sequence stars break the magnetic disk-locking, the faster would be their rotation when they reach the main sequence. Afterwards, during the main-sequence phase, the surface rotation rate (together with the rotation of the whole convective envelope) slows down thanks to the angular momentum loss associated with the stellar winds, but the core mostly preserves its fast rotation.

Rapid rotation affects the structure of the helium core, and delays the ignition of the helium core flash. This implies that, the faster the stellar core rotates, the more is the flash delayed, and the more time has the star to lose mass during the RGB evolution. 
Dispersion in core rotation within stars belonging to a GC was suggested, e.g. by
\citet{fusipecci_1978}, to interpret the mass dispersion among the stars populating the HB (that is, the HB more or less extended morphology).

Here we apply the key concept of this scenario in the context of the formation of multiple populations and propose that the different rotation rates and mass loss may be the result the different formation environments for 1G and 2Ge. In fact, 2G stars form in denser environments in the innermost regions of a more extended 1G system \citep[e.g.\,][]{dercole_2008, dercole2016, calura_2019}. A larger environment density implies more frequent star-to-star dynamical interactions able to destroy the stellar disks early enough, when the T\,Tauri  radius is significantly larger than the final main sequence radius. The star is then free to contract while preserving its angular momentum at the time  of detachment and, as the inertia momentum decreases, its rotation  rate increases \citep[see][]{tailo_2015}.

Hydro-dynamical simulations following the 2G formation show that the most extreme 2G stars (those less diluted with the environment gas, and thus preserving a larger helium content) are born in denser regions \citep[e.g.][]{calura_2019}, where the stars may have shorter disk lifetime, gain higher rotation rates and eventually suffer larger mass loss during the RGB evolution. On the other hand, less extreme 2G (those with a smaller  helium enhancement formed with the dilution of processed gas and pristine gas), form less concentrated than the most extreme populations \citep[see again][]{calura_2019} and are thus subject to less frequent interactions during the pre-main sequence phase resulting in slower rotation and a smaller difference in the RGB mass loss. This scenario thus provides a possible explanation for the observed  correlation between the $\delta Y_{max}$ and $\rm \Delta {\mu}_{e}$.

Figure\,\ref{pic:dmue_mcluster} shows that $\rm \Delta {\mu}_{e}$\ is also correlated with the clusters' present-day and initial masses. Although this correlation is not surprising in consideration of the fact that
\citet{milone_2018} ---their Fig.\,13--- found a very tight correlation between $\delta Y_{max}$\ and the clusters' masses, it provides an additional hint concerning the key ingredients for the formation of multiple populations.

In the general context of the framework based on the pre-main sequence early-disc loss, the link between $\rm \Delta {\mu}_{e}$, $\delta Y_{max}$\ and cluster mass further strengthen the possibility of a fundamental connection between formation, very early dynamics and helium abundance of multiple populations. 
The simulations by \citet{calura_2019} show that, in general, more extreme 2G populations form more concentrated than the less extreme (more diluted) 2G population and this, in turn, as discussed above, is consistent with the correlation found in this study between $\delta Y_{max}$\  and $\rm \Delta {\mu}_{e}$. The additional link with the cluster's mass indicates that low-mass clusters tend to form less extreme  populations (i.e. the 2Ge of low-mass clusters has smaller $\delta Y_{max}$\ as indicated by the correlation between  $\delta Y_{max}$\ and  the cluster's mass \citep{milone_2018}. This is a manifestation of an additional ingredient in the formation process: although this aspect certainly requires further investigation, we point out here that this trend is in general consistent with the model of pristine diluting gas re-accretion discussed in \citet{dercole2016}. In that framework, the dynamics of the early gas expulsion and re-accretion of pristine gas  can lead to a more rapid re-accretion in low-mass clusters and the formation of a less extreme and less concentrated population compared to the more extreme and undiluted population forming in more massive clusters.

We conclude that the correlations of $\rm \Delta {\mu}_{e}$\ with the global parameters of the host clusters (helium content of the extreme population, present and initial mass of GC) are then fossil traces of the formation process, that can be explained in the framework of the formation of the 2G in more or less compact regions of the clusters. A schematic representation of this model is given in Figure \ref{pic:schema}. 
Specific models of these cases, along the line of computation presented in \citet{tailo_2015}, are planned for a future work.

We note, anyway, that \citet{tailo_2015} work was motivated by the necessity of explaining the peculiar shape of the blue hook stars in $\omega$\,Cen. This required an extreme increase (up to $\sim 0.04 M_\odot$) of the core mass at the helium flash, and therefore extreme values of main sequence rotation rate.  In a more general scenario, we are considering a negligible increase of the core mass at flash, and a small increase of the luminosity at which the flash occurs, able to provide $\sim$ 0.01-0.05$M_\odot$\ more mass loss (in most cases), corresponding to a small 2--5\% increase in RGB lifetime.

 In the case of Fornax GCs, \cite{dantona_2013} show that a main sequence rotation rate of $\sim$70$\mu$Hz is enough to increase mass lost by $\sim 0.03 M_\odot$, in the approximation by \cite{mengel_1976} and \cite{renzini_1977}. The models adopted for the estimate were based on the crude approximation of shell angular momentum conservation, which must be revised, based on the asteroseismic data for the the core rotation of red giants presently available \citep{aerts_2019}. Nevertheless, the low asteroseismic rotation velocities measured are at high variance also with the most recent models including the known mechanisms for the transport of angular momentum \citep[e.g.][]{cantiello_2014}. There is only one star of low mass (0.84$M_\odot$) for which asteroseismic data have been analyzed so far, the Kepler early red giant KIC 7341231 \citep{deheuvels_2012}. Interestingly, the authors find a core rotation of $\omega$=0.7$\mu$Hz, a factor at least five times larger than the envelope rotation, but the rate is too low to be compatible with the available models. Nevertheless, it is quite possible that the evolution is different for the stars formed in dense stellar regions, such the 2G stars of GCs we are considering here, and there are no data available to rule out  velocities a factor 100 larger needed by  the model.

\section{Summary and conclusions}
\label{sec:conc}

We combined high-precision photometry from the UV Legacy survey of Galactic GCs \citep[][]{piotto_2015,nardiello_2018} and stellar population models to homogeneously analyse, for the first time, the HBs of a large sample of 46 clusters. 

We identified candidate 1G, 2G and 2Ge stars along the HB and inferred their mass losses and average HB masses by following the recipe from pilot papers on M\,3 and M\,4 by \citet{tailo_2019b, tailo_2019a}. In particular, we used helium abundances inferred by \citet{milone_2018} to fix the helium content of the 1G, 2G and 2Ge stars and break the degeneracy between helium and mass loss, which is one of the main challenges of HB studies.  

The main results include:
\begin{enumerate}[(i)]

\item Mass loss of 1G stars ranges from $\sim0.05$ to $\sim$0.25 $\rm M_{\odot}$ and correlates with the iron abundance of the host GCs. The resulting linear relation, which is described by Equation\,\ref{eq:mu1_ge_rel}, is not reproduced by the Reimers' mass loss law for fixed $\rm \eta_{R}$.
We use our determinations of mass loss to derive the {\it empirical law} that describes mass loss in 1G stars of GCs, described by Equation\,\ref{eqn:etar_1}.

\item The strong correlation between $\rm \mu_{1G}$ and [Fe/H] corroborates the evidence that metallicity is the main parameter of the HB morphology. The finding that the 1G of M13-like GCs  have higher $\rm \mu_{1G}$ values than 1G stars in M3-like GCs with similar metallicities, suggests that mass loss, after metallicity is a second parameter that determines the HB shape.
 As an alternative, 1G stars of M13-like GCs are enhanced by $\sim 0.01-0.03$ in helium mass fraction and share the same mass losses as the 1G of M3-like clusters. Otherwhise.
 following \citet{dantona_2008}, we speculate that M13-like clusters have entirely lost their true 1Gs. Hence, the stars that we call 1G are second-generation stars slightly enhanced in helium.    

\item 2Ge stars of all studied GCs lose more mass than the corresponding 1Gs, thus confirming the conclusion by \citet{tailo_2015,tailo_2017, tailo_2019b, tailo_2019a} on $\omega$\,Cen, NGC\,6441, NGC\,6388, M\,4 and M\,3. 
 The mass loss difference between 2Ge and 1G stars, $\rm \Delta \mu_{e}$ correlates with the maximum internal helium variation and the mass of the host GC. Previous papers provided empirical evidence that the internal helium variation associated with multiple populations correlates with the color extension of the HB.
  We show that helium variations alone do not entirely reproduce the observations, and that enhanced mass loss of 2Ge stars, in addition to helium, is needed to explain the observed HBs. This finding provide further evidence that mass loss, together with helium variation, are second parameters of the HB morphology. 
  
  \item Results on mass loss may provide information on the formation environment of the distinct populations.  
   The scenario proposed by \citet{tailo_2015} suggests that the accretion disks of pre-main-sequence 2Ge stars are disrupted at early stages by dynamical interactions in the dense environment of the innermost cluster regions. As a consequence, 2Ge stars exhibit faster rotation rates of the core and prolonged life as red giants, which result in the mass-loss increase.     
  Our finding of high mass loss in 2Ge stars relative to the 1G is consistent with this scenario and provide evidence that 2Ge stars  formed in the dense cluster centers.  

\end{enumerate}

\section*{Acknowledgements}
We thank Holger Baumgardt for providing the initial masses of GCs. We also thank the anonymous referee for her/his accurate report and on point comments which helped improve the original manuscript.  
This work has received funding from the European Research Council (ERC) under the European Union's Horizon 2020 research innovation programme (Grant Agreement ERC-StG 2016, No 716082 'GALFOR', PI: Milone, http://progetti.dfa.unipd.it/GALFOR), and the European Union's Horizon 2020 research and innovation programme under the Marie Sklodowska-Curie (Grant Agreement No 797100, beneficiary Marino). MT, APM and ED acknowledge support from MIUR through the FARE project R164RM93XW SEMPLICE (PI: Milone). MT and APM  have been supported by MIUR under PRIN program 2017Z2HSMF (PI: Bedin).
%The Acknowledgements section is not numbered. Here you can thank helpful
%colleagues, acknowledge funding agencies, telescopes and facilities used etc.

\section*{Data Availability}
The photometric catalogues underlying this article are available in the HST UV legacy survey public database \citep{piotto_2015,nardiello_2018}, at http://groups.dfa.unipd.it/ESPG/treasury.php. The models obtained via the ATON 2.0 code are not yet available to the public but they can be provided upon request.

%%%%%%%%%%%%%%%%%%%%%%%%%%%%%%%%%%%%%%%%%%%%%%%%%%



\bibliographystyle{mnras}
\bibliography{HBGC}

%%%%%%%%%%%%%%%%%%%%%%%%%%%%%%%%%%%%%%%%%%%%%%%%%%

%%%%%%%%%%%%%%%%% APPENDICES %%%%%%%%%%%%%%%%%%%%

\appendix
\section{The analysed HB sample}

\begin{figure*}
    \centering
    %placeholders
    \includegraphics[width=0.66\columnwidth]{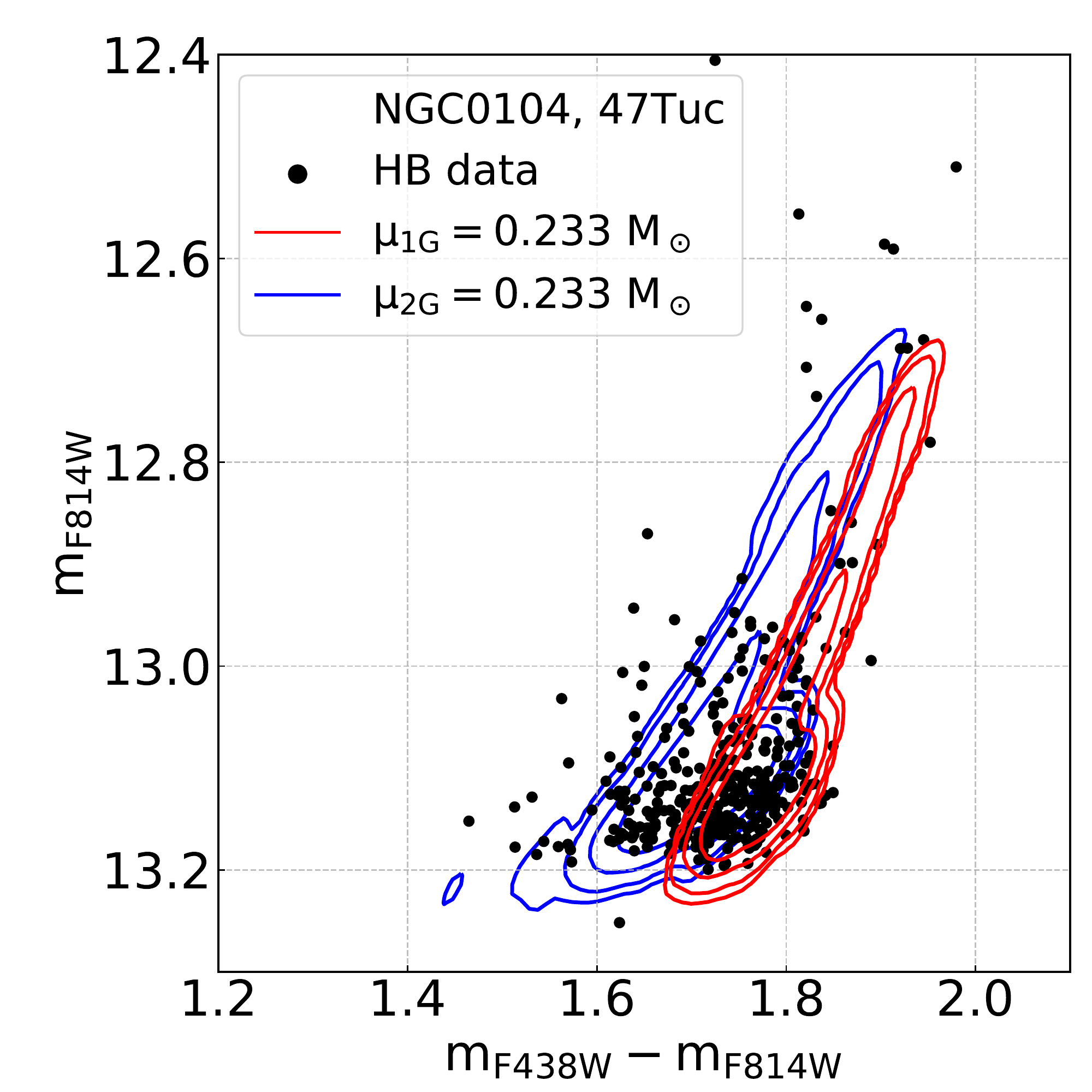}
    \includegraphics[width=0.66\columnwidth]{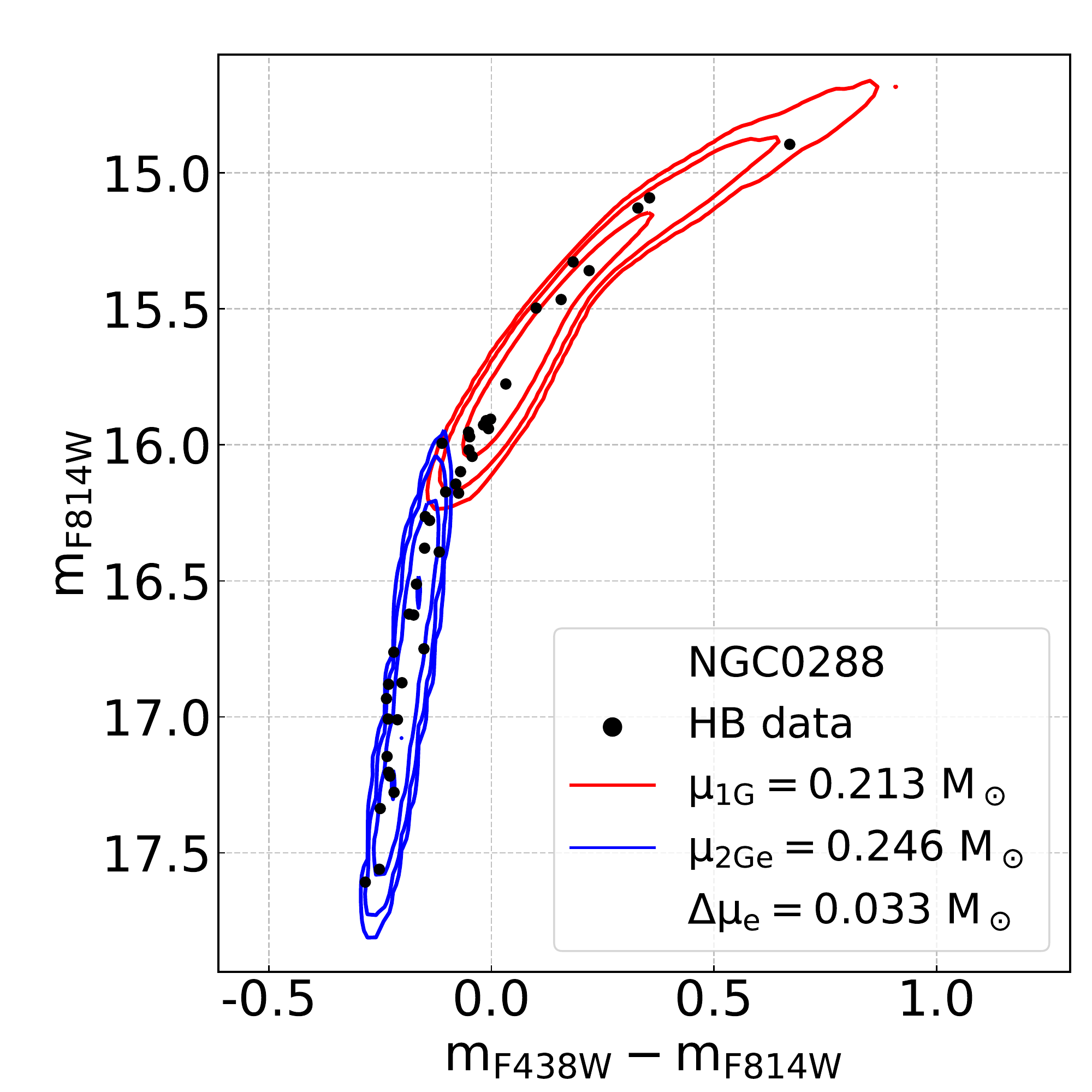}
    \includegraphics[width=0.66\columnwidth]{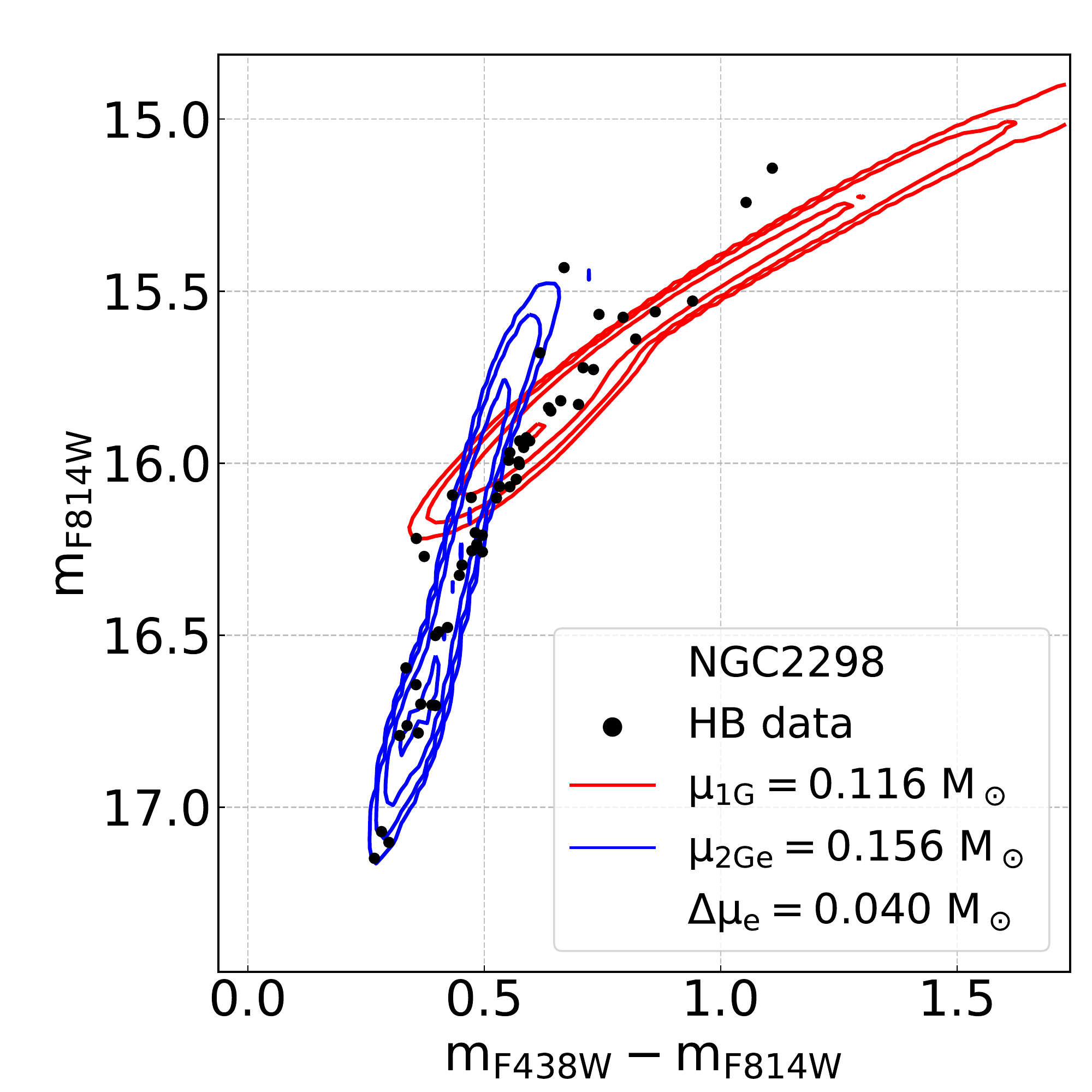}

    \includegraphics[width=0.66\columnwidth]{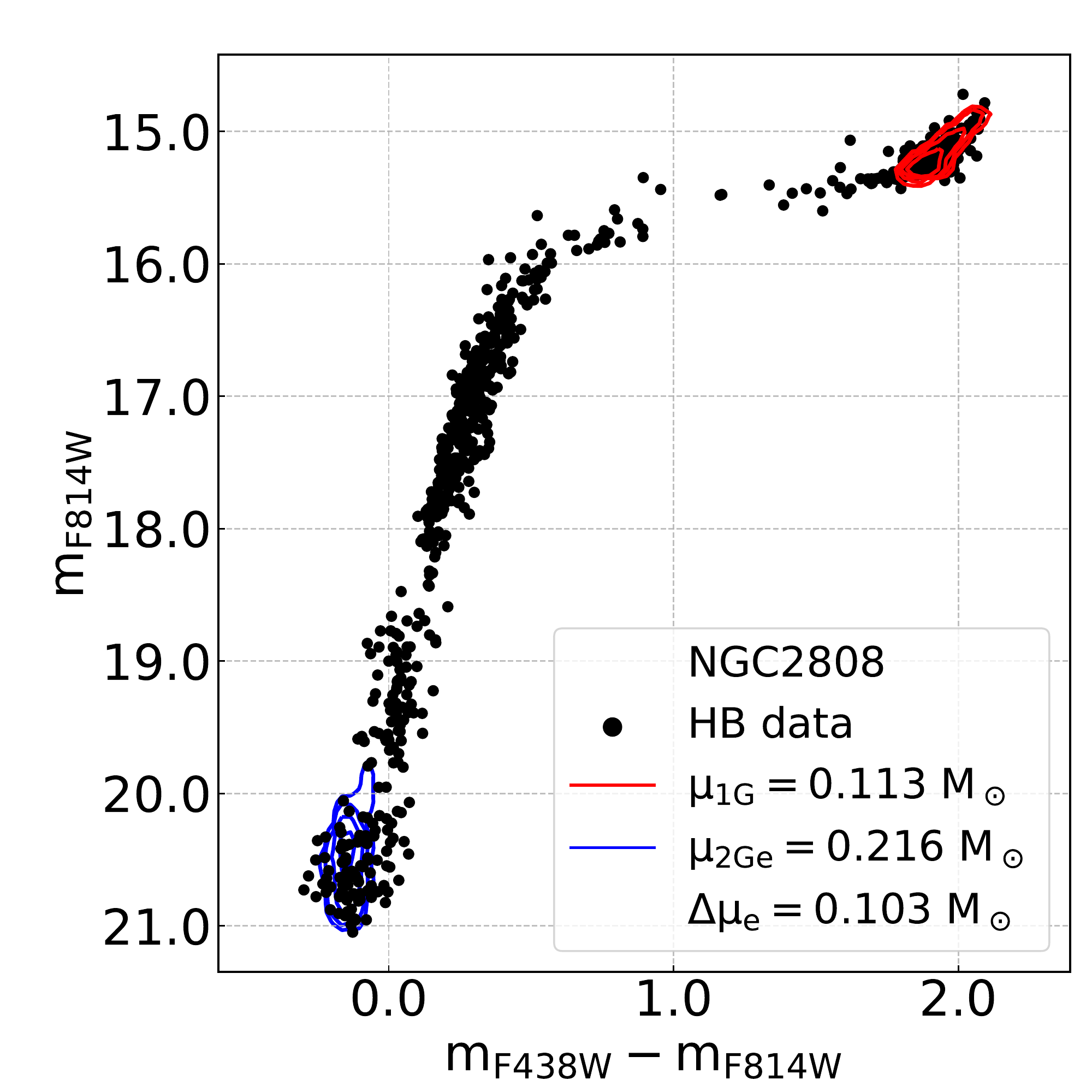}
    \includegraphics[width=0.66\columnwidth]{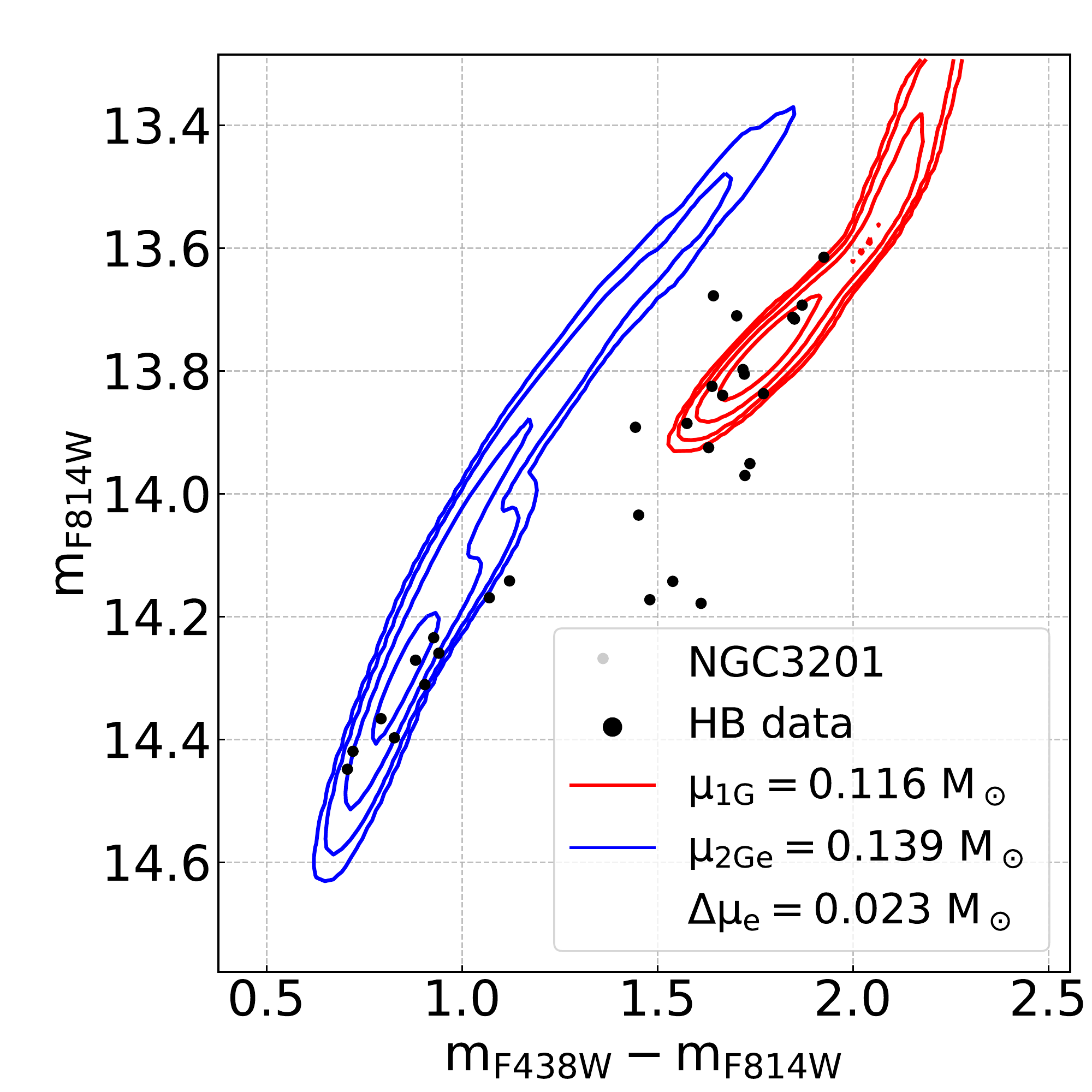}
    \includegraphics[width=0.66\columnwidth]{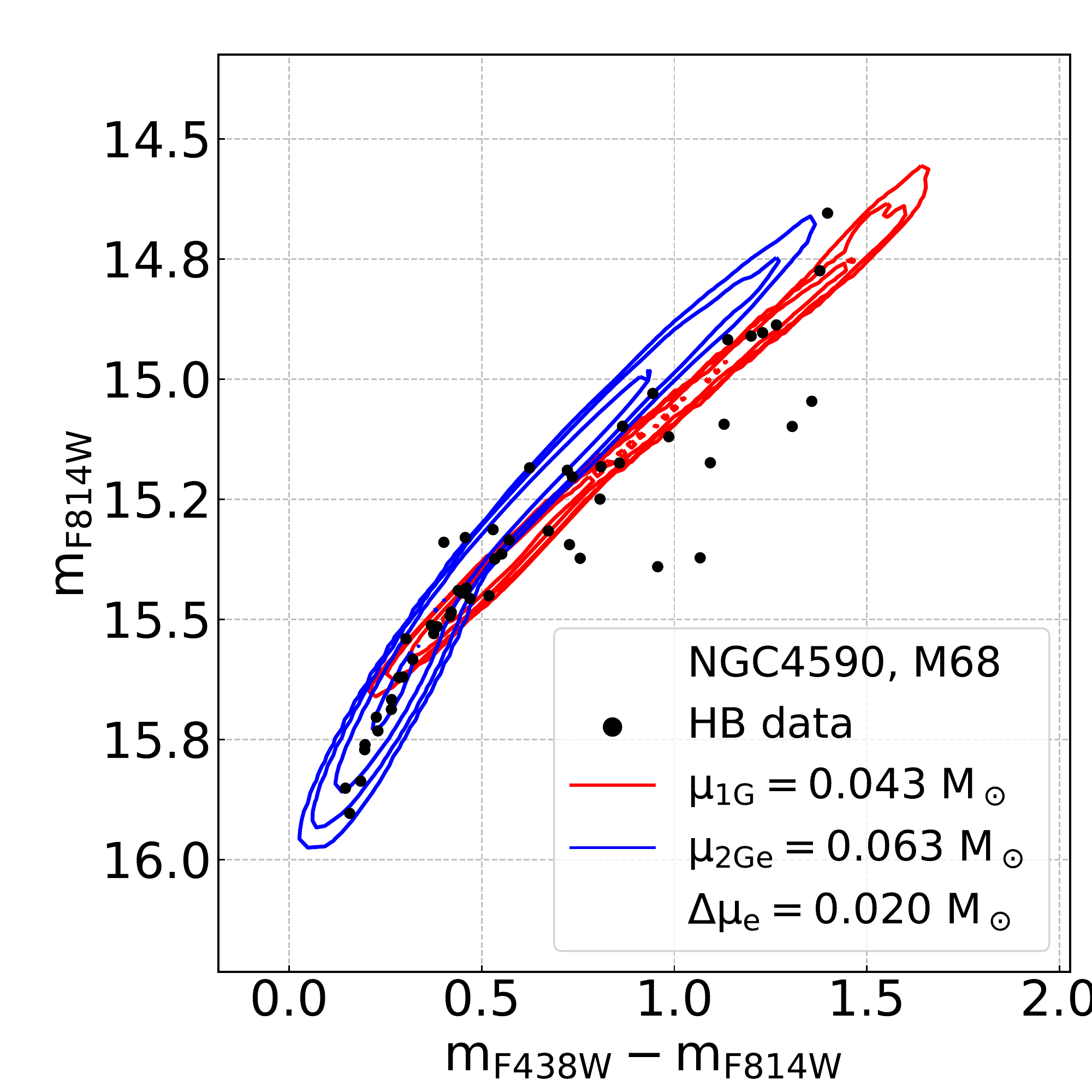}

    \includegraphics[width=0.66\columnwidth]{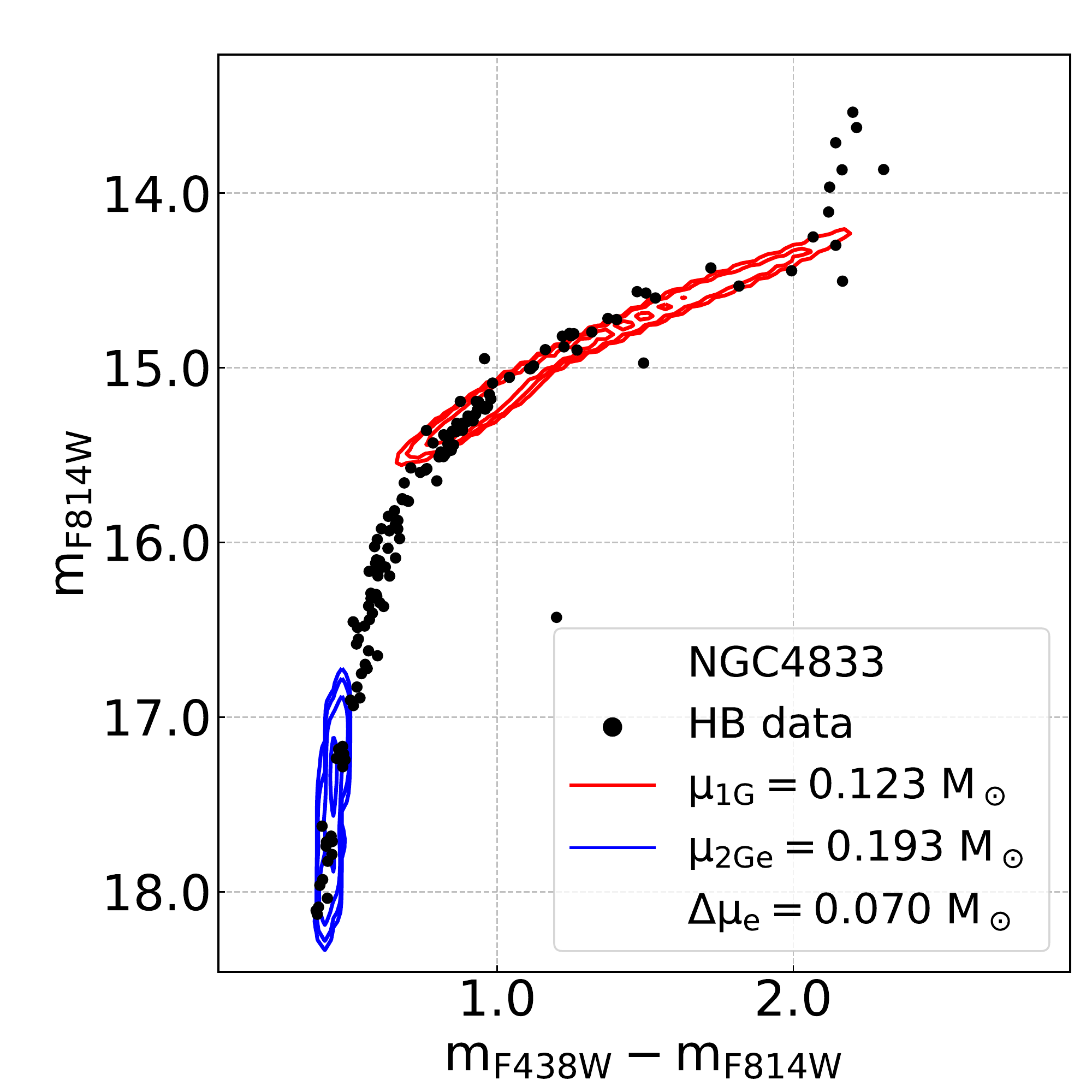}
    \includegraphics[width=0.66\columnwidth]{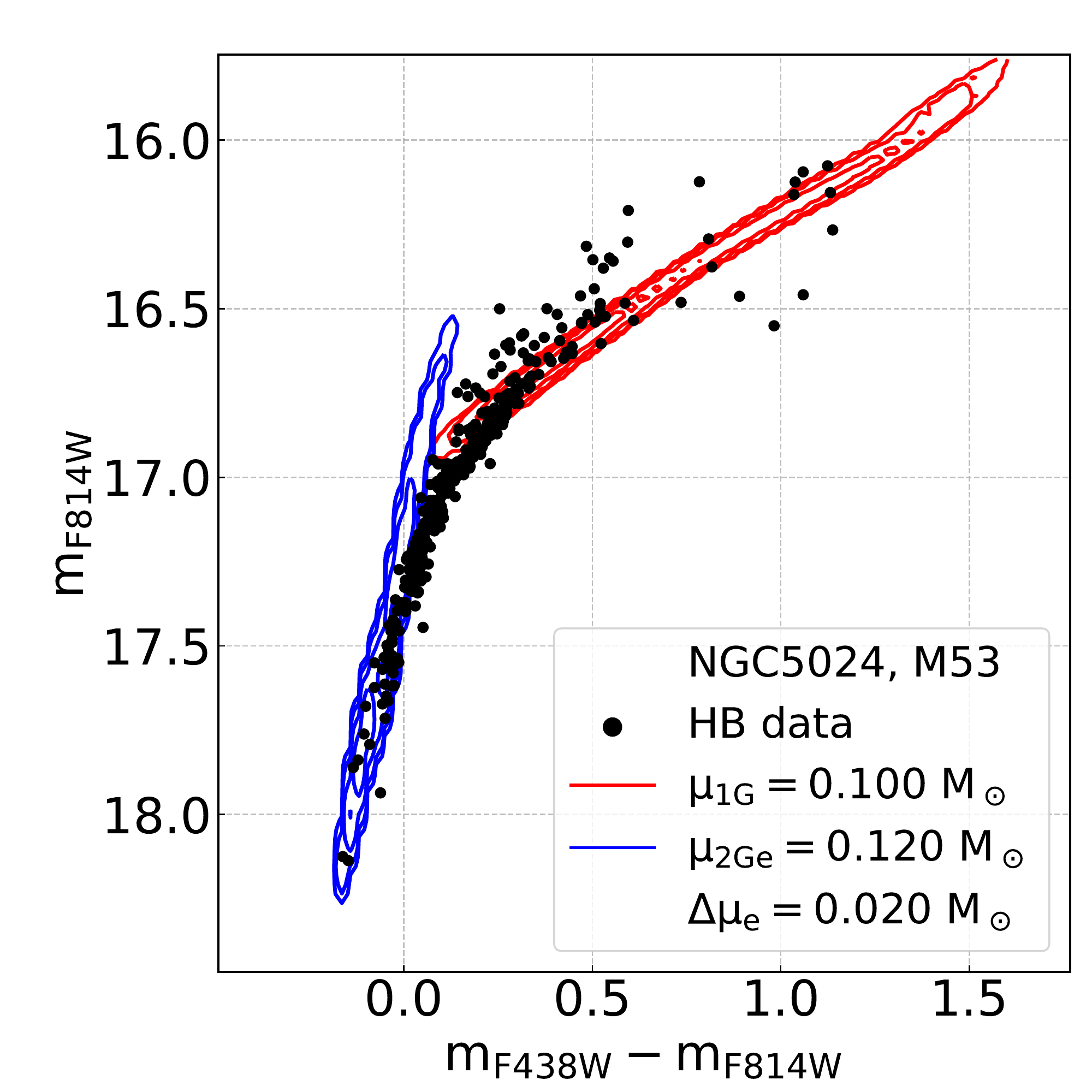}
    \includegraphics[width=0.66\columnwidth]{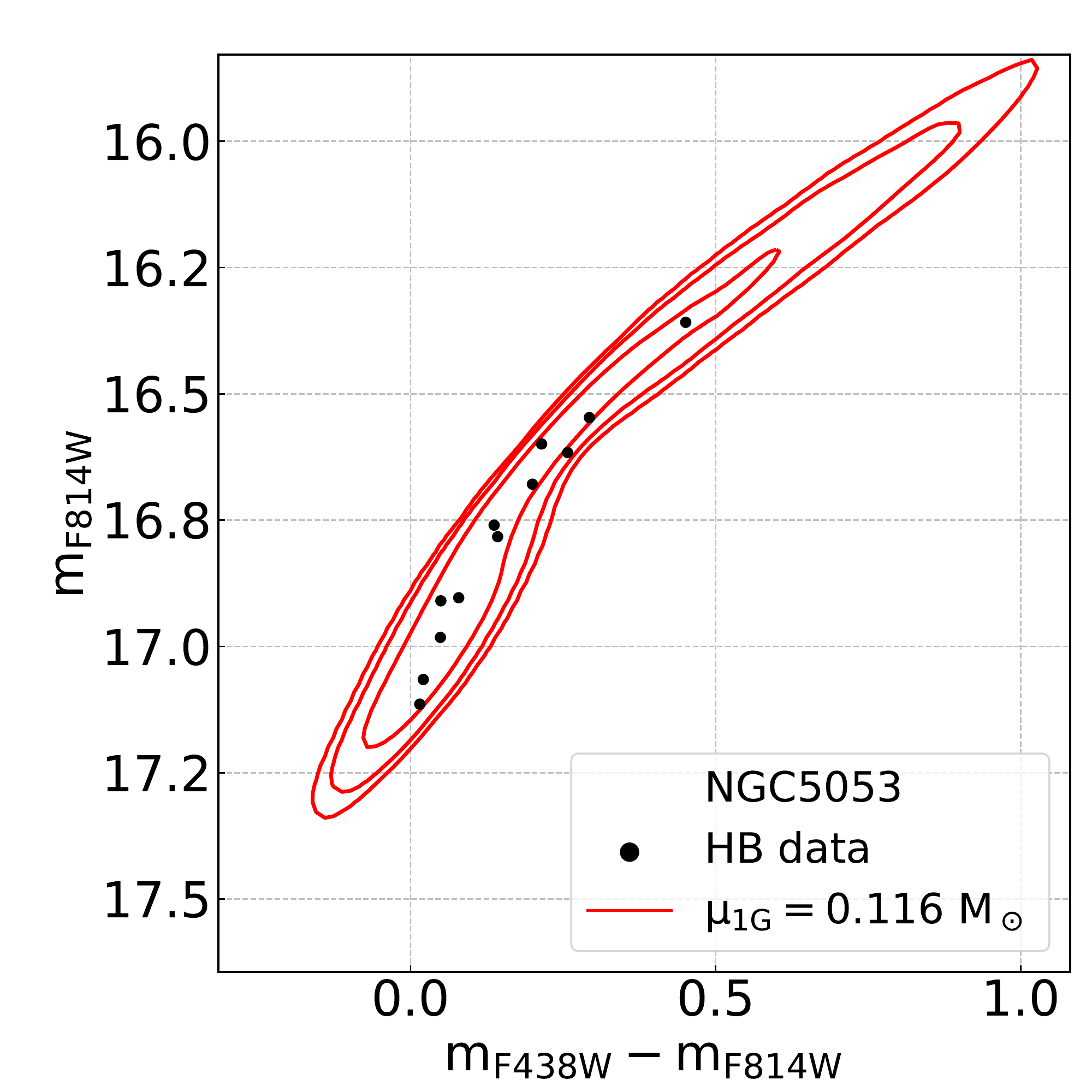}

    \includegraphics[width=0.66\columnwidth]{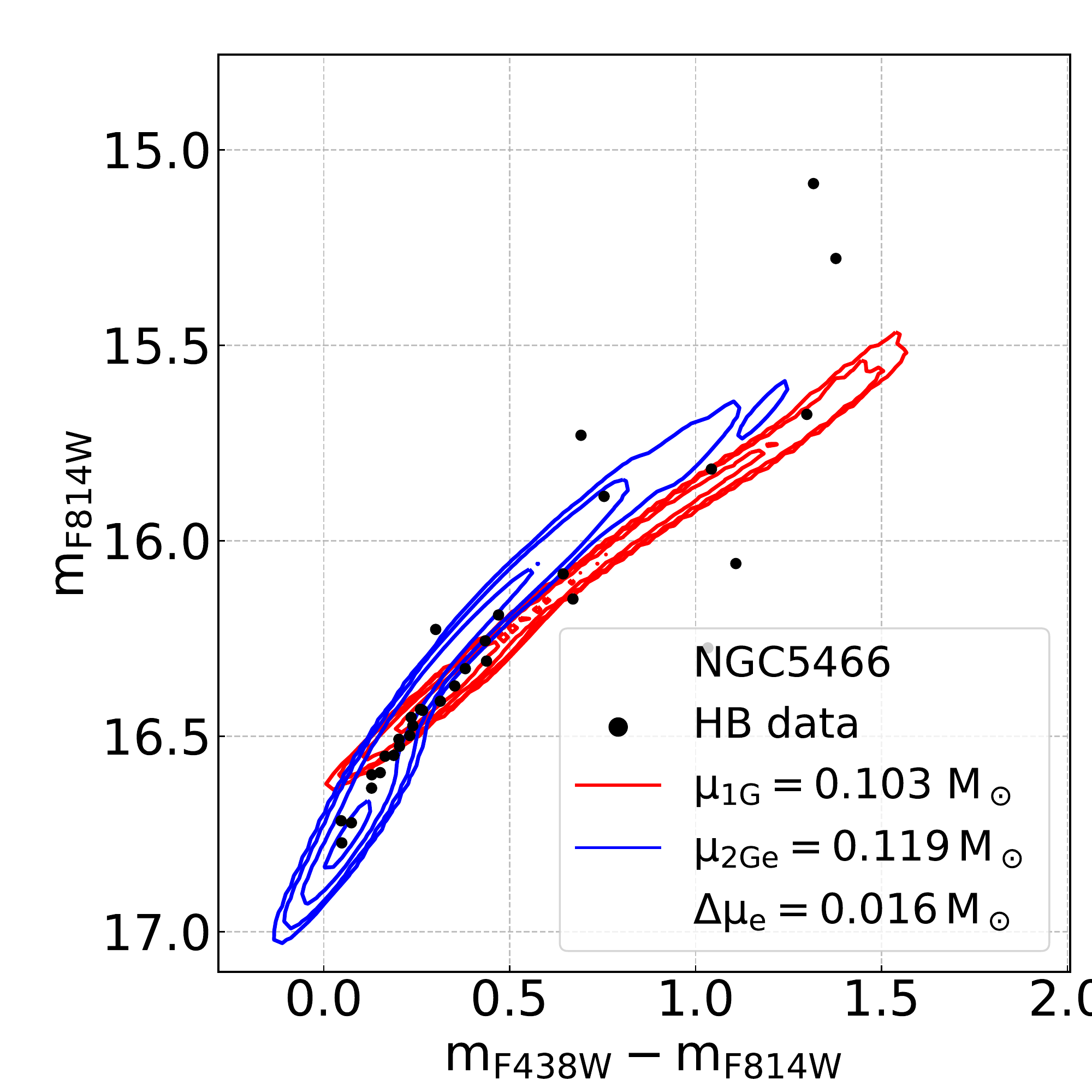}
    \includegraphics[width=0.66\columnwidth]{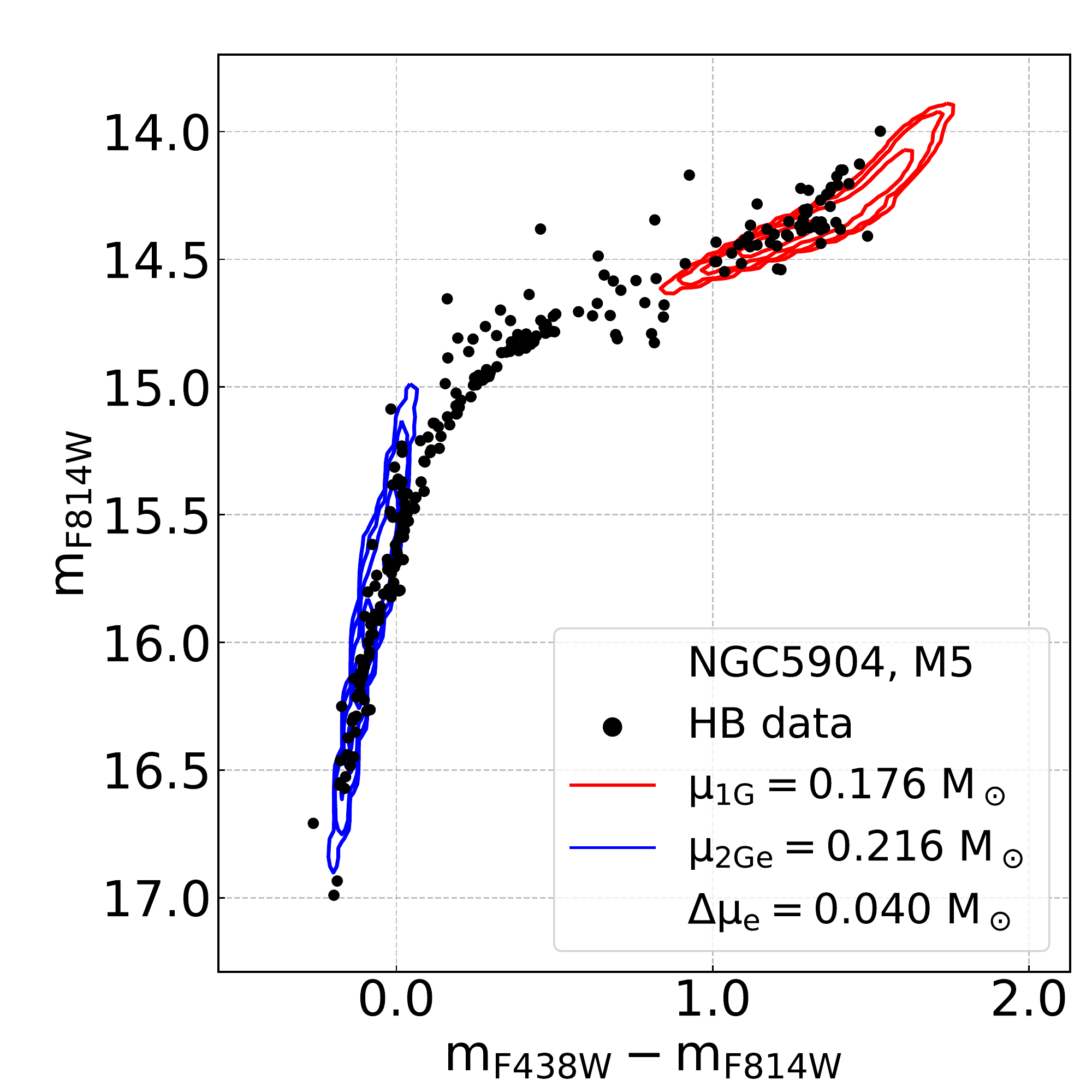}
    \includegraphics[width=0.66\columnwidth]{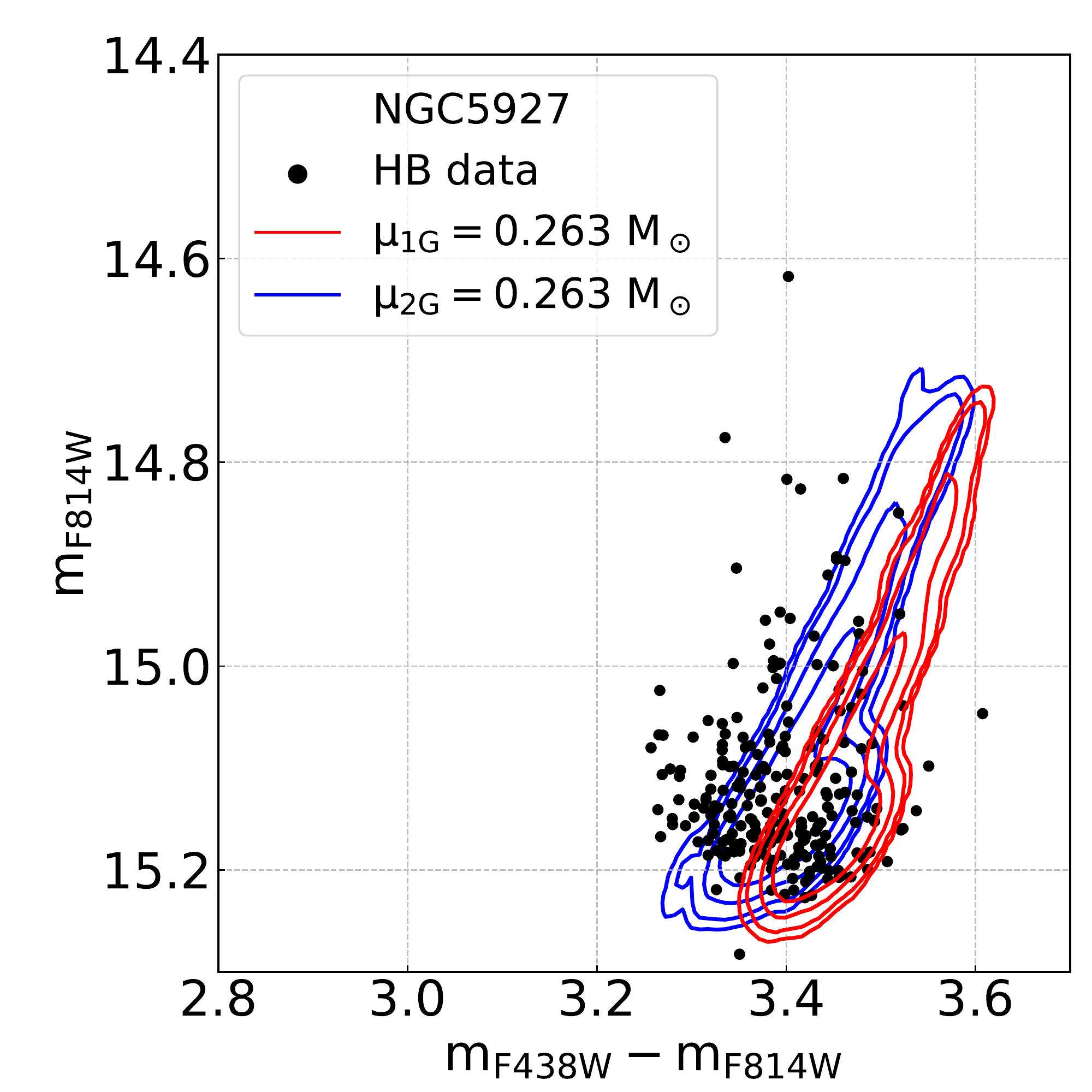}

    \caption{
    The analysed GCs sorted by catalogue number. We superimpose  the best-fit simulations of 1G and 2Ge or 2G HB stars on the observed CMDs of NGC\,0104 (47\,Tuc), NGC\,0288, NGC\,2298, NGC\,2808, NGC\,3201, NGC\,4590 (M\,68), NGC\,4833, NGC\,5024 (M\,53), NGC\,5053, NGC\,5466, NGC\,5909 (M\,5) and NGC\,5927. The average mass losses of 1G and 2Ge (or 2G) stars are provided in the insets together with the corresponding mass-loss difference.
    }
    \label{pic:showcase_1}
\end{figure*}

\begin{figure*}
    \centering
    %placeholders
    \includegraphics[width=0.66\columnwidth]{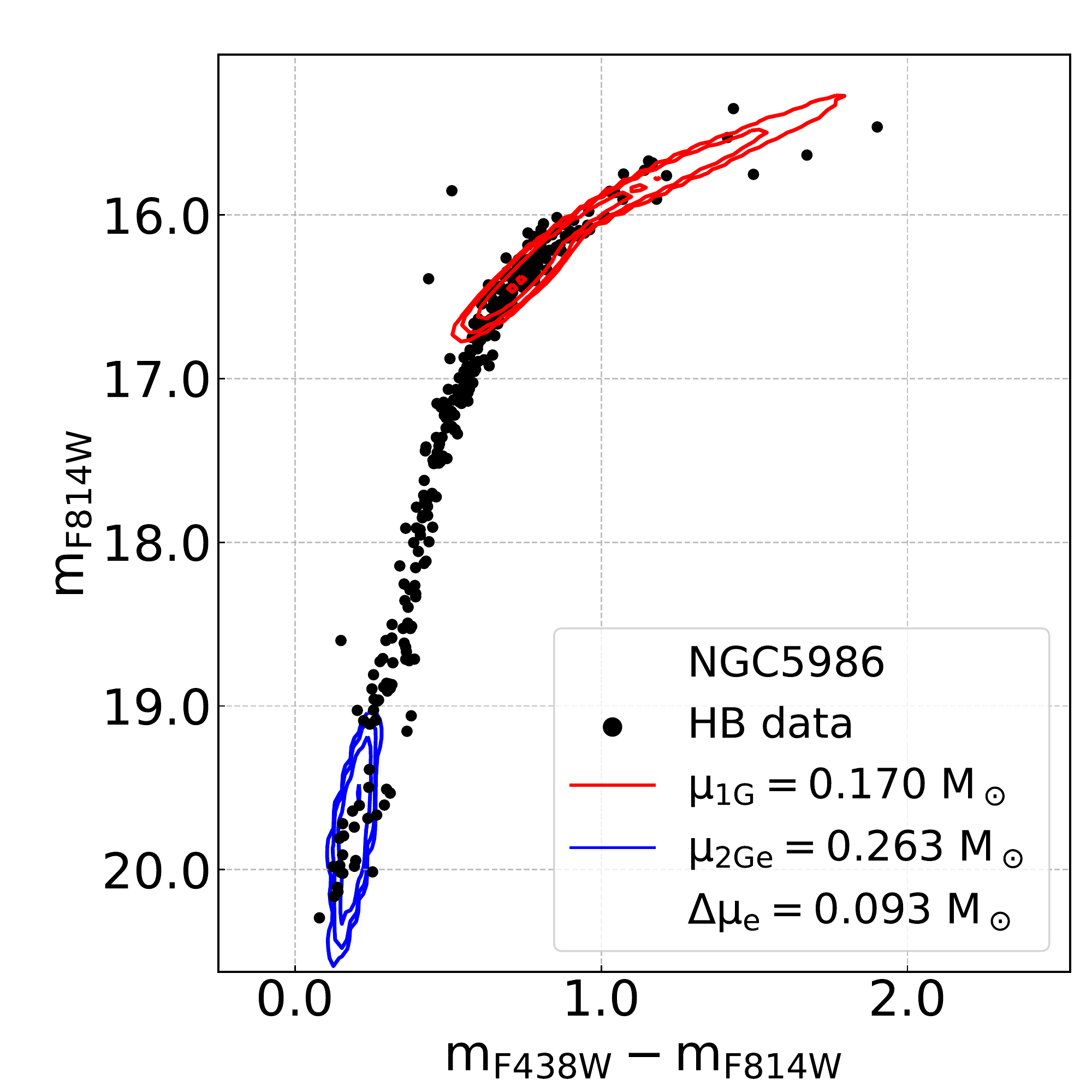}
    \includegraphics[width=0.66\columnwidth]{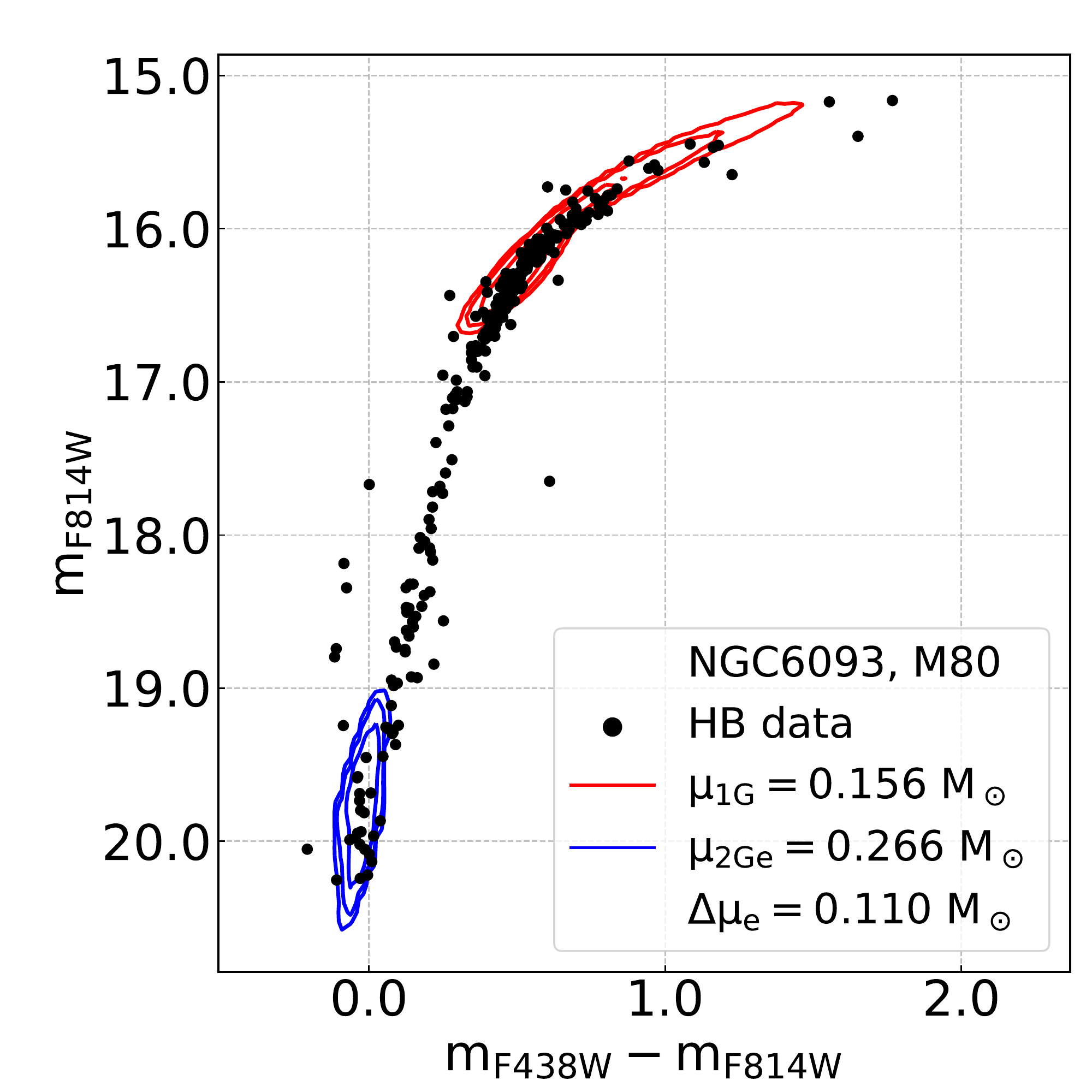}
    \includegraphics[width=0.66\columnwidth]{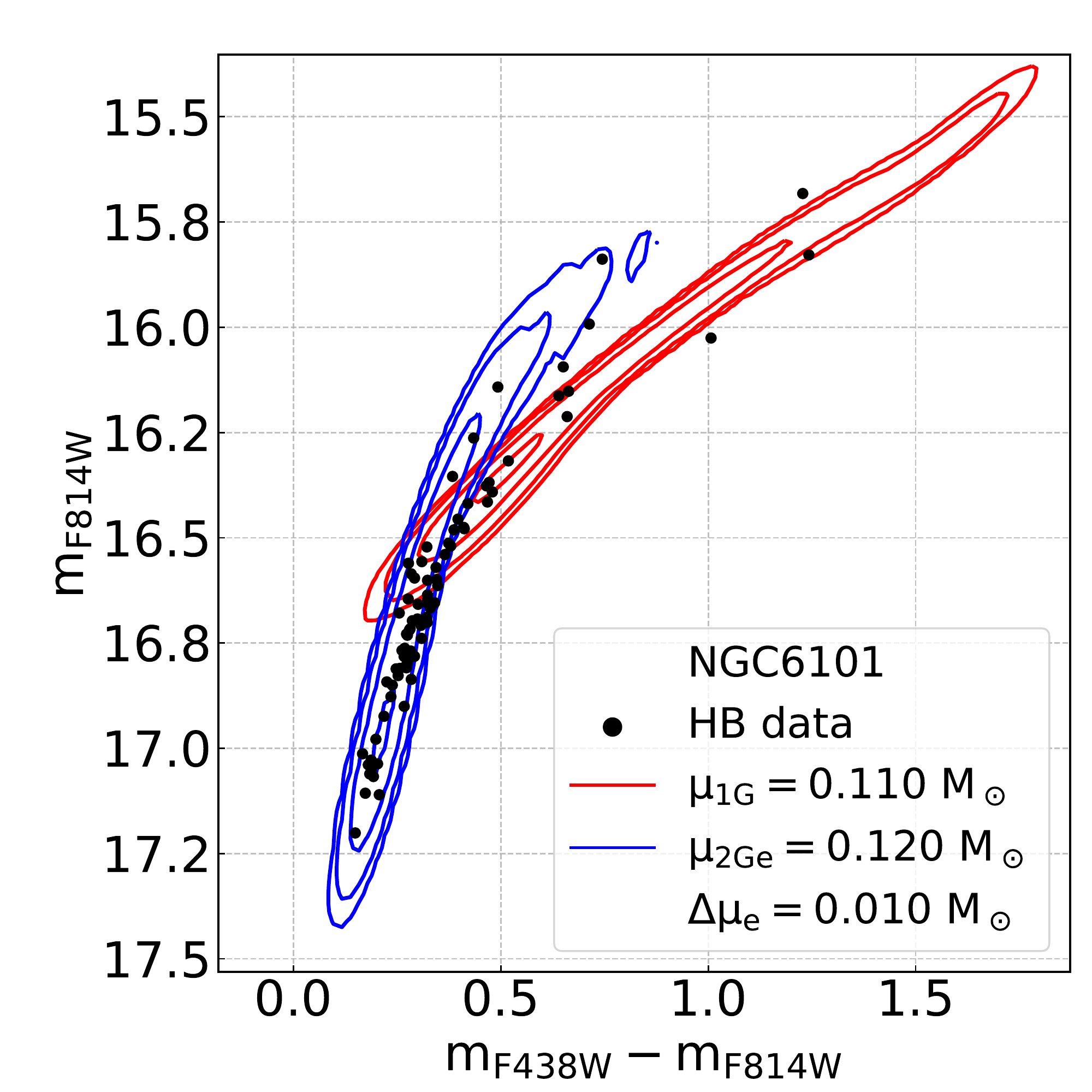}

    \includegraphics[width=0.66\columnwidth]{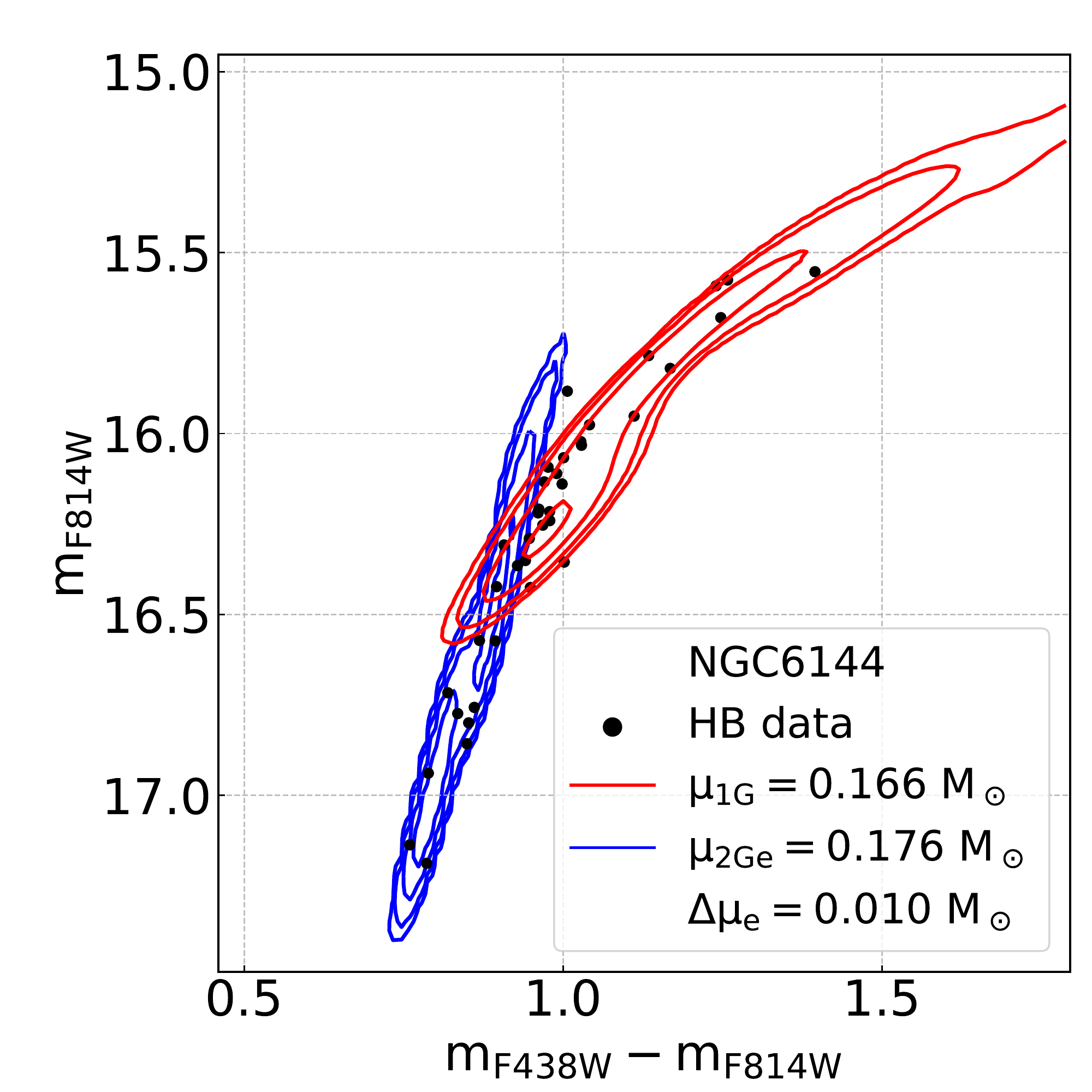}
    \includegraphics[width=0.66\columnwidth]{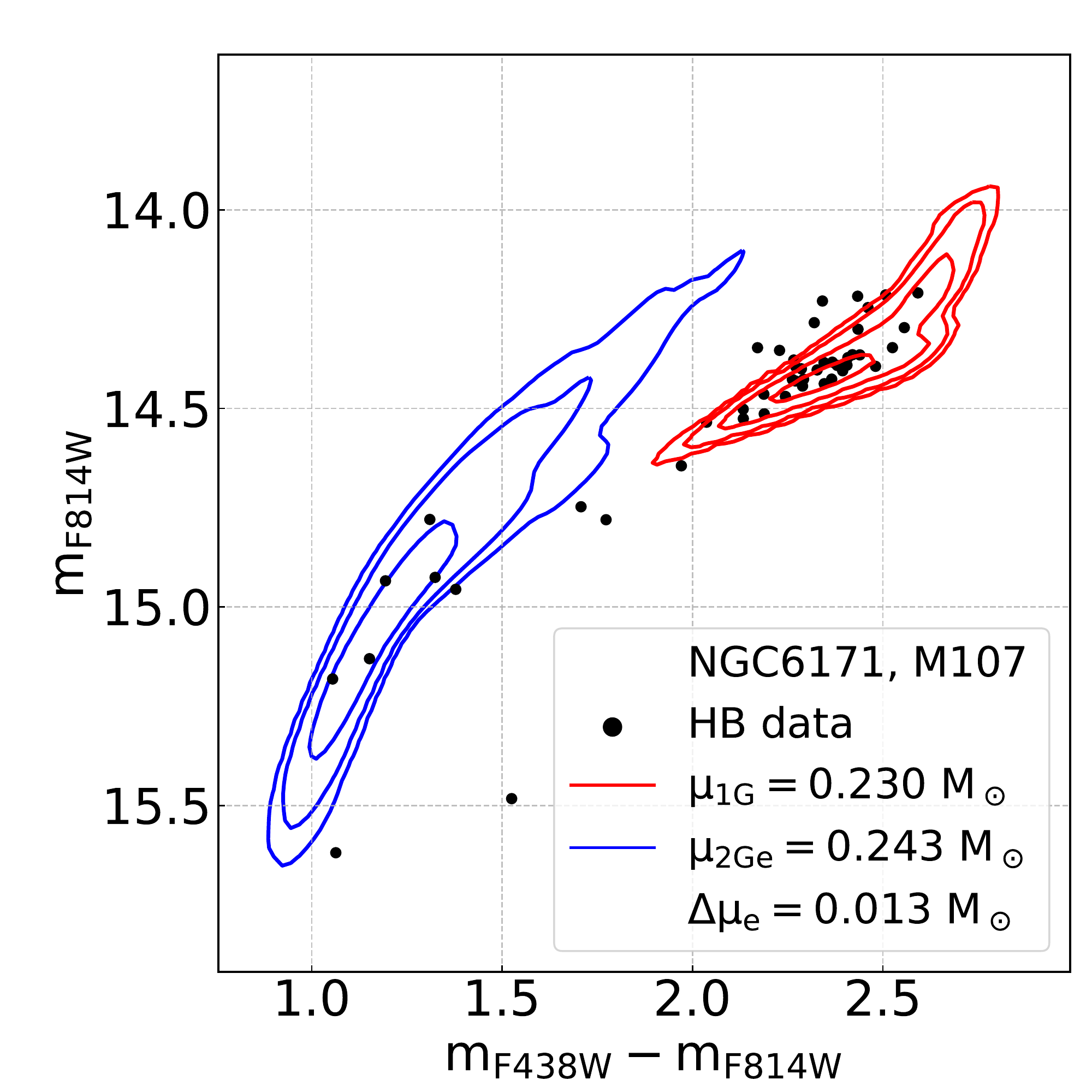}
    \includegraphics[width=0.66\columnwidth]{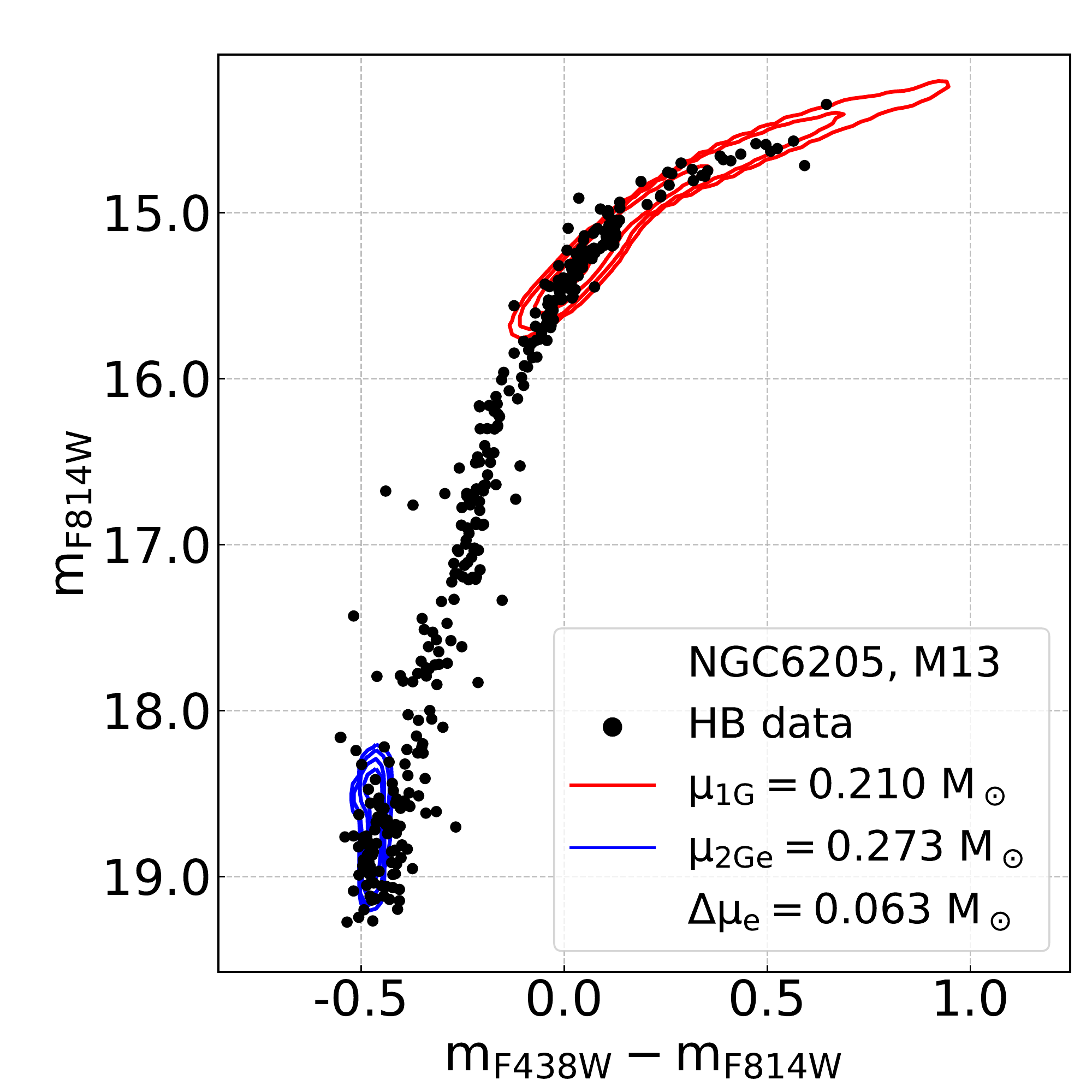}

    \includegraphics[width=0.66\columnwidth]{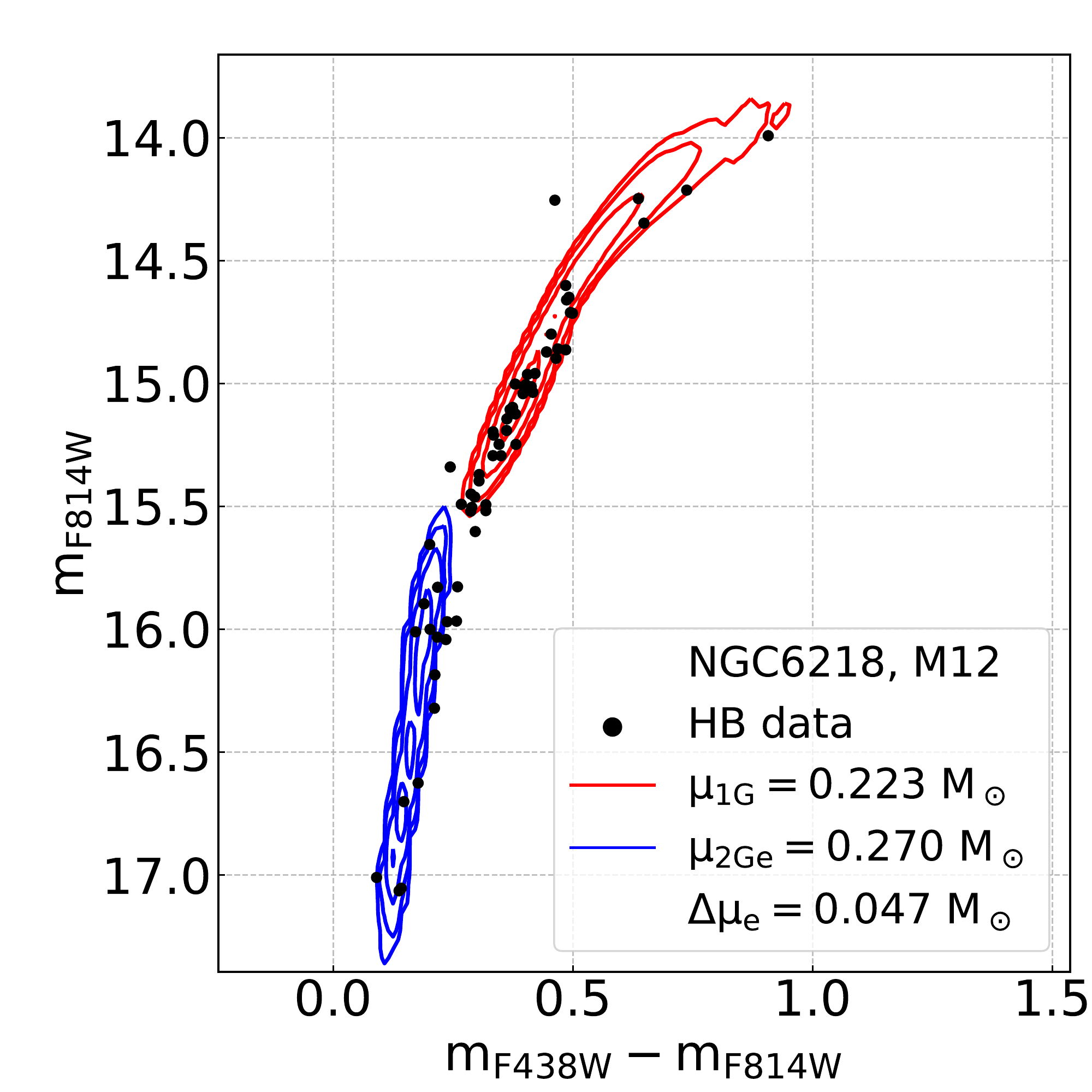}
    \includegraphics[width=0.66\columnwidth]{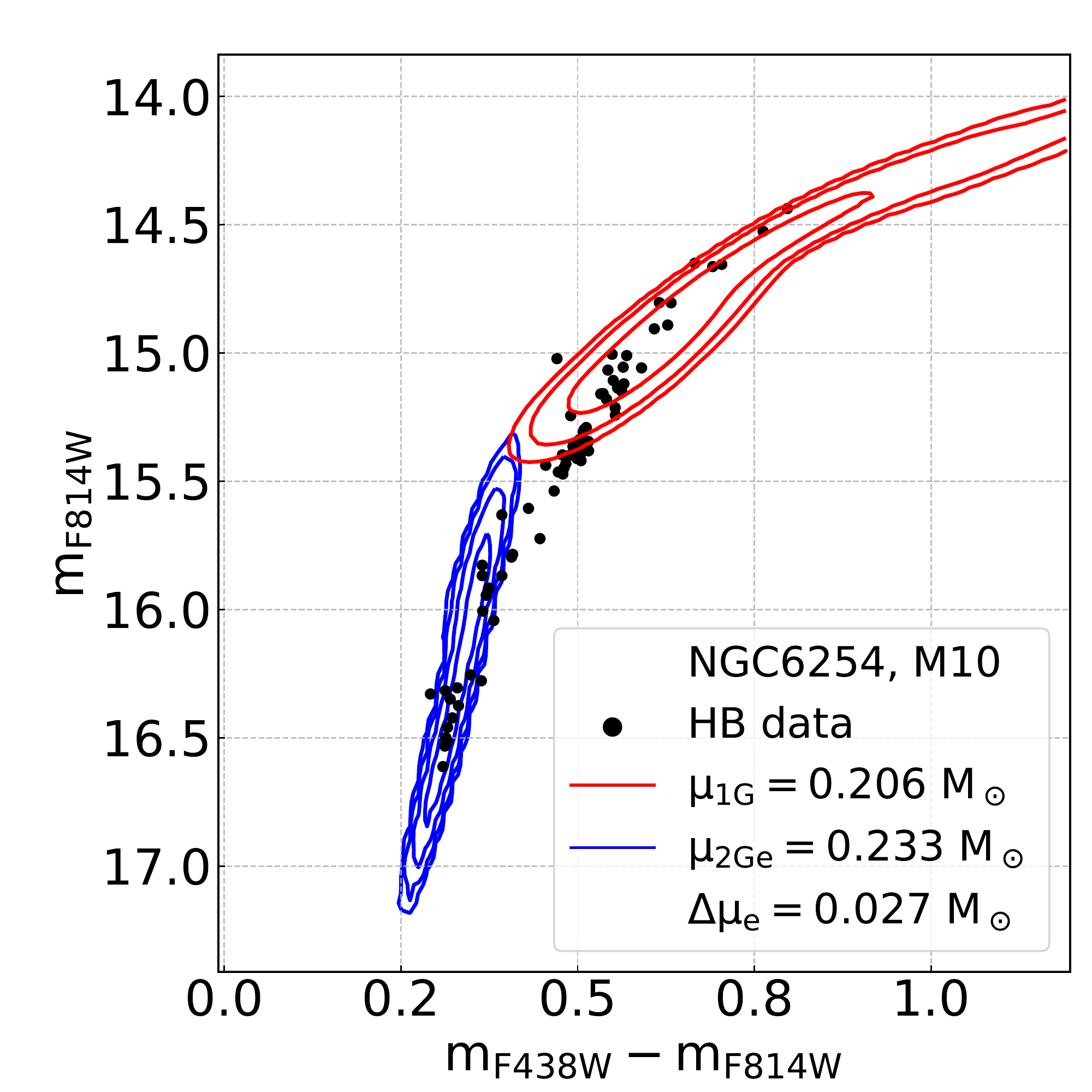}
    \includegraphics[width=0.66\columnwidth]{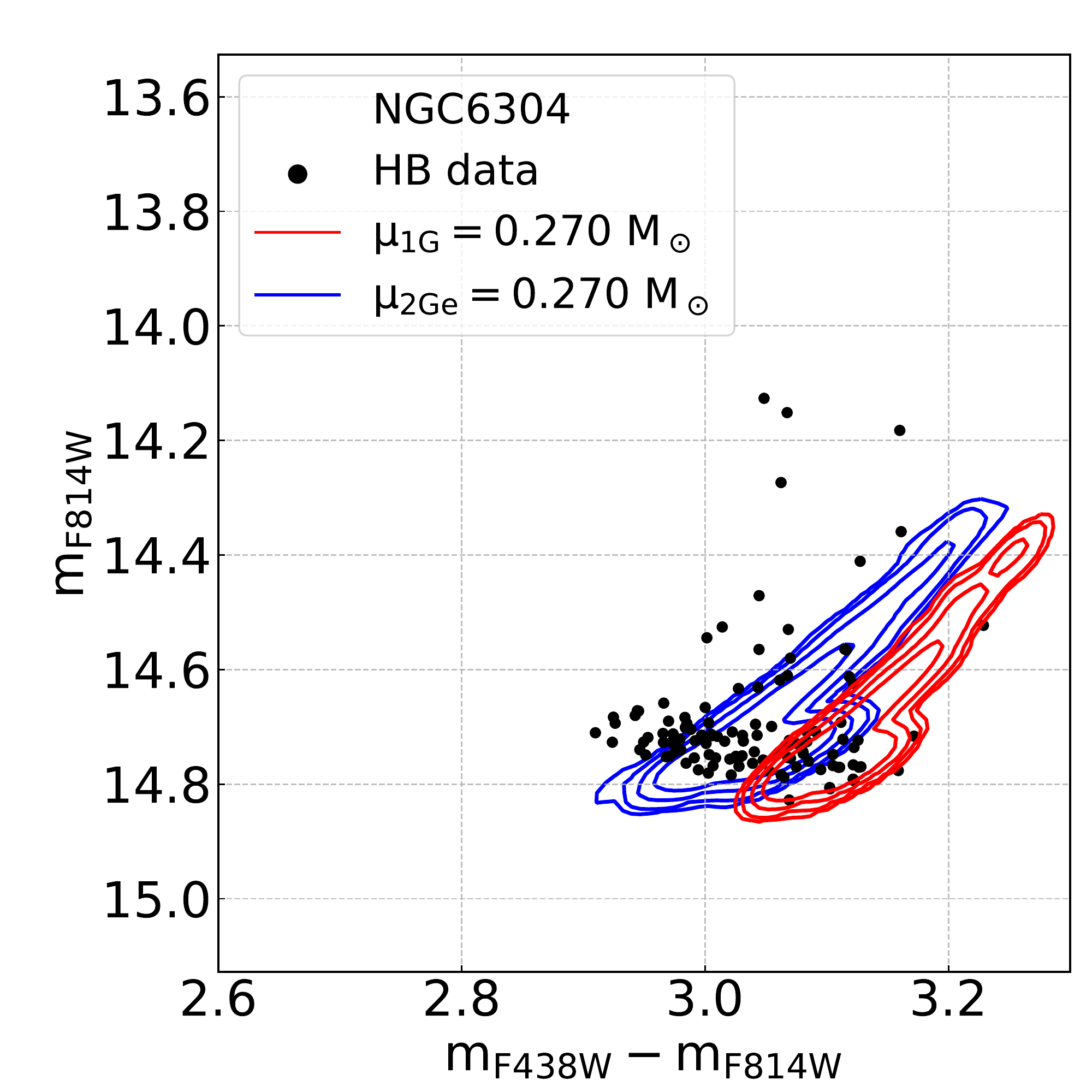}

    \includegraphics[width=0.66\columnwidth]{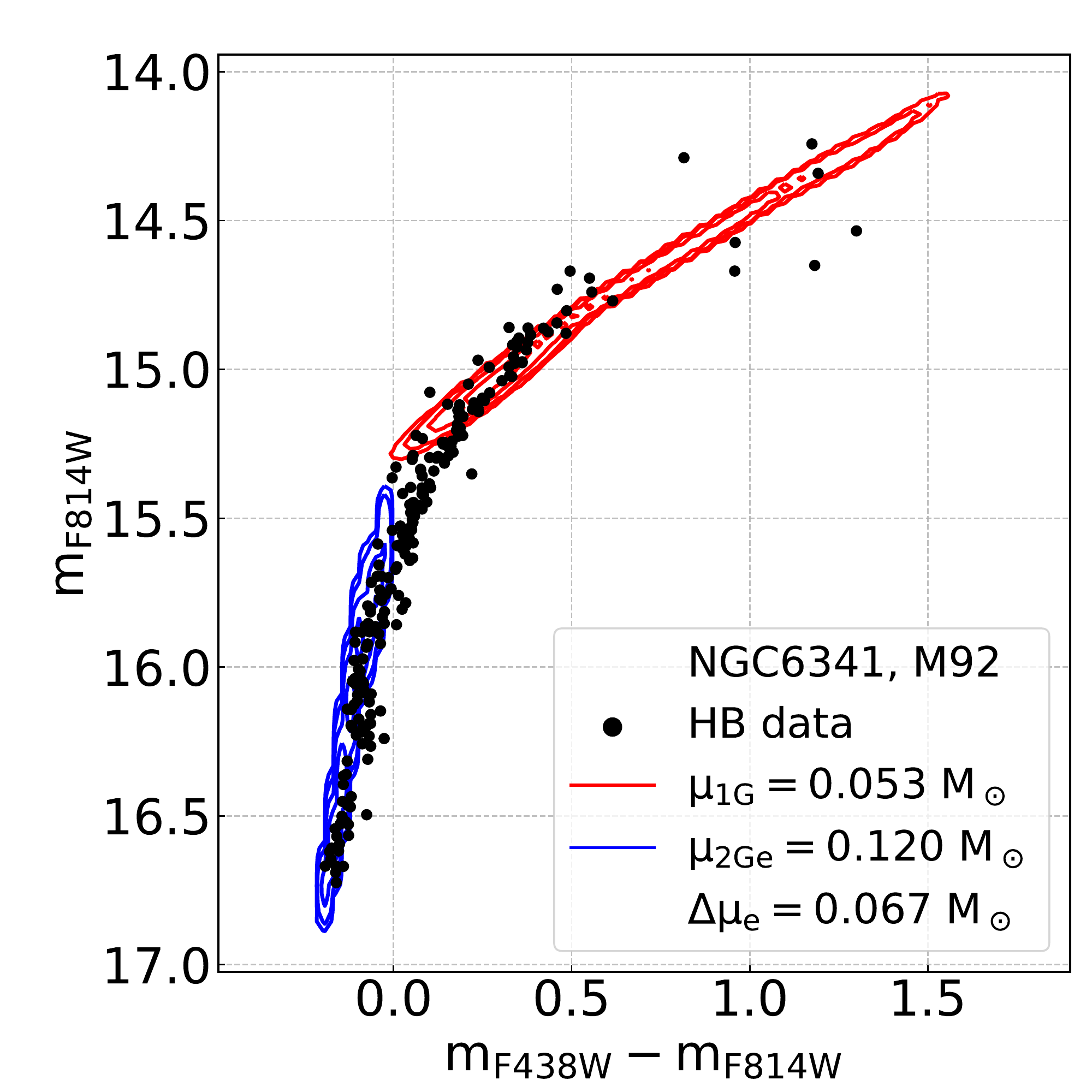}
    \includegraphics[width=0.66\columnwidth]{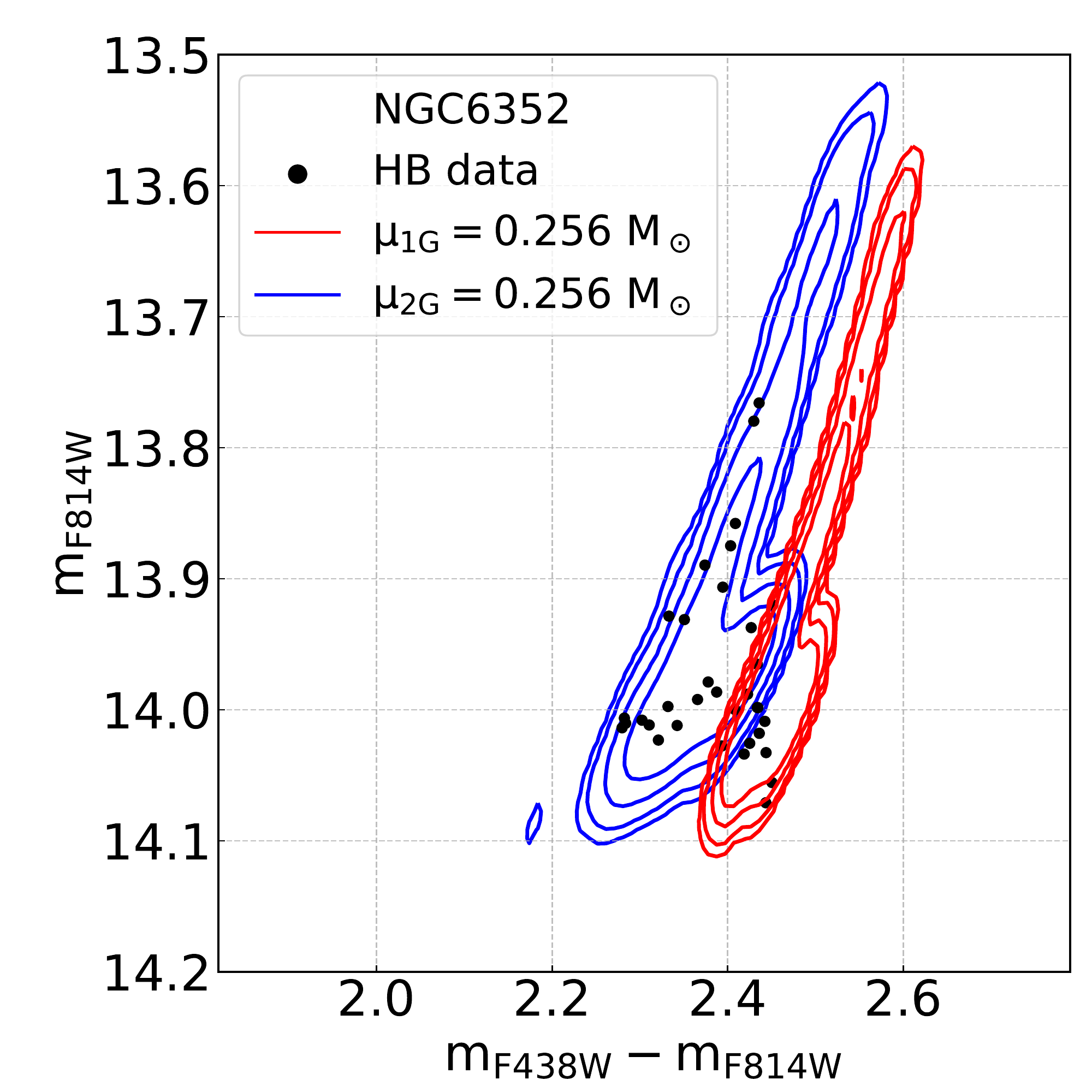}
    \includegraphics[width=0.66\columnwidth]{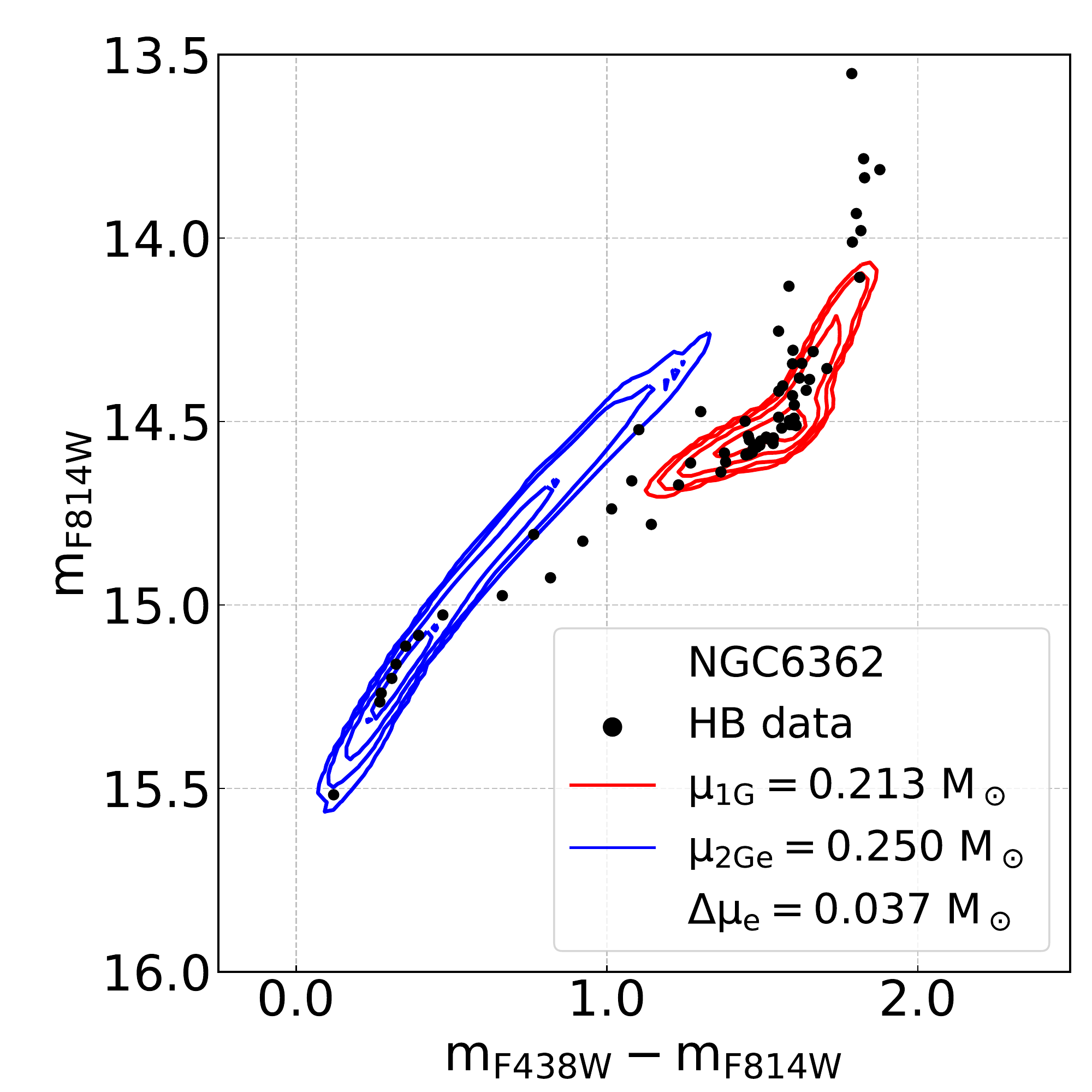}

    \caption{
    As Figure \ref{pic:showcase_1} for NGC\,5986, NGC\,6093 (M\,80), NGC\,6101, NGC\,6144, NGC\,6171 (M\,107), NGC\,6205 (M\,13), NGC\,6218 (M\,12), NGC\,6254 (M\,10), NGC\,6304, NGC\,6341  (M\,92), NGC\,6352 and NGC\,6362.
    }
    \label{pic:showcase_2}
\end{figure*}

\begin{figure*}
    \centering
    %placeholders
    \includegraphics[width=0.66\columnwidth]{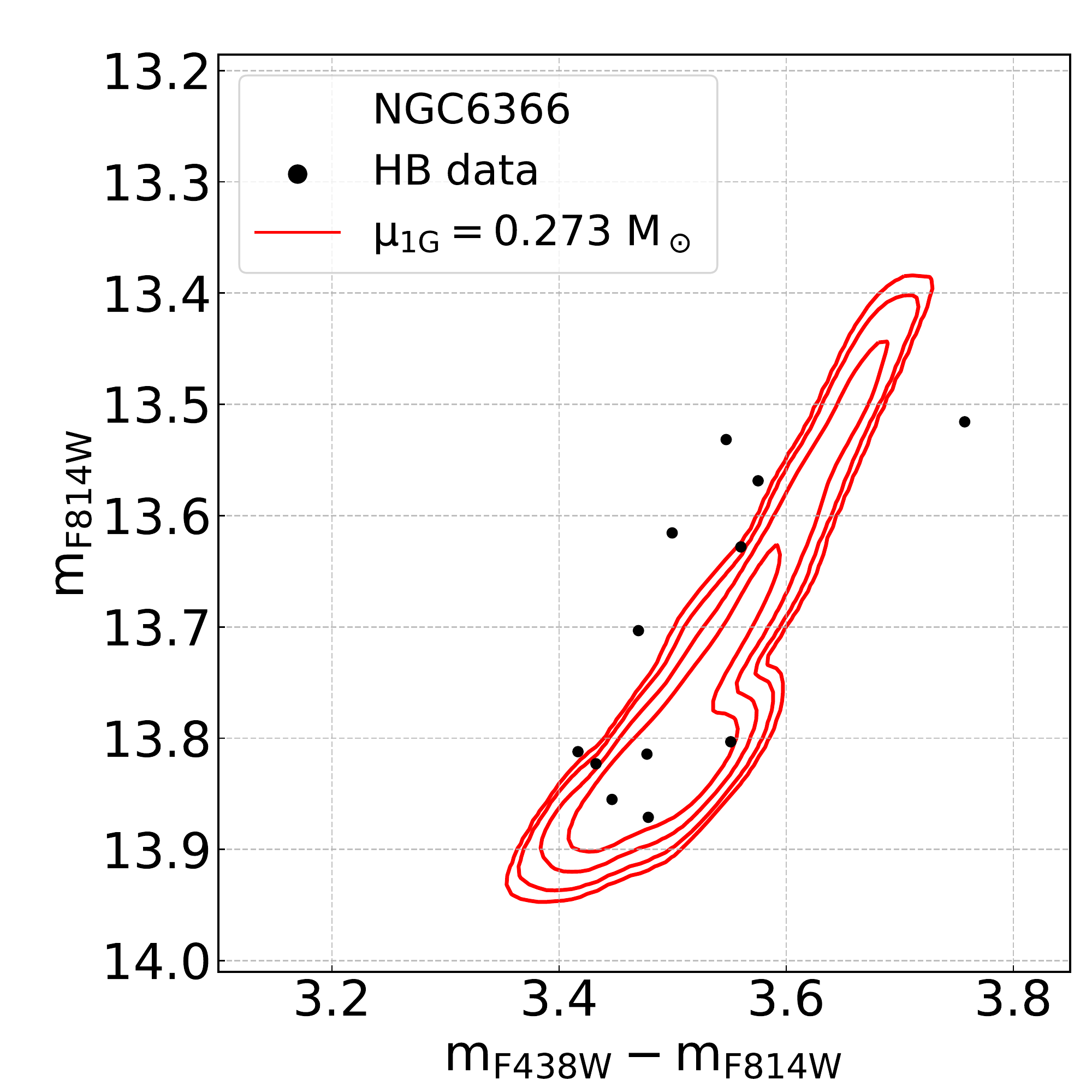}
    \includegraphics[width=0.66\columnwidth]{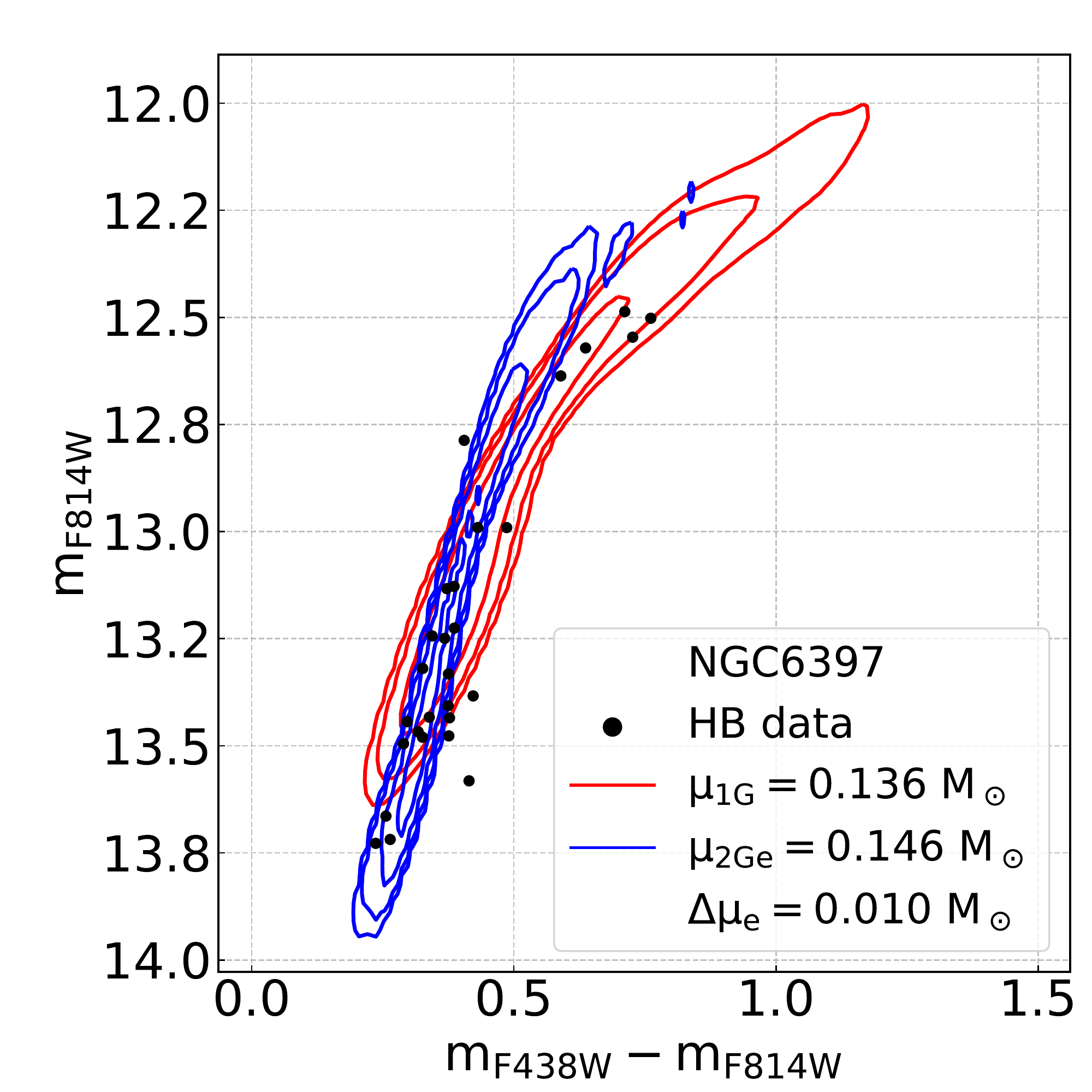}
    \includegraphics[width=0.66\columnwidth]{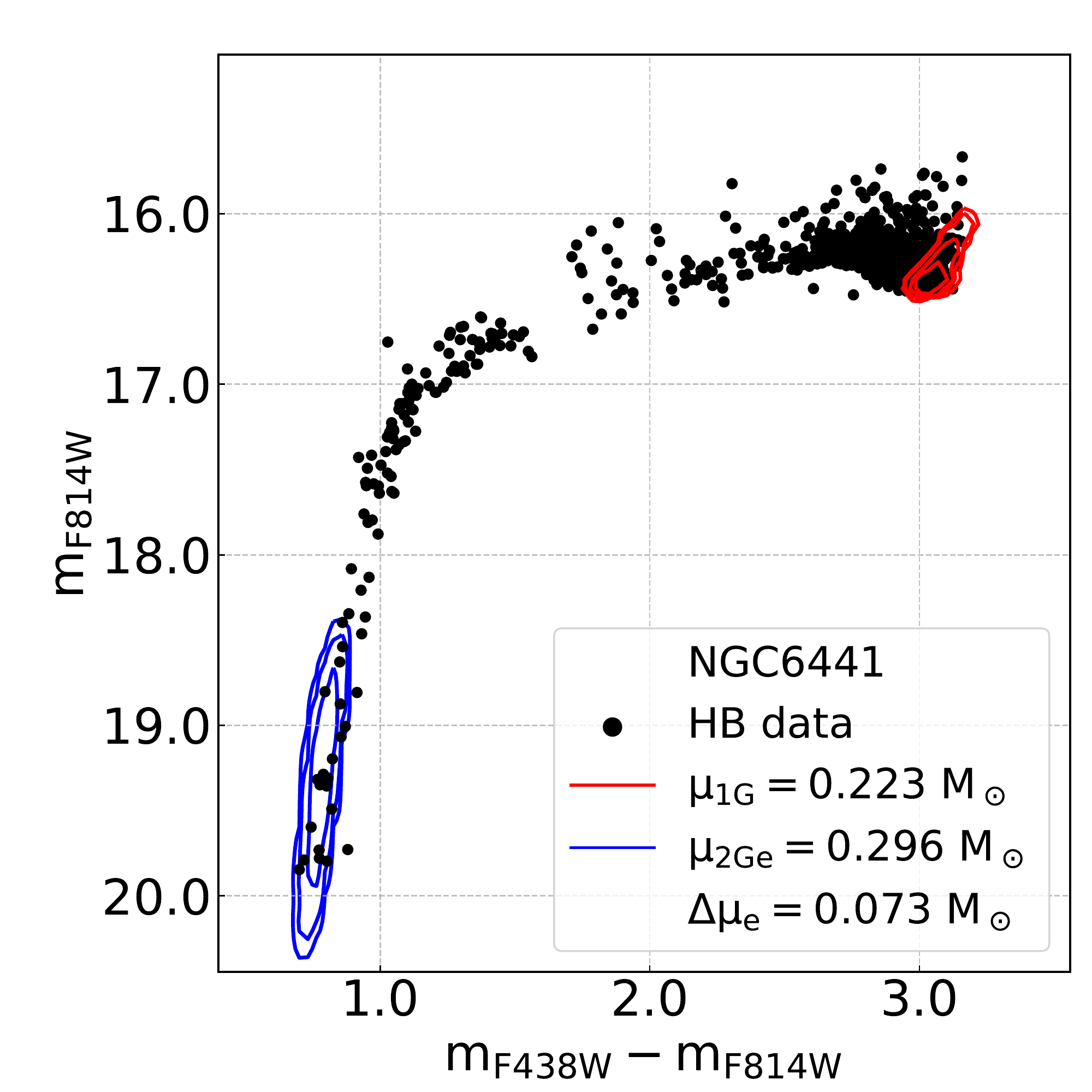}

    \includegraphics[width=0.66\columnwidth]{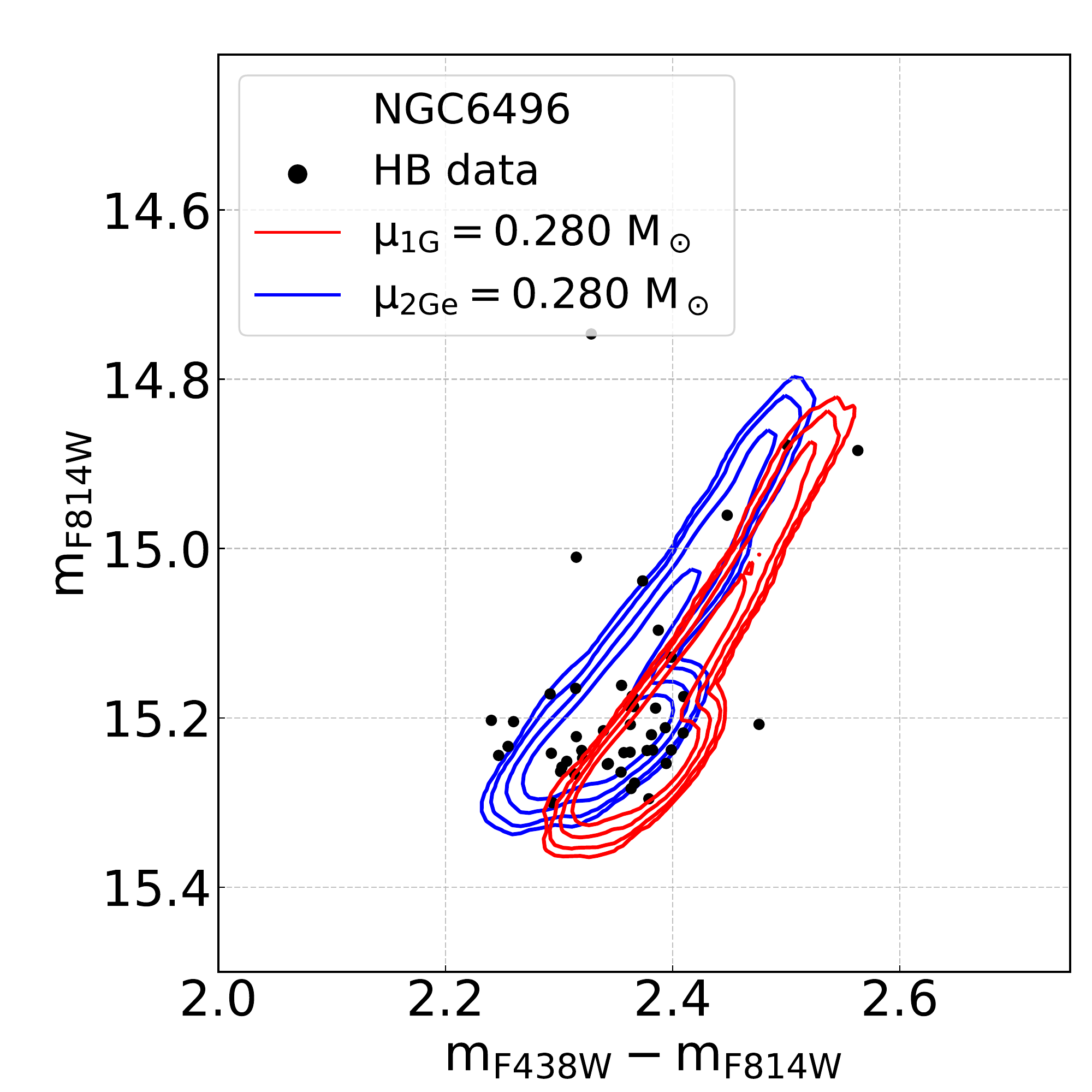}
    \includegraphics[width=0.66\columnwidth]{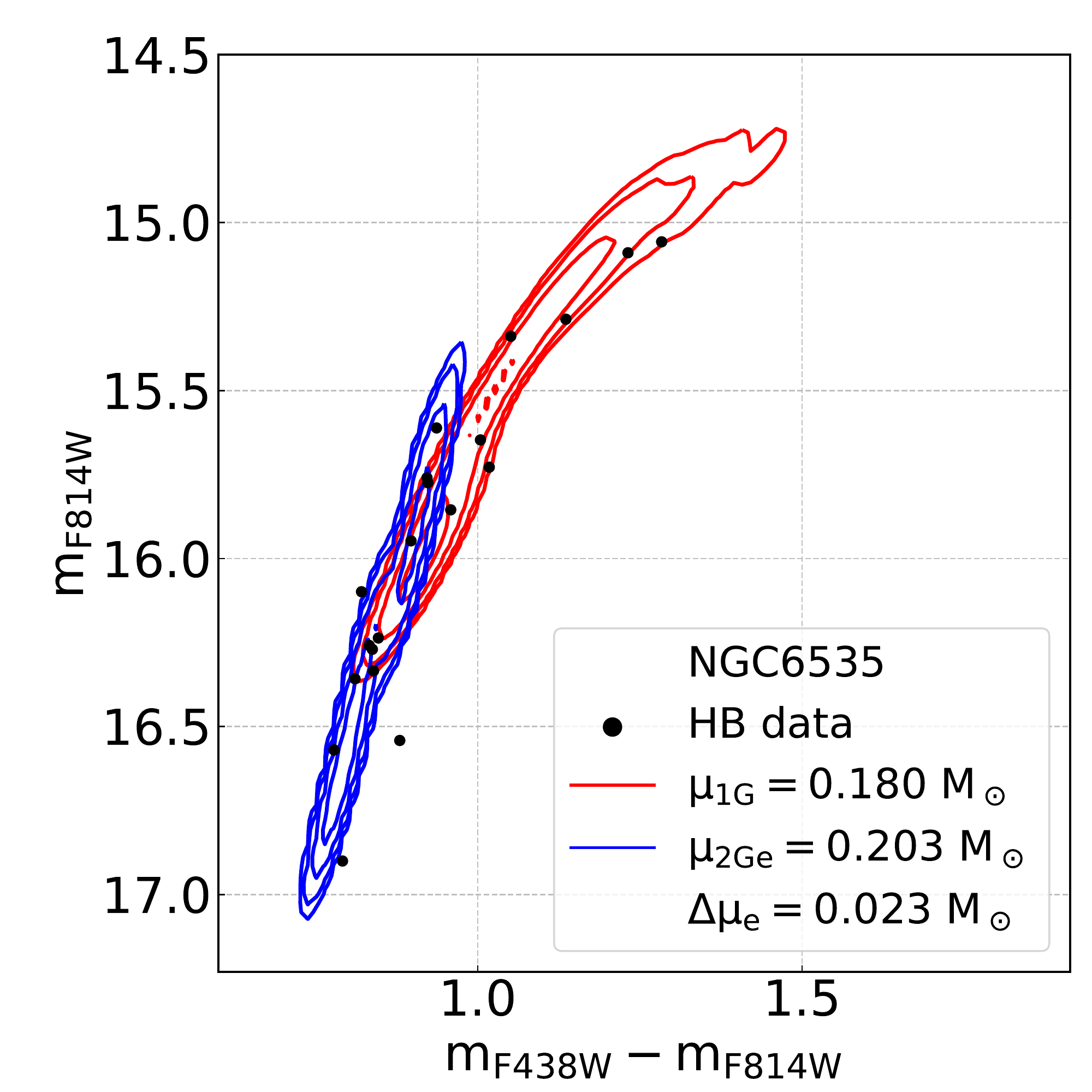}
    \includegraphics[width=0.66\columnwidth]{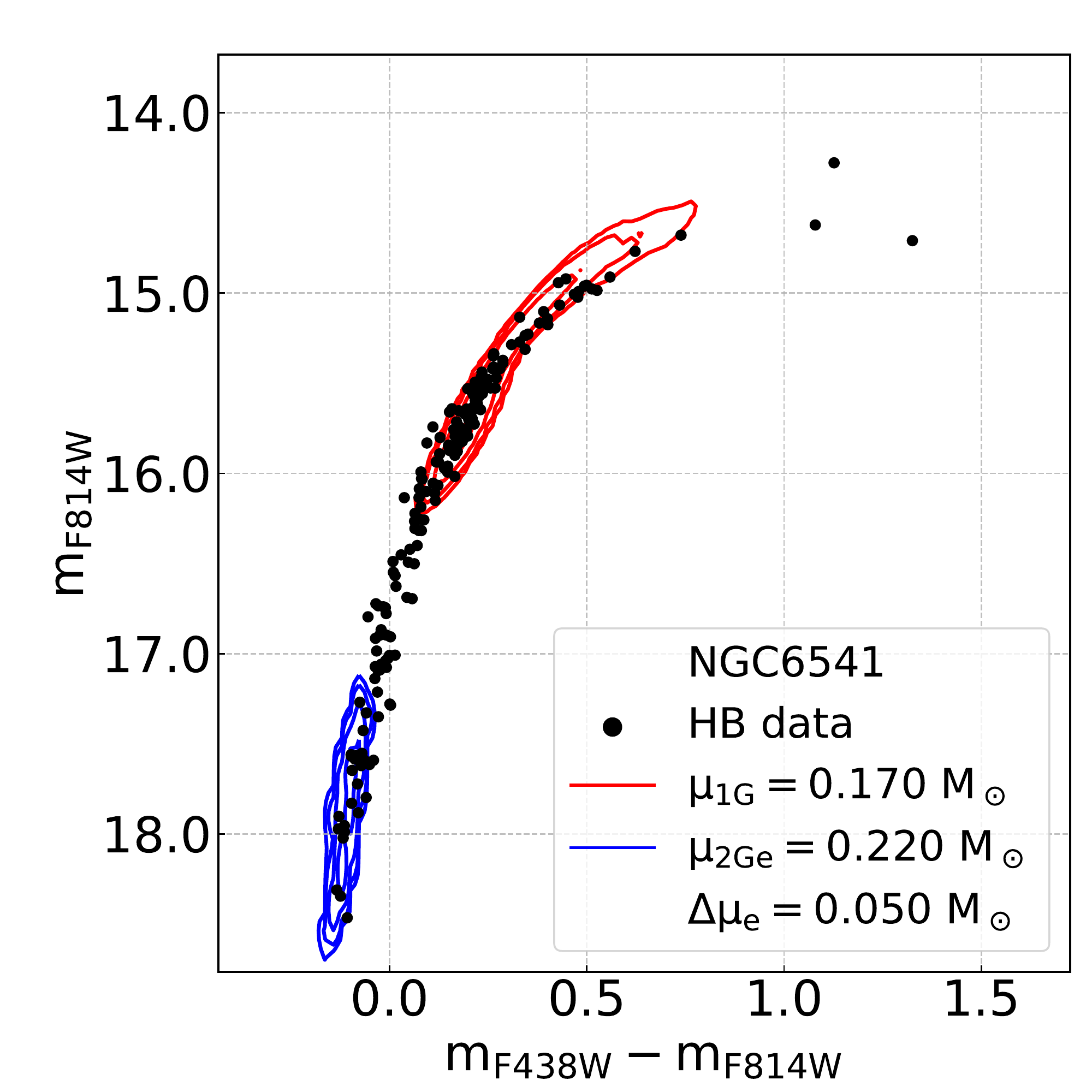}

    \includegraphics[width=0.66\columnwidth]{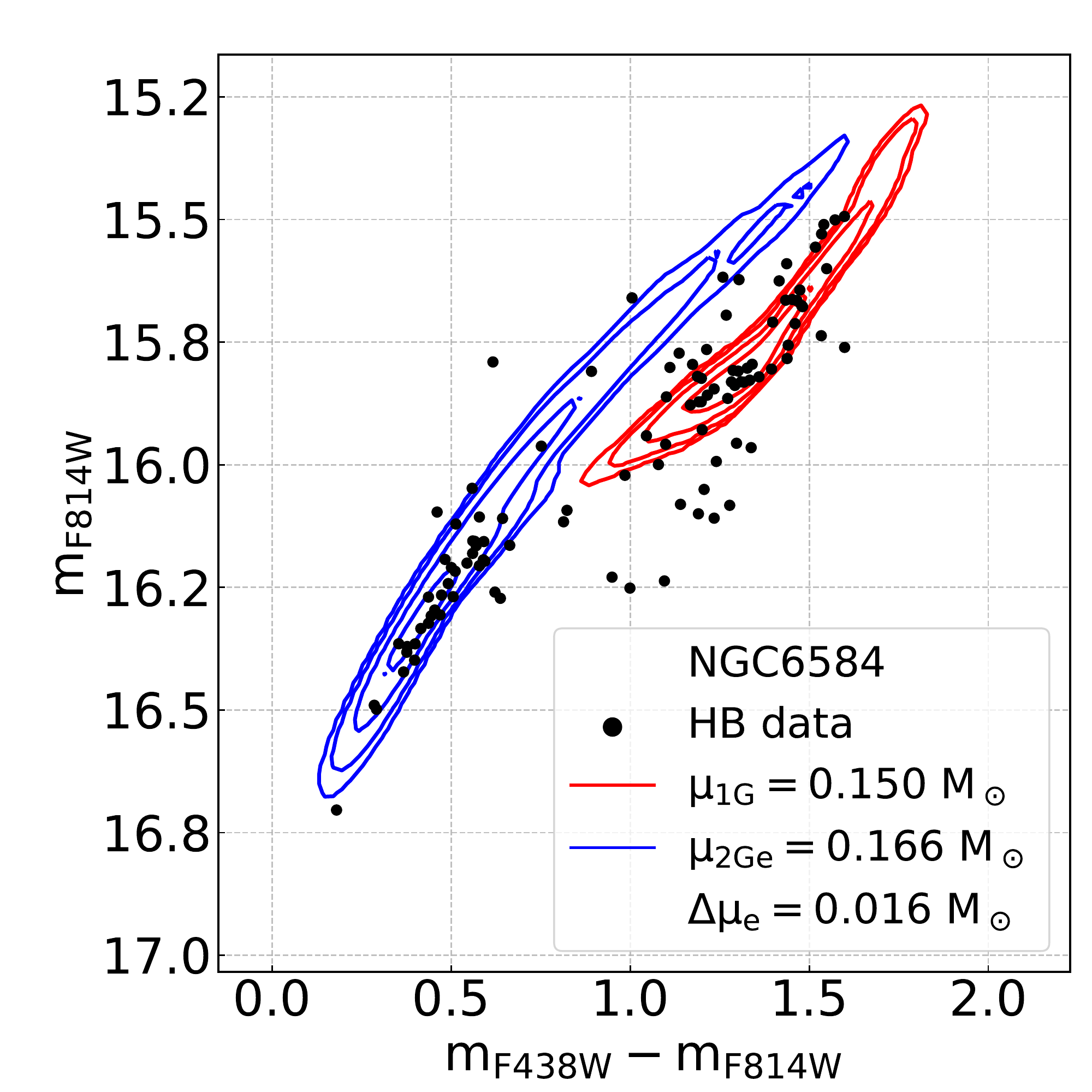}
    \includegraphics[width=0.66\columnwidth]{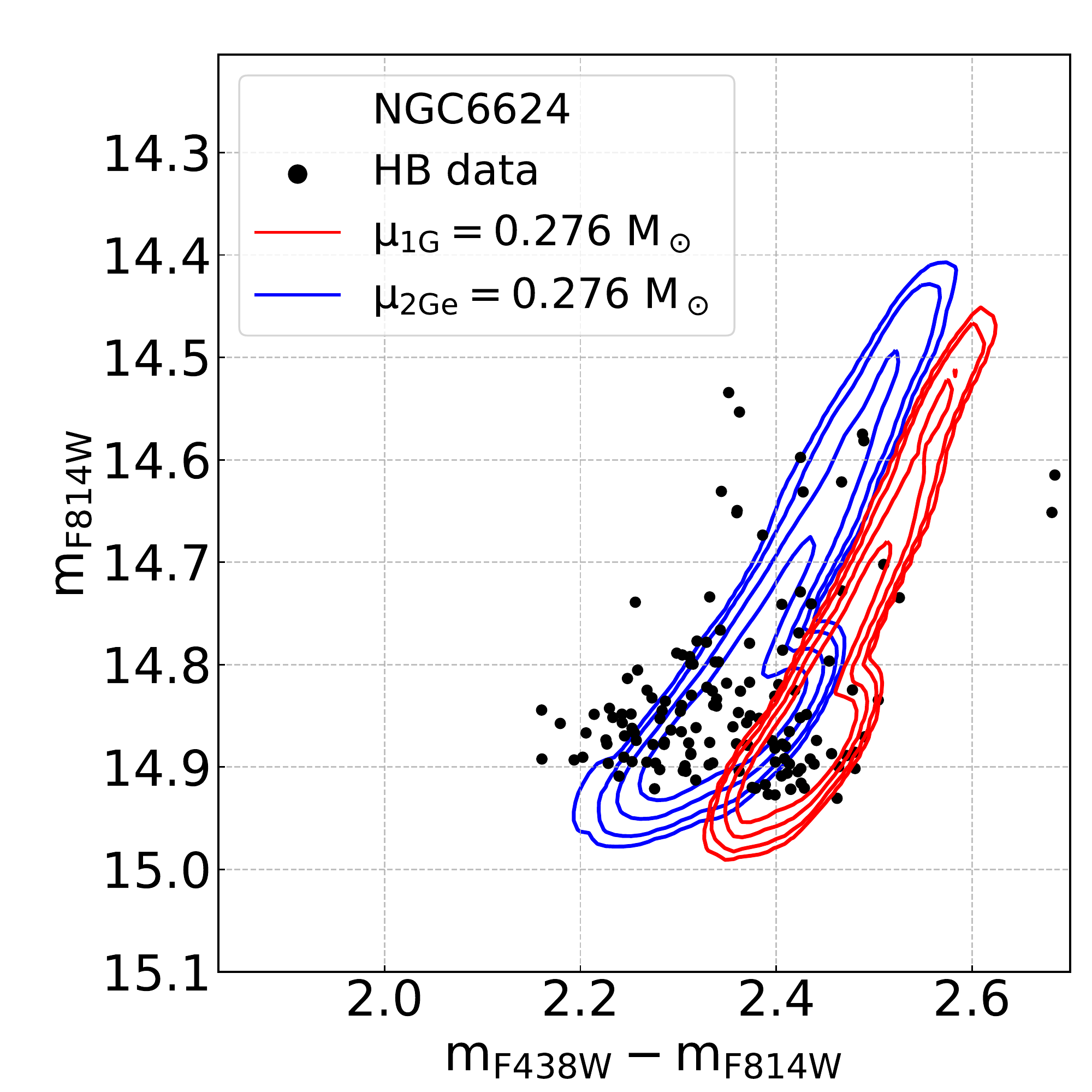}
    \includegraphics[width=0.66\columnwidth]{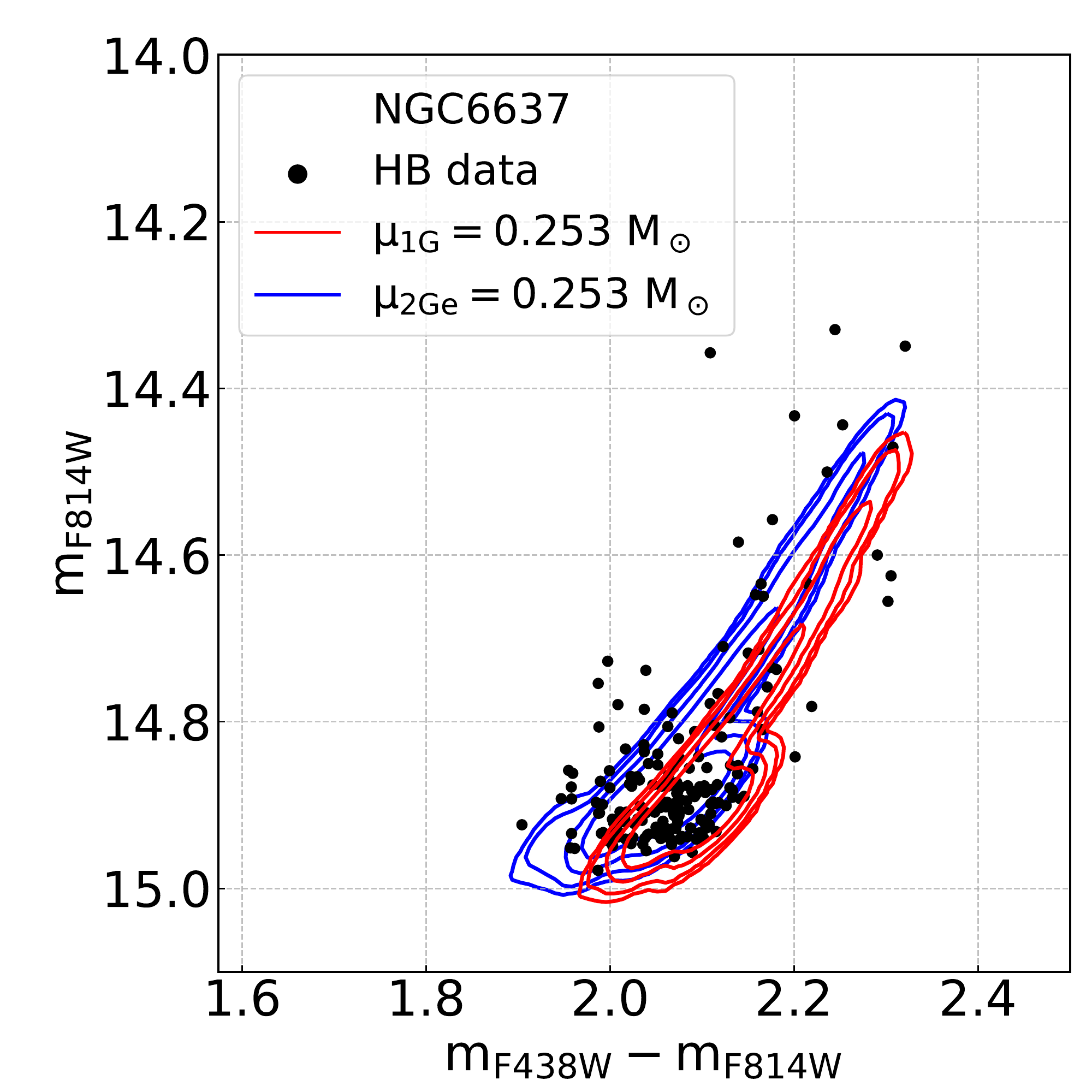}

    \includegraphics[width=0.66\columnwidth]{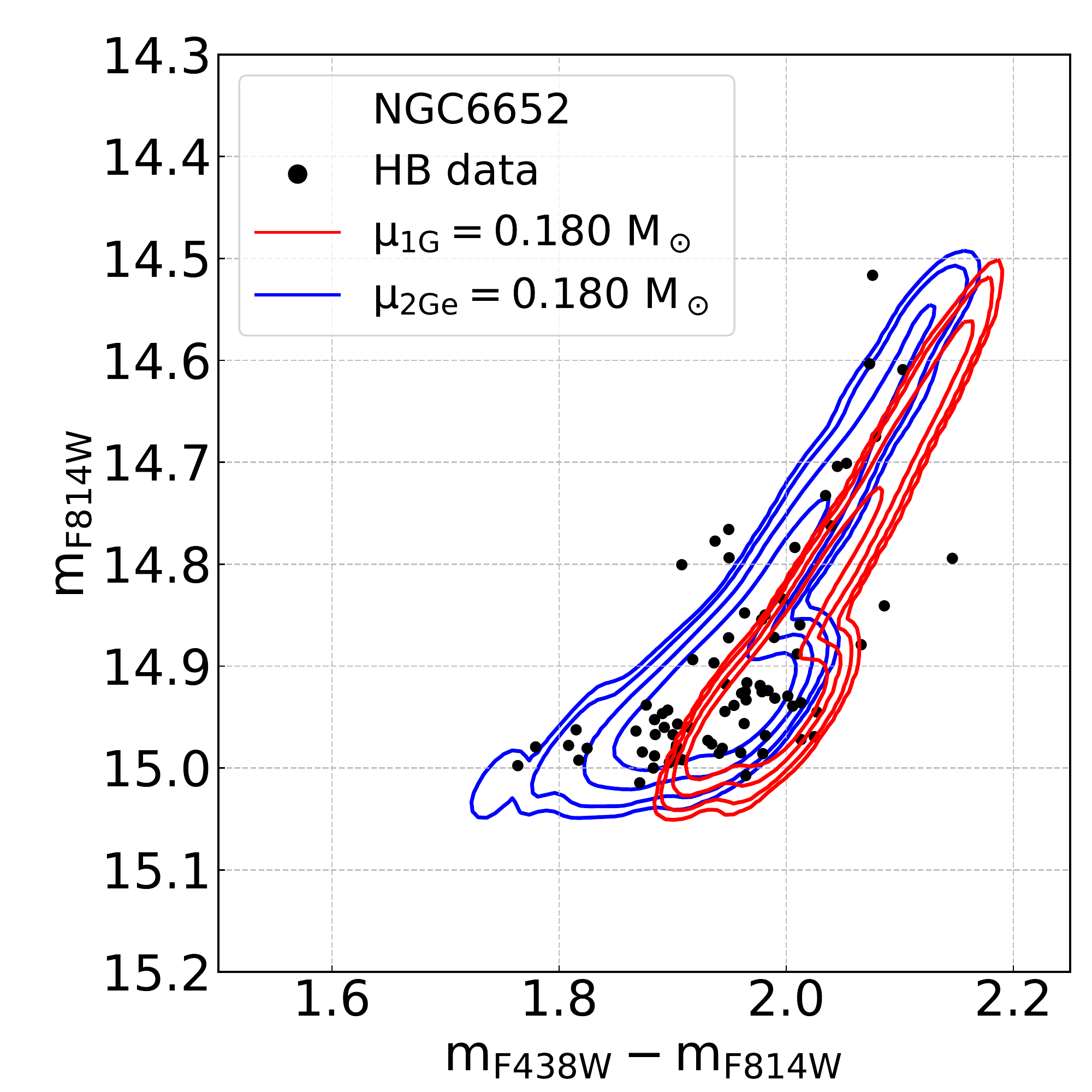}
    \includegraphics[width=0.66\columnwidth]{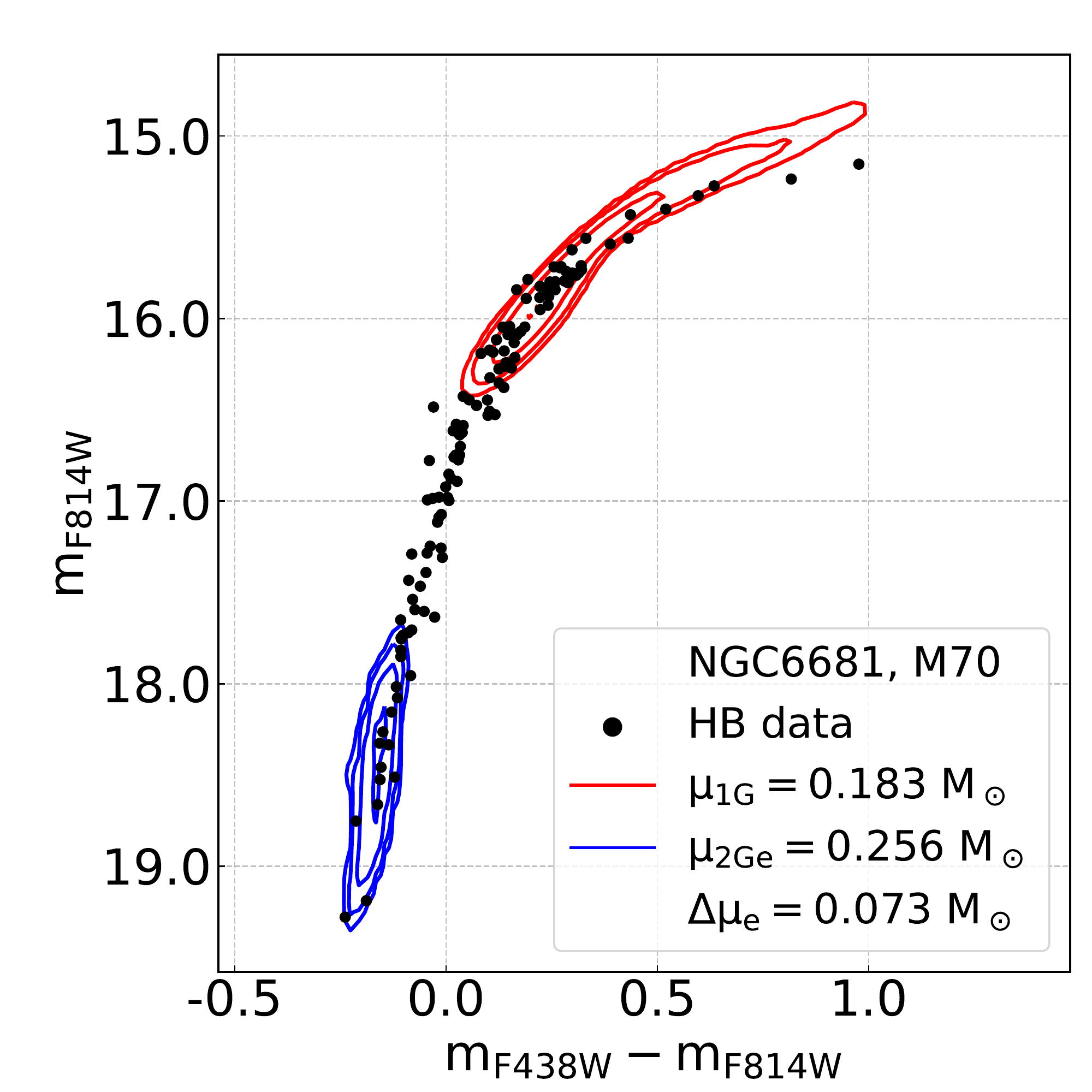}
    \includegraphics[width=0.66\columnwidth]{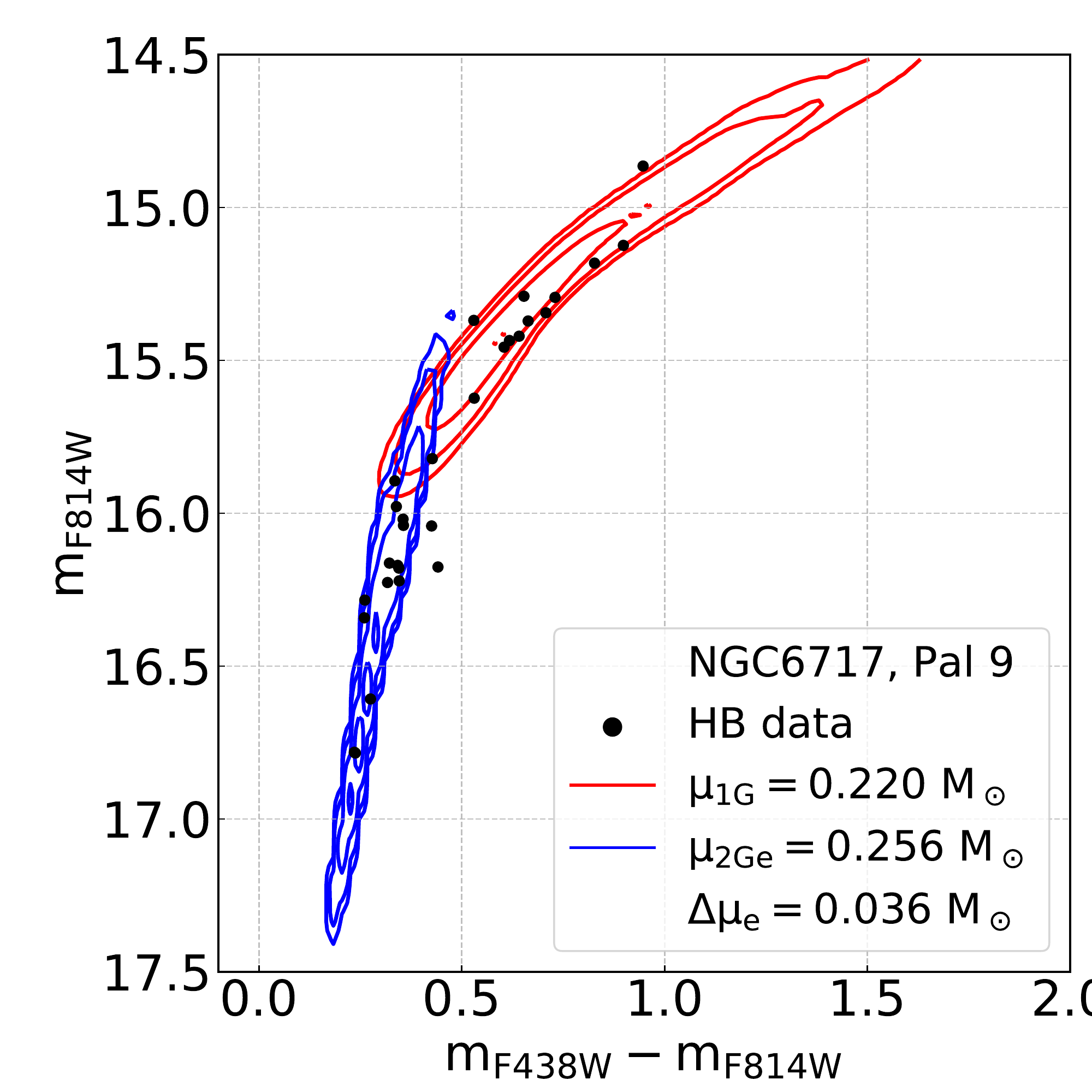}

    \caption{
    As Figure \ref{pic:showcase_1} for NGC\,6366, NGC\,6397, NGC\,6441, NGC\,6496, NGC\,6535, NGC\,6541, NGC\,6584, NGC\,6624, NGC\,6637, NGC\,6652, NGC\,6681 (M\,70) and NGC\,6717 (Pal\,9).
    }
    \label{pic:showcase_3}
\end{figure*}

\begin{figure*}
    \centering
    %placeholders
    \includegraphics[width=0.66\columnwidth]{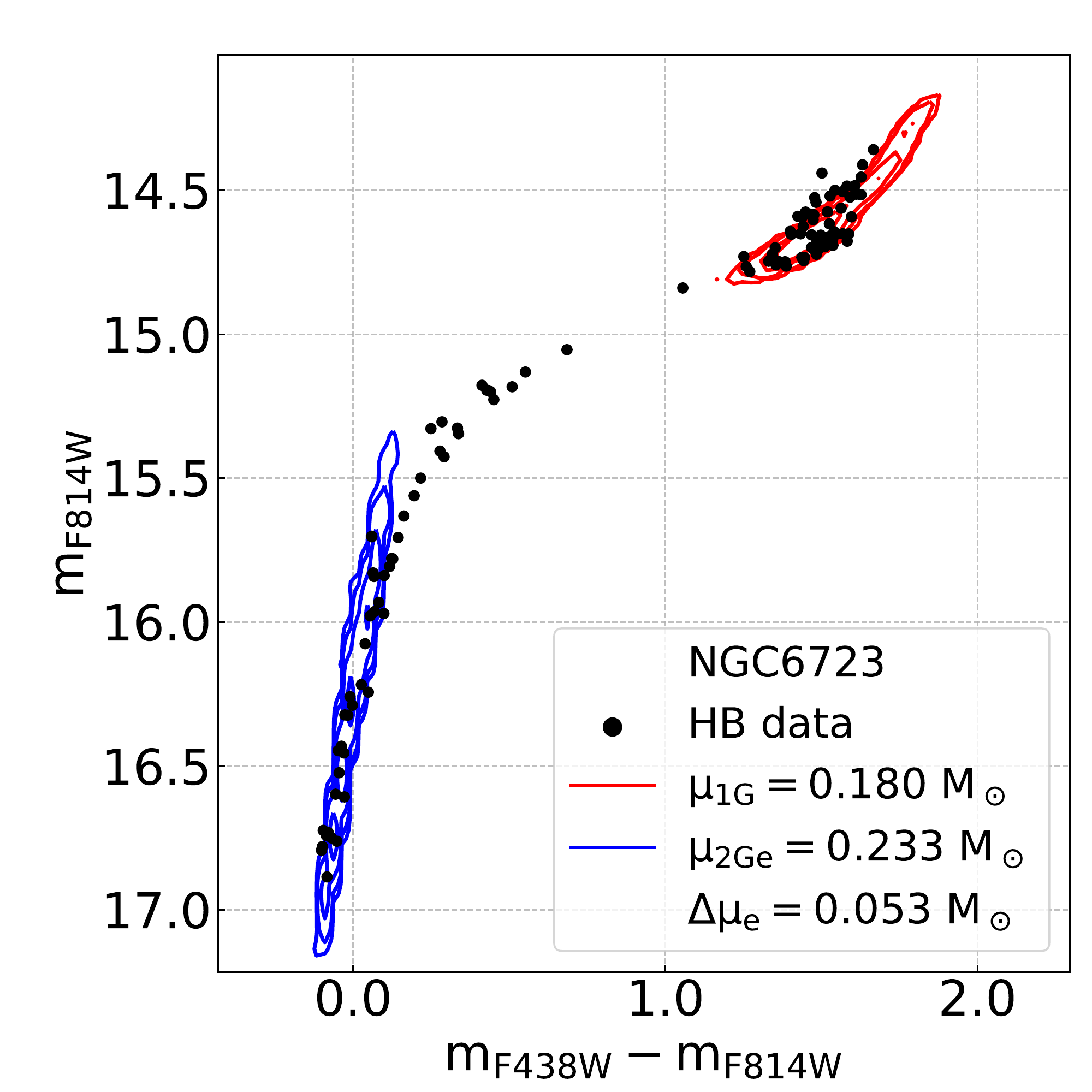}
    \includegraphics[width=0.66\columnwidth]{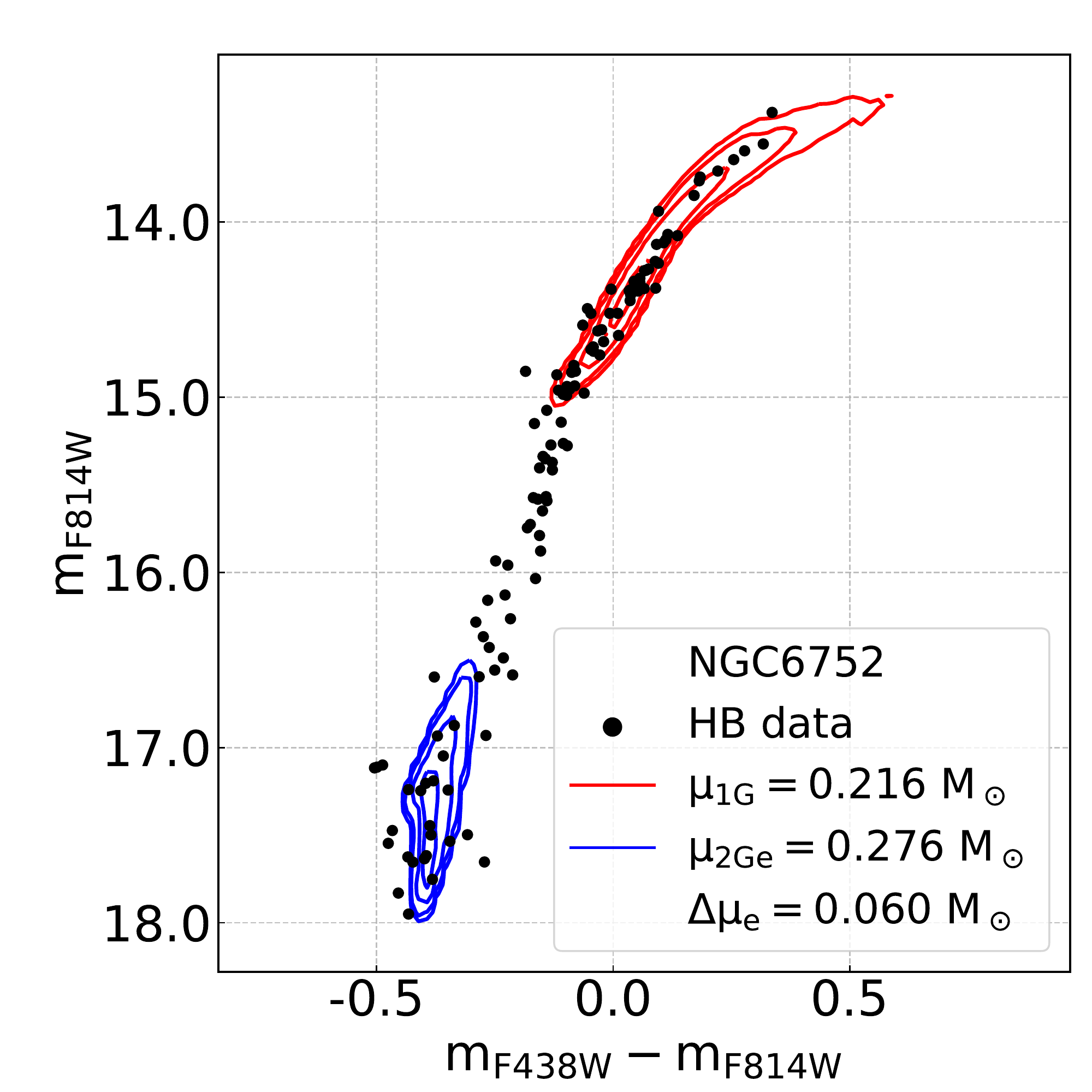}
    \includegraphics[width=0.66\columnwidth]{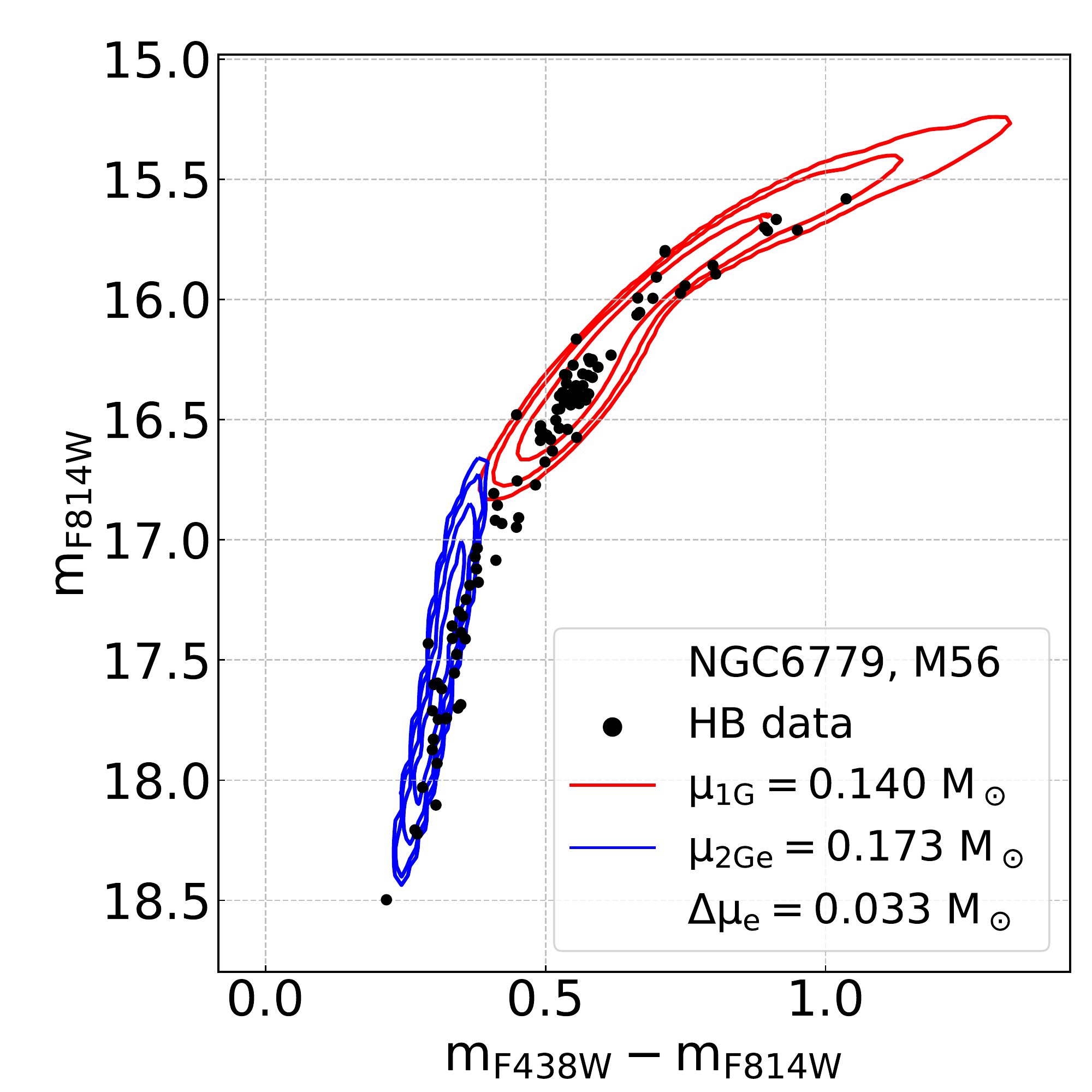}
    \includegraphics[width=0.66\columnwidth]{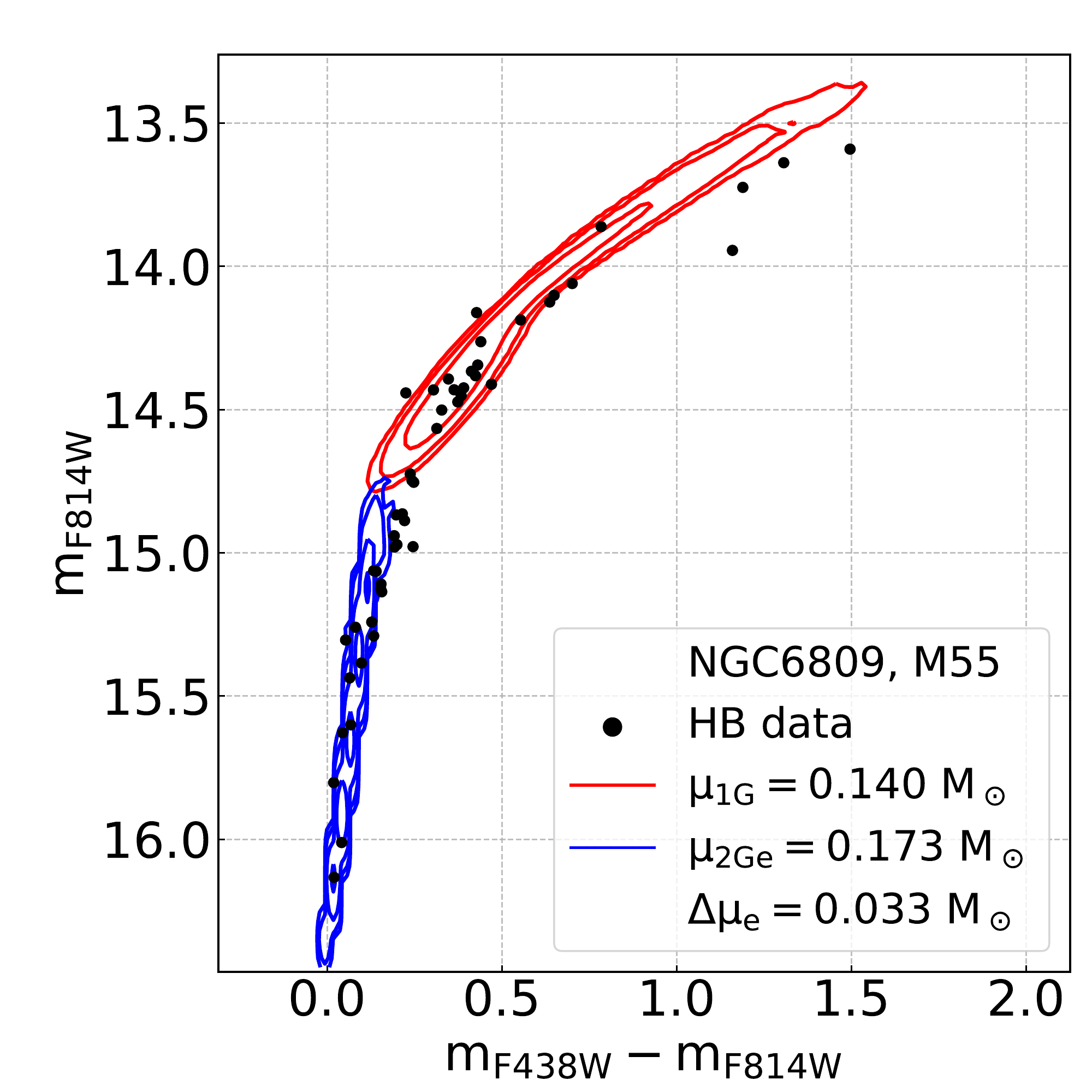}
    \includegraphics[width=0.66\columnwidth]{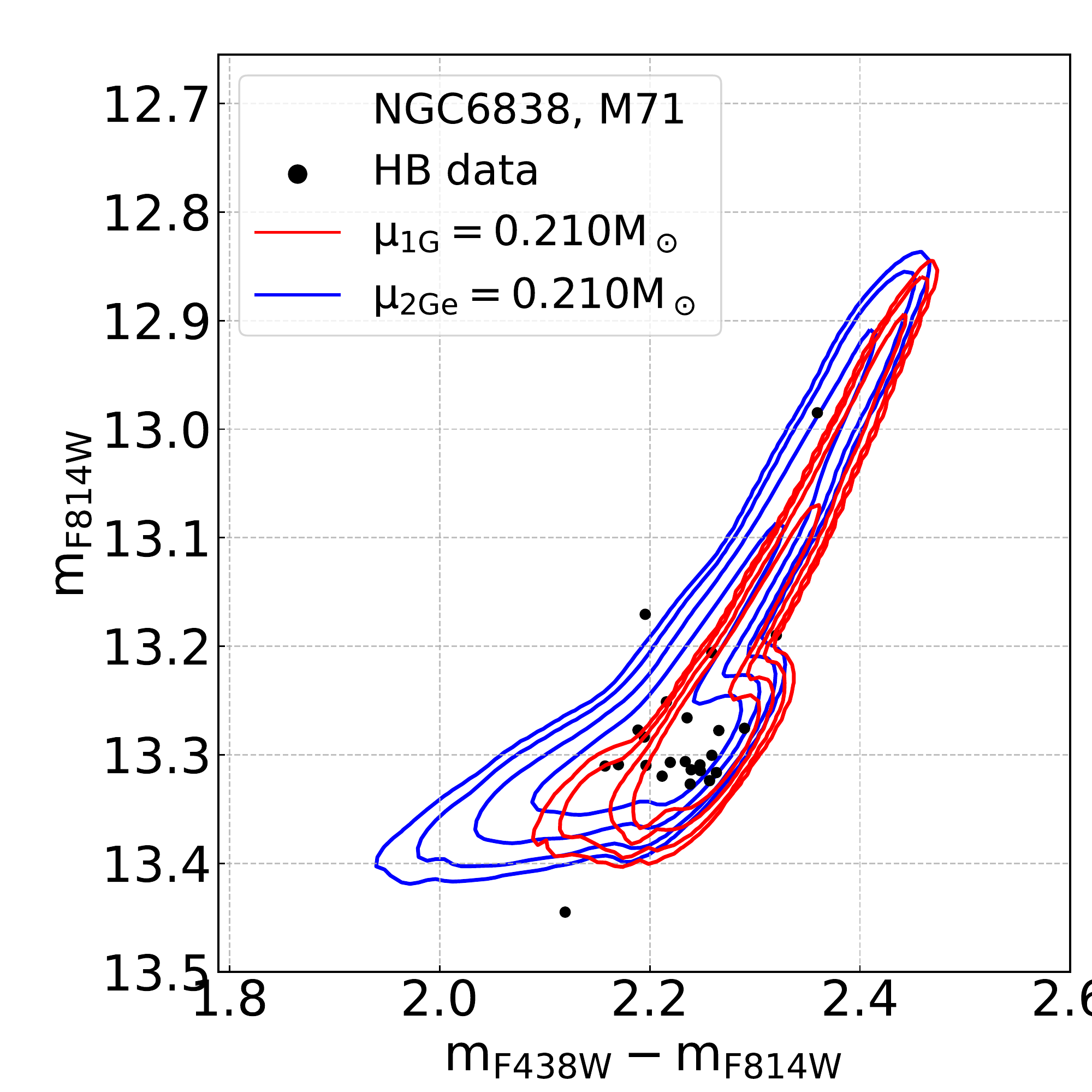}
    \includegraphics[width=0.66\columnwidth]{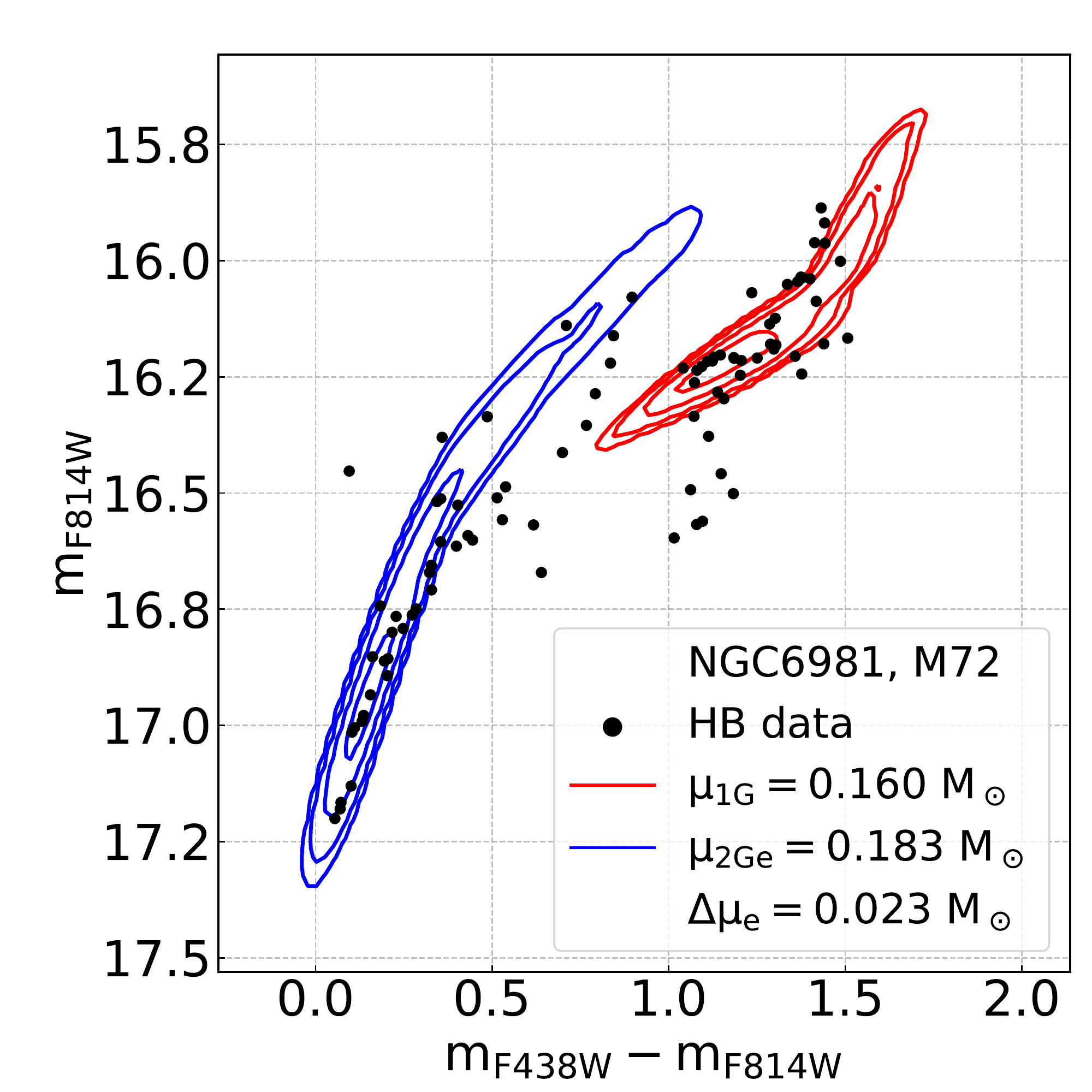}
    \includegraphics[width=0.66\columnwidth]{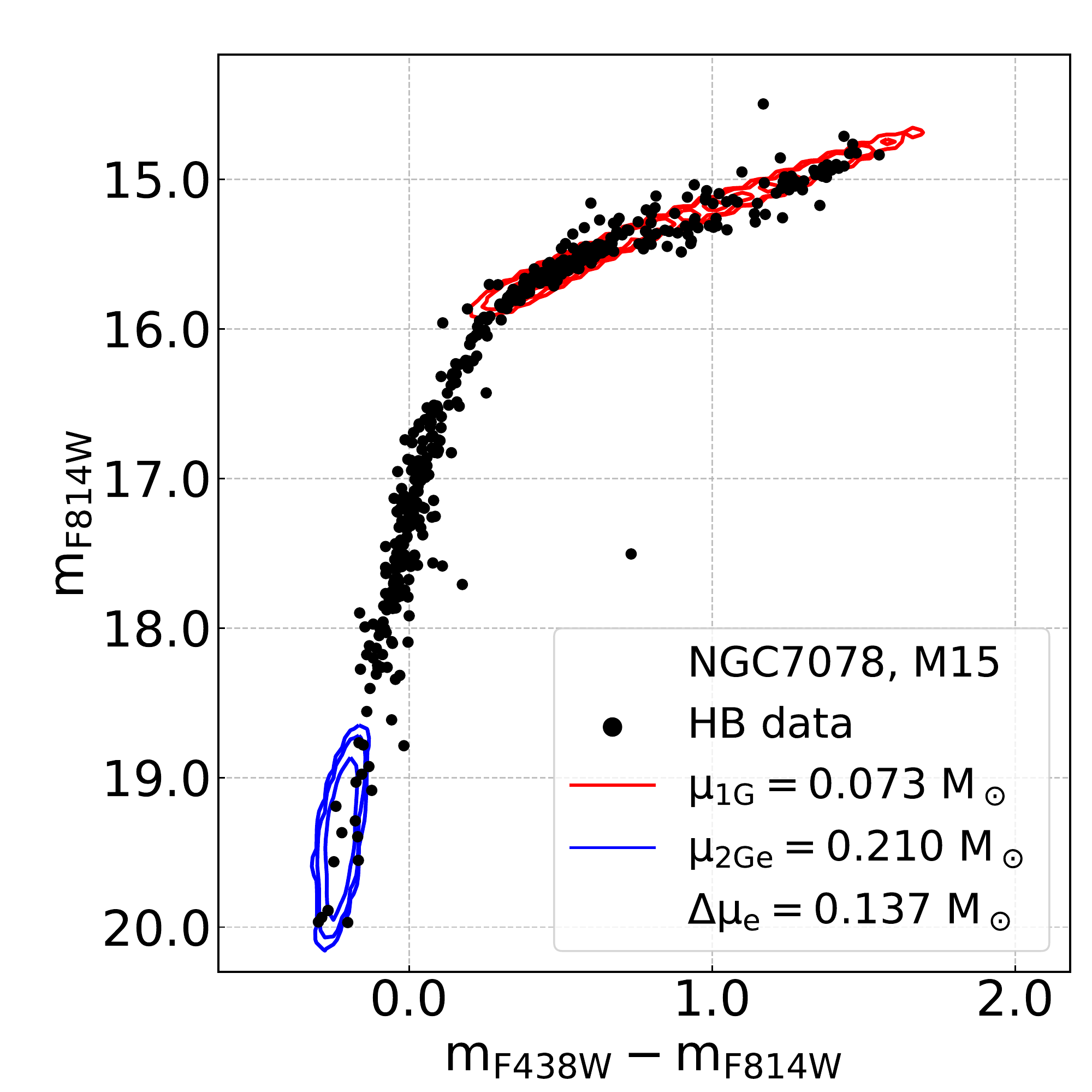}
    \includegraphics[width=0.66\columnwidth]{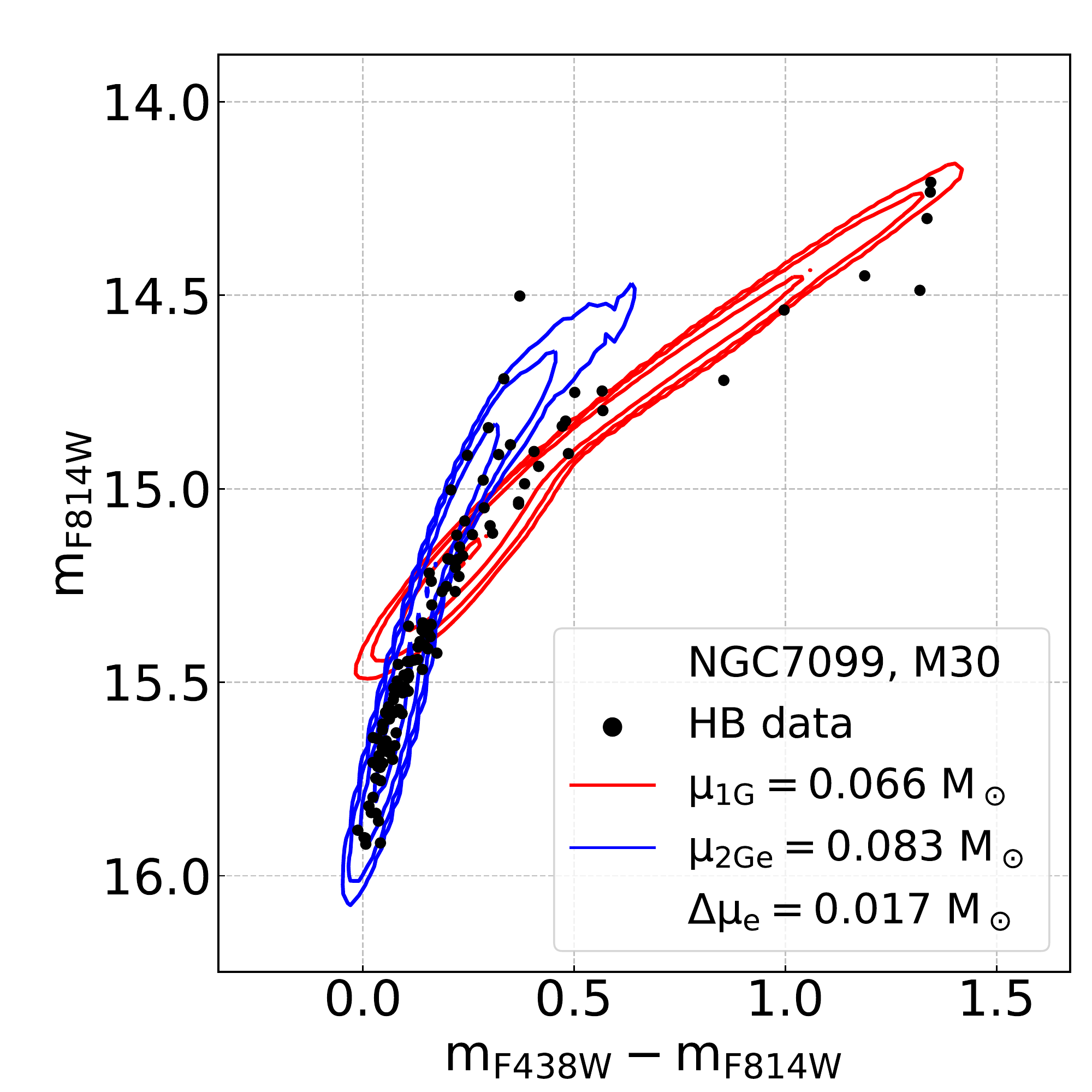}

    \caption{
    As Figure \ref{pic:showcase_1} for NGC\,6723, NGC\,6752, NGC\,6779 (M\,56), NGC\,6809 (M\,55), NGC\,6838 (M\,71), NGC\,6981 (M\,72), NGC\,7078 (M\,15), NGC\,7099.
    }
    \label{pic:showcase_4}
\end{figure*}

We show in Figures \ref{pic:showcase_1} to \ref{pic:showcase_4} the HBs of the 44 GCs analysed in this work.
Each panel in the collection represents the CMD of the HB stars in each GC, in the $\rm m_{F438W}-m_{F814W}$ vs. $\rm m_{F814W}$ bands. We represent the best fit simulations for the 1G and the 2Ge as the red and blue contour plots, respectively. If the GC has only red HB stars (as in the case of NGC\,6637), the blue contour plot refers to the 2G as a single group. The label in each panel reports the value of $\rm \mu$ for the two stellar populations and, when relevant, the difference between these value ($\rm \Delta \mu_e$). The different GCs are listed following their NGC number for an easier identification.
For the cases of NGC\,6121 and NGC\,5272, which bring our total sample to 46 GCs, we refer to \cite{tailo_2019a} and \cite{tailo_2019b} respectively.  

\label{sec:app_showcase}

\section{Bi-variate relations with GC parameters}
\label{sec:app_parameters}

\begin{table*}
    \centering
	\caption{Summary of the GC parameters we exploit in this work, together with the symbol we associate them, and their source. The sources references are the following: (a) \citet{milone_2018}; (b) \citet{marinf_2009}; (c) \citet{dotter_2010}; (d) \citet{VandenBerg_2013}; (e) \citet{mackey_2005}; (f) \citet[][in its 2003 and 2010 version]{harris_1996}; (g) \citet{milone_2014}; (h) \citet{milone_2012c}; (i)  \citet{milone_2017}; (l) \citet{lagioia_2019}; (m) \citet{torelli_2019}; (n) \citet{baumgardt_2018}; (o) \citet{baumgardt_2019}
	}
    \begin{adjustbox}{width=2\columnwidth,center}
    \begin{tabular}{lc|lc}
    \hline
    \hline
    Parameter name&Abbr.&Parameter name&Abbr.\\
    \hline
    Helium enhancement of the 2G (a)&$\rm \delta Y_{2G,1G}$&Helium enhancement of the 2Ge (a)&$\rm \delta Y_{2G,1G}$\\
    Relative Age (b) &Age&Age (c)&--\\
    Age (d)&--&Horizontal branch ratio (e)&HBR\\    
    Iron over Hydrogen ratio (f)&[Fe/H]&Reddening(f)&E(B-V)\\        
    Integrated visual absolute magnitude (f)&$\rm M_V$&Concentration (f)&c\\    
    Luminosity density at the center (f)&$\rm \rho_0$&Central Surface brightness (f)&$\rm SB_0$\\    
    Projected ellipticity of isophotes (f) &$\rm \epsilon$&Specific RRLyrae density (f) &$\rm S_{RR}$\\   
    Distance of the rHB from the RGB (g)&$\rm L_1$&HB length (g)&$\rm L_2$\\         
	Binary (q>0.5) fraction in the core (h)&$\rm f_{bin,c}$&\makecell[l]{Binary (q>0.5) fraction between the core\\ and the half mass radius (h)}&$\rm f_{bin,hm}$\\
    Binary (q>0.5) fraction outside the half mass radius (h)&$\rm f_{bin,ohm}$&1G star ratio (i)&$\rm N_{1G}/N_{T}$\\
    RGB width in $\rm m_{F275W}-m_{F814W}$ (i) &$\rm W_{F275W}$&RGB width in $\rm C_{F275W,F336W,F438W}$ (i)&$\rm W_{C,F275W}$\\
	\makecell[l]{RGB width in $\rm C_{F275W,F336W,F438W}$\\ without [Fe/H] dependency (i)}&$\rm \Delta W_{C,F275W}$&RGB width in $\rm C_{F336W,F438W,F814W}$ (l)&$\rm W_{C,F336W}$\\
	\makecell[l]{RGB width in $\rm C_{F336W,F438W,F814W}$\\ without [Fe/H] dependency (l)}&$\rm \Delta W_{C,F336W}$& Ratio between CND in I and V-I bands (m)&$\rm \tau_{HB}$\\
    Mass of the cluster (n) &M&Initial mass of the cluster (o)&$\rm M_i$\\    
    Mass Luminosity ratio in the V band (n) &M/L&Core radius (n)&$\rm r_c$\\    
    Projected half light radius (n)&$\rm r_{hl}$&Half mass radius (n)&$\rm r_{hm}$\\    
    Tidal radius (n)&$\rm r_t$&Mass function slope (n)&MF slope\\    
    Core density (n)&$\rm \rho_{c}$& Density within the half mass radius (n)&$\rm \rho_{hm}$\\    
    Relaxation time at half mass (n)&$\rm T_{rh}$& Mass fraction of the dark remnants (n)&$\rm F_{remn}$\\    
    Central velocity dispersion (n)&$\rm \sigma_{0}$&Central escape velocity (n)&$\rm v_{esc}$\\    
    Mass segregation coefficient in the core (n)&$\rm \eta_{c}$& \makecell[l]{Mass segregation coefficient\\ within the half mass radius (n)}&$\rm \eta_{c}$\\ 
    Distance from the galactic center (o)&$\rm R_{GC}$&Mean heliocentric velocity (o)&$\rm RV$\\    
    Apogalacticon radius (o)&$\rm R_{apog}$&Perigalacticon radius (o)&$\rm R_{perig}$\\     
    Position components (o)&$\rm (X,Y,Z)$&Velocity components (o)&$\rm (U,V,W)$\\         
    \hline
    \hline
	\end{tabular}
	\end{adjustbox}
	\label{tab:sources}
\end{table*}

\begin{figure*}
    \centering
    %placeholders
    \includegraphics[width=1.8\columnwidth,trim={0cm 0cm 0cm 0cm}]{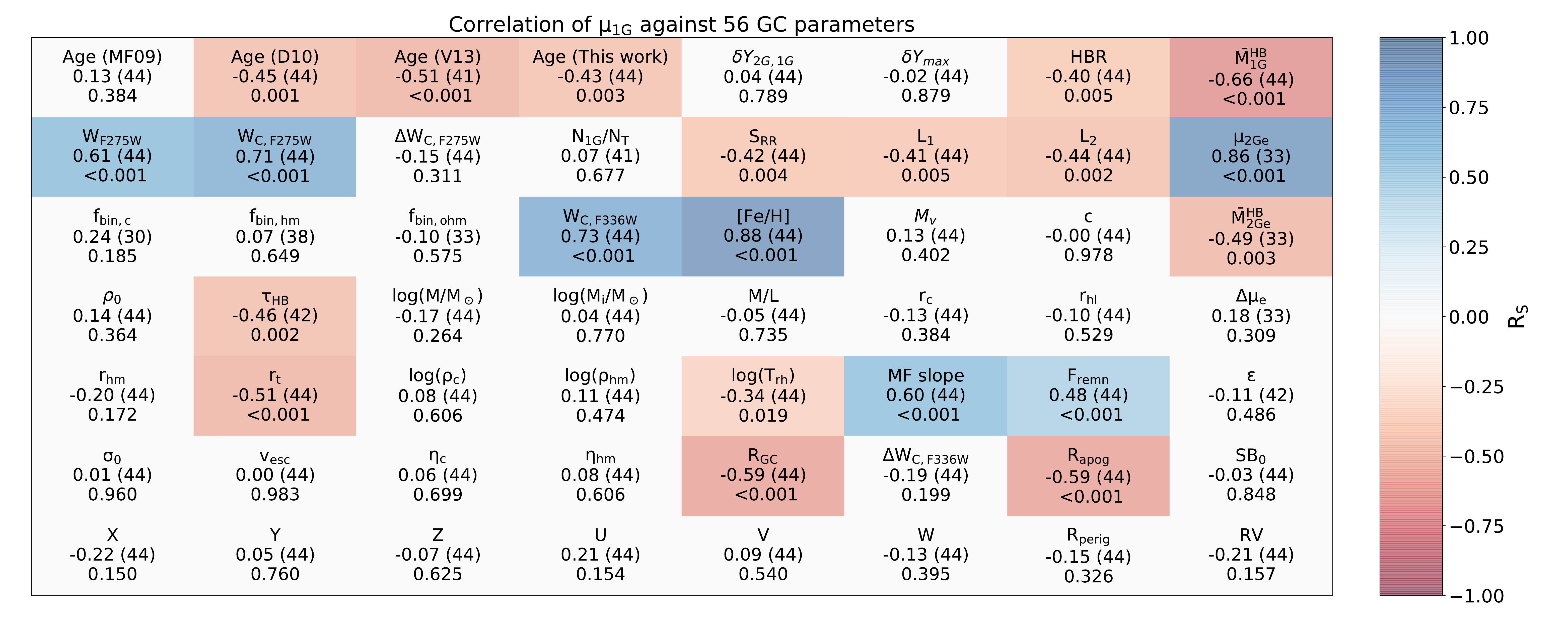}
    \caption{Correlation of $\rm \mu_{1G}$ against 56 GC parameters. Each cells in the figure represents a parameter. The first row reports the name of parameter (see \S\,\ref{sec:app_parameters} and Table \ref{tab:sources}). The second row reports the value of the Spearman rank coefficient ($\rm R_s$) and the number of degrees of freedom. The last row in the cells reports the p-value for the test. Each cell is color coded according to the value of $\rm R_s$. No colour has been assigned to non-significant correlations (p>0.05).}
    \label{pic:map_mu1g}
\end{figure*}

\begin{figure*}
    \centering
    %placeholders
    \includegraphics[width=1.8\columnwidth,trim={0cm 0cm 0cm 0cm}]{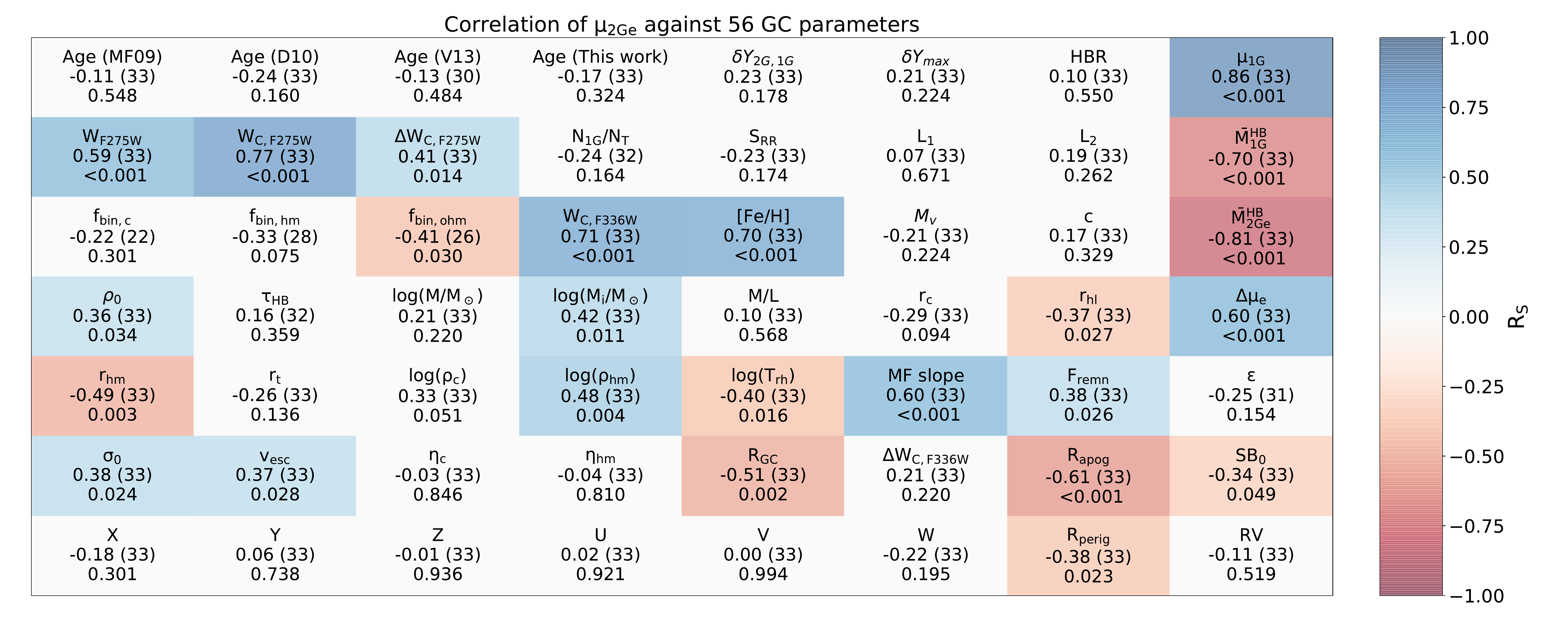}
    \caption{As Figure \ref{pic:map_mu1g} but for $\rm \mu_{2Ge}$.}
    \label{pic:map_mu2ge}
\end{figure*}

\begin{figure*}
    \centering
    %placeholders
    \includegraphics[width=1.8\columnwidth,trim={0cm 0cm 0cm 0cm}]{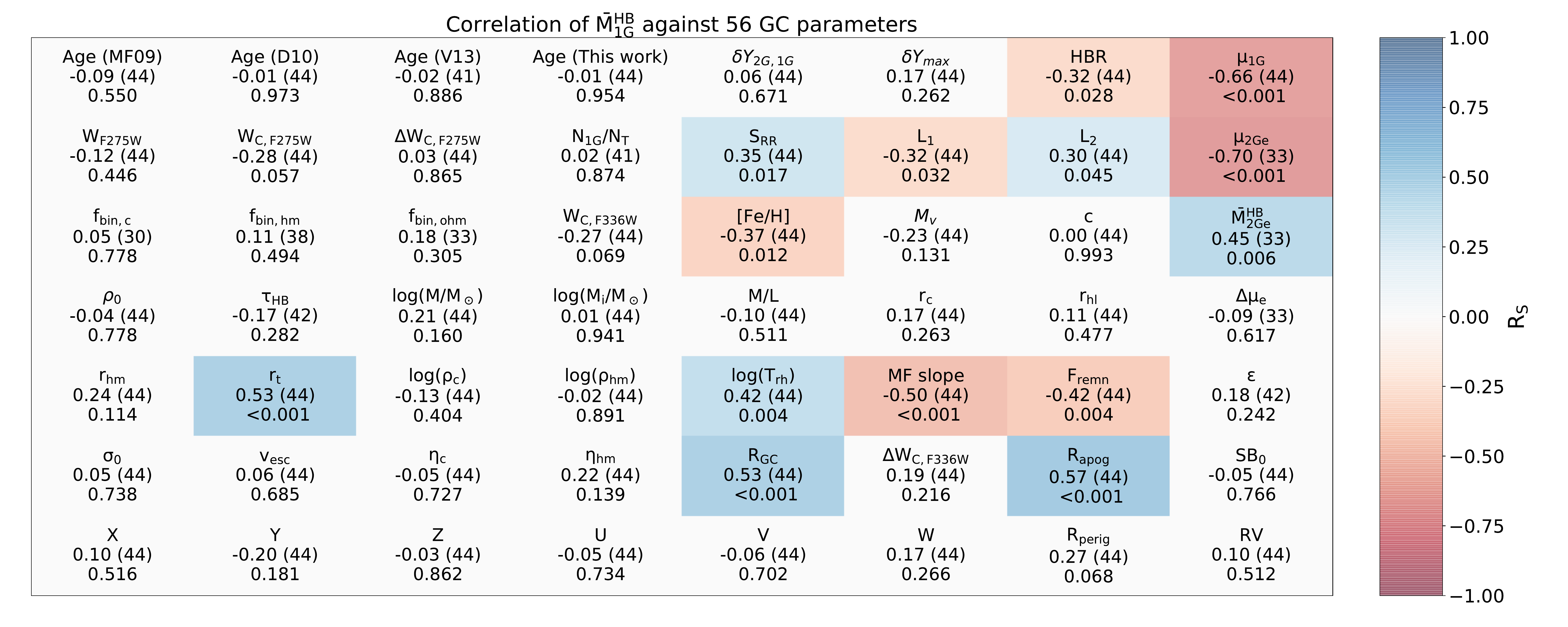}
    \caption{As Figure \ref{pic:map_mu1g} but for $\rm \bar{M}^{HB}_{1G}$.}
    \label{pic:map_mhb1g}
\end{figure*}

\begin{figure*}
    \centering
    %placeholders
    \includegraphics[width=1.8\columnwidth,trim={0cm 0cm 0cm 0cm}]{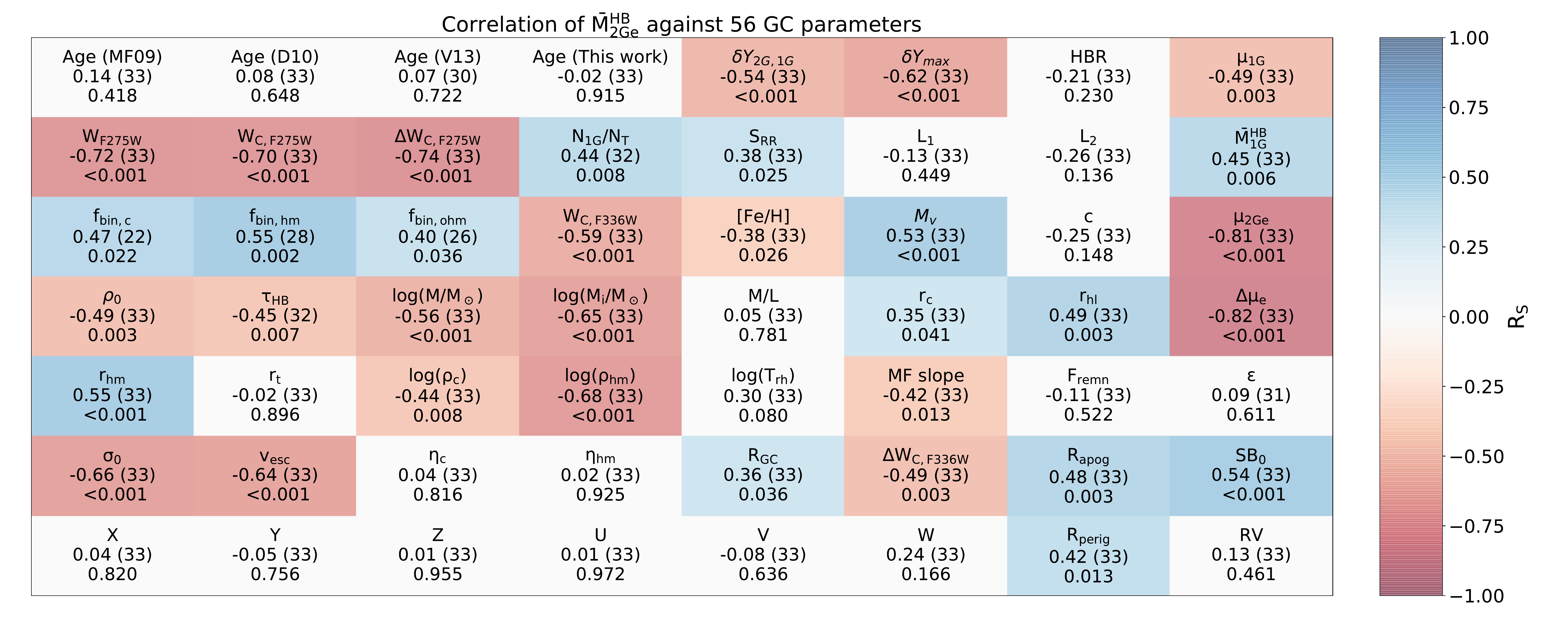}
    \caption{As Figure \ref{pic:map_mu1g} but for $\rm \bar{M}^{HB}_{2Ge}$.}
    \label{pic:map_mhb2ge}
\end{figure*}

\begin{figure*}
    \centering
    %placeholders
    \includegraphics[width=1.8\columnwidth,trim={0cm 0cm 0cm 0cm}]{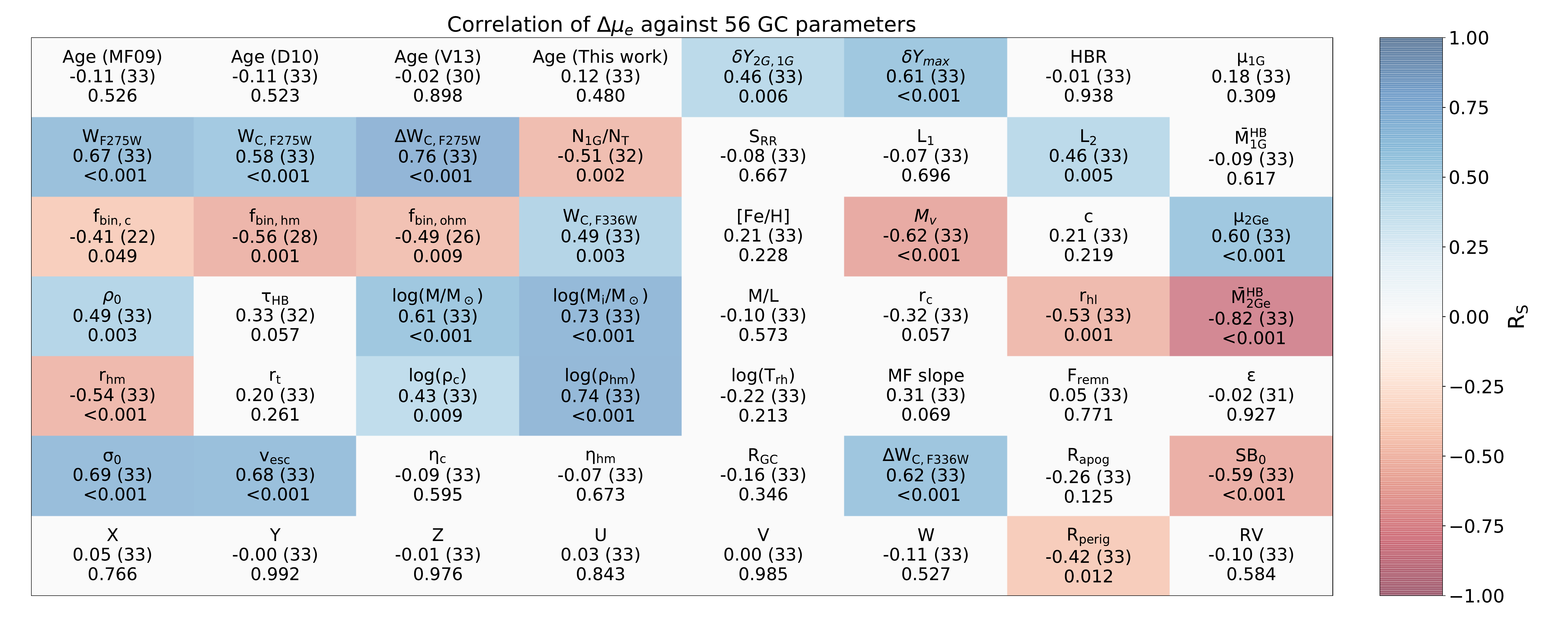}
    \caption{As Figure \ref{pic:map_mu1g} but for $\rm \Delta \mu_e$.}
    \label{pic:map_dmue}
\end{figure*}

In the main paper we investigated the correlations between mass loss, average HB mass of 1G and 2Ge stars and some parameters of the host GCs. For completeness, we provide in the following the results of a bi-variate analysis that involves the 56 GC parameters that are listed in Table \ref{tab:sources}.  Specifically, we provide the results on the correlations between $\rm \mu_{1G}$,  $\rm \mu_{2Ge}$, ($\rm \bar{M}^{HB}_{1G}$, $\rm \bar{M}^{HB}_{2Ge}$, $\rm \Delta \mu_e$) and the cluster parameters in correlation maps provided in Figures~\ref{pic:map_mu1g} (~\ref{pic:map_mu2ge}, ~\ref{pic:map_mhb1g}, ~\ref{pic:map_mhb2ge}, ~\ref{pic:map_dmue}). 

For each couple of parameters, we provide the Spearman rank coefficient ($\rm R_S$) and the p-value, which are indicative of the significance of the correlation, and the number of degrees of freedom.  The color of each cell Figures \ref{pic:map_mu1g} to \ref{pic:map_dmue} is indicative of the value of $\rm R_S$ as indicated by the color-bar. The correlations that, according to their p-value have low significance (p$>$0.05).
 
The strongest correlation described in Figure \ref{pic:map_mu1g} involves $\rm \mu_{1G}$ and the iron abundance as discussed in the main text. We also find significant correlations with the RGB width in the F275W$-$F814W color and the $C_{\rm F275W,F336W,F438W}$ and $C_{\rm F336W,F438W,F814W}$ pseudo-colors, which are quite expected because these quantities correlates with cluster metallicity \citep{milone_2017, lagioia_2019}.  $\rm \mu_{1G}$ anti-correlates with cluster ages as a consequence of the age-metallicity relation.
Other correlations involve $\rm R_{GC}$, $\rm R_{apog}$, the slope of the MF slope and $\rm r_t$ ($\rm R_S=-0.51$). As illustrated in Figure~\ref{pic:map_mu2ge}, similar conclusion can be extended to the mass loss of 2Ge stars. 

The average mass loss of 1G stars show some correlations with $\rm R_{apog}$, $\rm R_{GC}$, MF slope and $r_t$ (Figure \ref{pic:map_mhb1g}).
The mass loss of 2Ge stars exhibit strong anti-correlations with  $\rm \Delta W_{C,F275W}$, $\rm  W_{F275W}$ and $\rm W_{C,F275W}$ ($R_S \lesssim -0.7$). Additional correlations and anticorrelations involve $\rm \delta Y_{2G,1G}$, $\rm \delta Y_{max}$, $\rm f_{bin,hm}$, $\rm W_{C,F336W}$, $\rm M_V$, present-day mass and the initial mass of the host cluster, $\rm \rho_{hm}$, $\rm \sigma_0$, $\rm SB_0$ and $\rm v_{esc}$. Noticeable, these quantities correlate or anticorelate with each other \citep[e.g.][]{milone_2017, milone_2020, lagioia_2019}.
Similar conclusion can be extended to  $\rm \Delta \mu_e$ (Figure \ref{pic:map_mhb2ge}).

\section{The age--metallicity relation and the impact of age on the HB morphology}
\label{sec:app_age_rel}

\begin{figure*}
    \centering
    %placeholders
    \includegraphics[width=1.8\columnwidth,trim={3cm 0cm 3cm 0cm}]{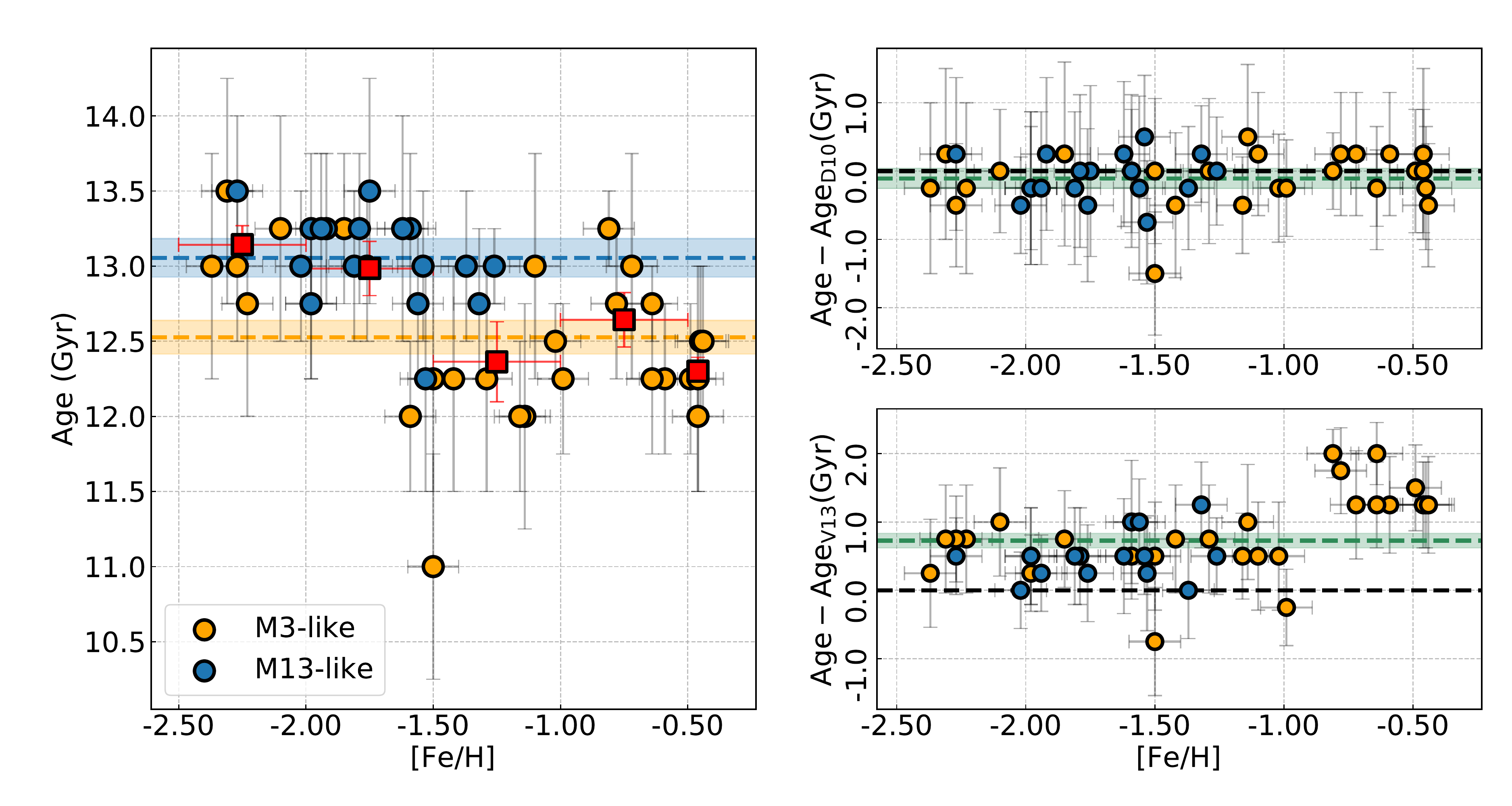}
    \caption{\textit{Left panel:} Age as function of the [Fe/H] value for the GC in our sample. M3- and M13-like GCs are represented as orange and blue dots, respectively. The red squares represent the average age values of different [Fe/H] intervals (see text) while the two dashed lines are the average age of the M3- and M13-like clusters, together with their $1\sigma$ intervals. \textit{Right panels:} Comparison of the results of this work with the ones from \citet[][D10, top]{dotter_2010} and \citet[][V13, bottom]{VandenBerg_2013}. The green dashed lines represent the average age difference between two sets, together with their $\rm 1\sigma$ intervals.}
    \label{pic:a_fe_rel}
\end{figure*}

In the following, we further investigate the impact of age and metallicity on the HB morphology.
A visual inspection at the left panel of Figure \ref{pic:a_fe_rel} reveals that, based on the ages derived in this paper, the metal poor GCs are, on average, older than the metal rich ones. To quantify age variation as a function of metallicity, we defined five intervals of iron abundance, ranging from [Fe/H]=$-$2.5 to solar in bins of 0.5 dex and estimate the average age of GC in each bin (red squares in the left panel of Figure \ref{pic:a_fe_rel}). Clearly, the average age decreases towards higher metallicity with clusters having [Fe/H]$>$-1.5 being, on average, $\sim 0.75$Gyr younger of their metal-poorer counterparts.

The age-metallicity relation derived in this work is similar to the one by \citet{dotter_2010} for the same GCs. Moreover, we find that GCs with large Galactocentric distances ($\rm R_{GC}>$8kpc) are older than those with $\rm R_{GC}<$8kpc, thus confirming previous conclusion that the age-metallicity relation is composed of two branches \citep[e.g.][]{forbes_2010, dotter_2011,kruijssen_2019}. 
This result is quite expected. Indeed, , as shown in the upper-right panel of Figure~\ref{pic:a_fe_rel}, when we compare our ages with those by \citet{dotter_2010} we find a good agreement (at $1-\sigma$ level).  On the contrary, our ages are systematically older by $\sim$0.75 Gyr than those provided by \citet{VandenBerg_2013} and the age difference, which is significant at more than 3-$\sigma$, increases with metallicity.

\begin{figure*}
    \centering
    %placeholders
    \includegraphics[width=1.8\columnwidth,trim={3cm 0cm 3cm 0cm}]{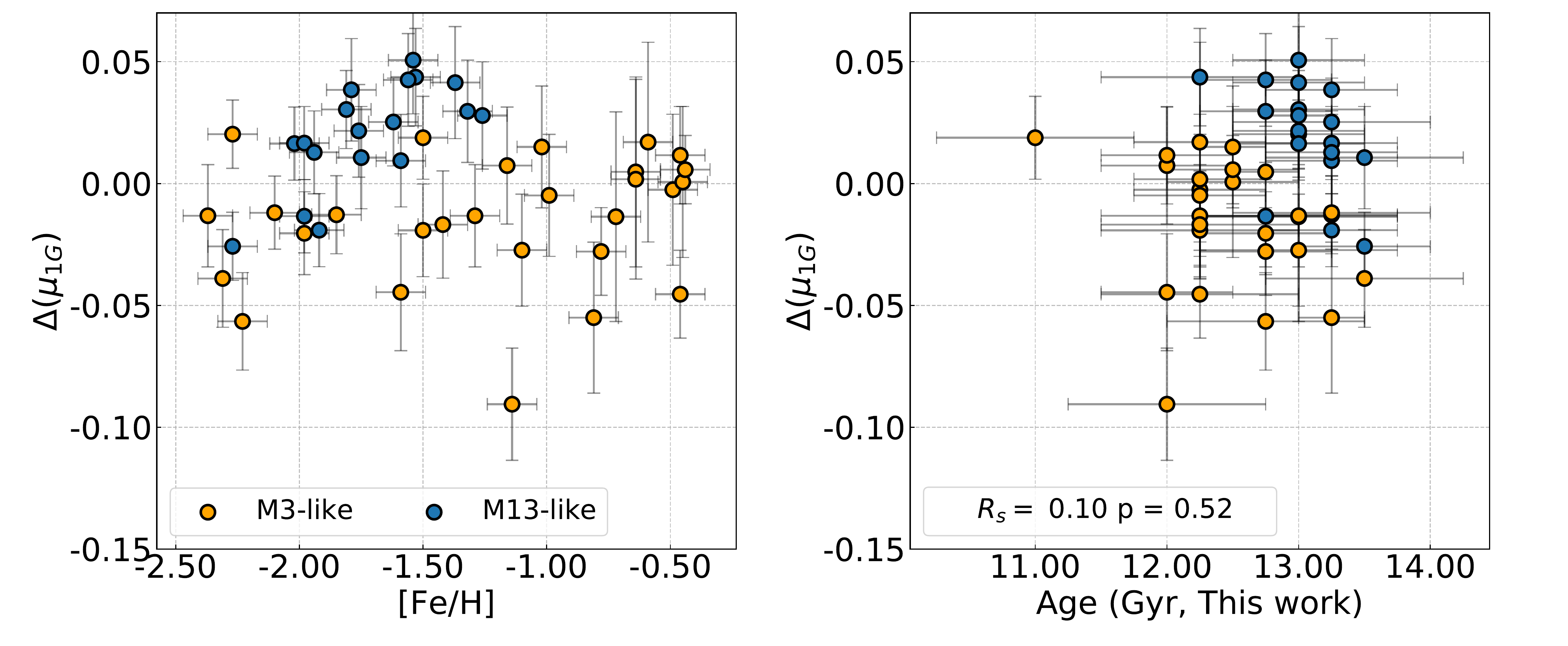}
    \caption{
    Residuals relative to Eq. 1 of the main text ($\rm \Delta(\mu_{1G})$) as function of [Fe/H] (left panel) and age (right panel).}
    \label{pic:a_residuals}
\end{figure*}

Figure \ref{pic:a_fe_rel} reveals that M\,3-like and M\,13-like GCs have different average ages (12.5$\pm\, 0.1$ and 13.0$\pm\, 0.1$ Gyr, respectively). To further investigate the relation between age and mass loss, we show in the right panel of \ref{pic:a_residuals} the mass-loss residuals to Eq.~1 ($\rm \Delta(\mu_{1G})$) against [Fe/H] and age. 
Although there is no significant correlation between these two quantities (R$_{\rm s}=0.1$), possibly as a consequence of their relatively-large error bars, we conclude that M\,13-like GCs are older and lose more mass than M\,3-like GCs and such differences are significant at more than 4-$\sigma$ level.
We verify that similar conclusions are inferred by using the ages by \citet{dotter_2010,dotter_2011}, \citet{VandenBerg_2013} and \citet{marinf_2009}.

However, the evidence that M\,3-like and M\,13 like GCs exhibit different HB shapes \citep[][see their Figure~2]{milone_2014} and have different mass losses does not indicate that age is responsible for the HB morphology. 
The reason to exclude age as a second parameter of the HB morphology, is that the simulated HBs used to infer the mass loss properly take into account cluster ages. This fact is clearly illustrated in Figure \ref{pic:mtip_1g_params}, where we  show how each value of $\rm M^{Tip}_{1G}$ that we inferred in this work is tailored for a specific cluster and is derived by using the corresponding age. 
Moreover,  the comparison between observed and simulated HBs represented in Figure \ref{pic:mtip_1g_params} reveals that difference in $\rm M^{Tip}_{1G}$ induced by an age difference of 0.5\,Gyr is $\sim$0.01\,$\rm M_\odot$. 
 Such a small amount of mass is not enough to compensate the differences in the 1G HB mass observed between the M\,3- and M\,13-like GCs, even if our simulations had not taken age into account.

We conclude that age does not play a major role in determining the HB morphology of Galactic GCs, thus challenging the conclusion by several  papers, where age is considered a second parameter of the HB morphology in GCs \citep[e.g.\,][]{sandage1967, gratton2010,dotter_2010,VandenBerg_2013, milone_2014}. Anyway, the small  age difference found between the M3- and the M13-like GCs leaves some residual space to argue that it may cause (through the small difference in evolving mass) a second-order effect on the mass lost by their RGB stars, linked to the complex influence of pulsation and magneto-acoustic processes, still largely unknown, on mass loss \citep[e.g.][]{mcdonald_2007}.

\section{ Standard helium flash models}
\label{app:standard_flash}
\begin{figure*}
    \centering
    %placeholders
    \includegraphics[width=1.00\columnwidth,trim={1cm 0cm 0cm 0cm}]{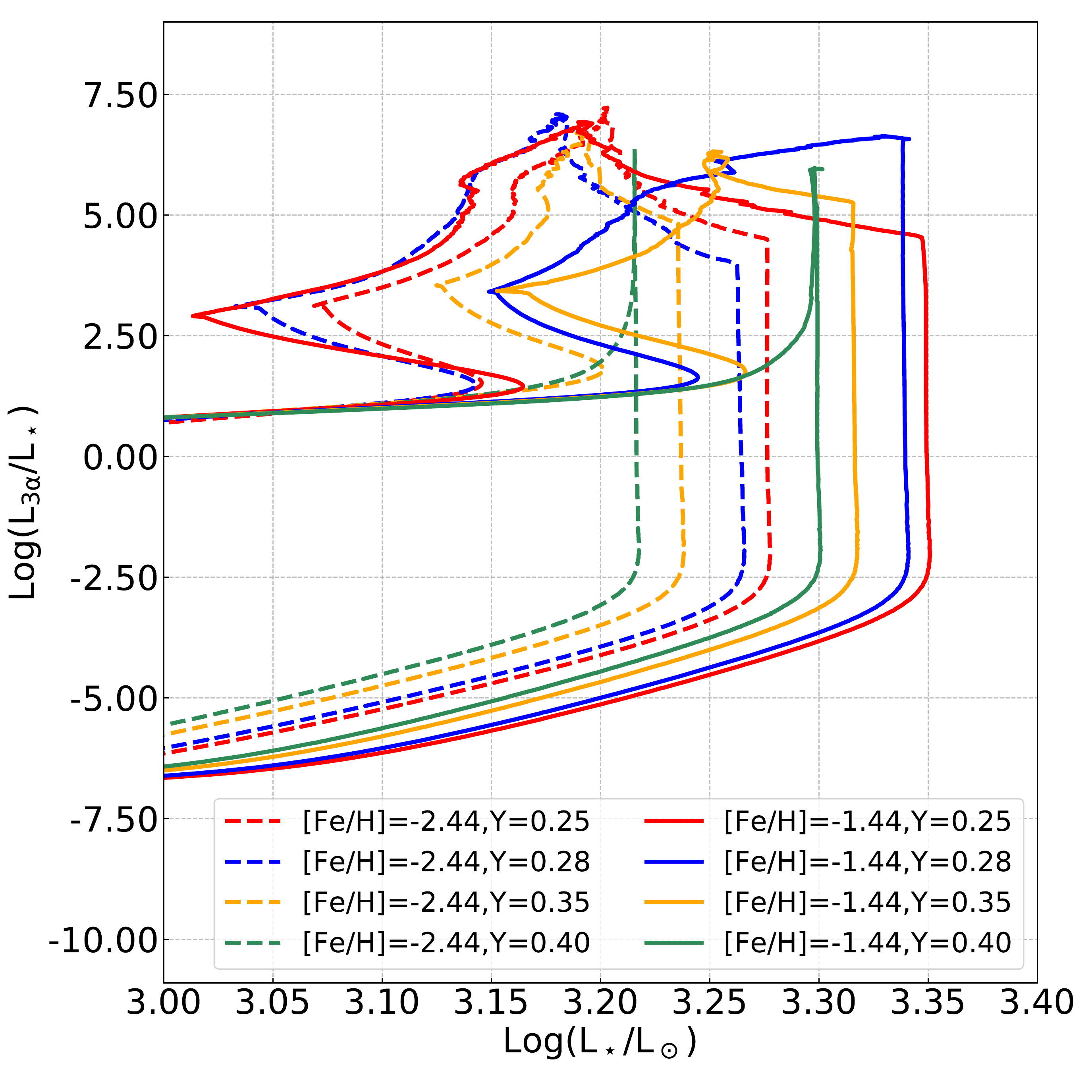}
    \includegraphics[width=1.00\columnwidth,trim={1cm 0cm 0cm 0cm}]{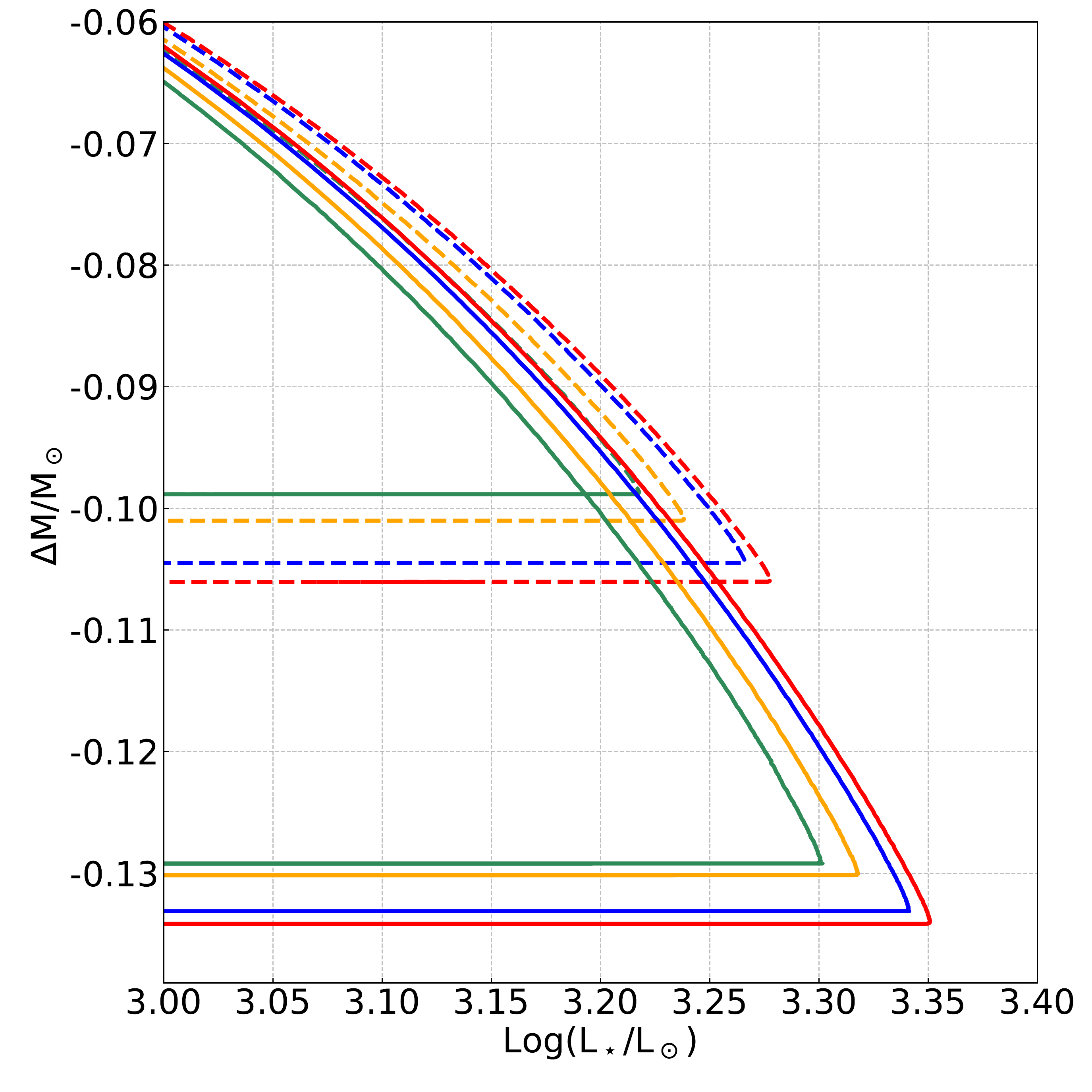}
    \caption{
    \textit{Left panel:} Fraction of the stellar luminosity coming from 3 alpha reactions, tracer of the helium flash process, as function of the total luminosity, tracer of the evolutionary stage. 
 	\textit{Right panel:} Total mass lost ($\rm \Delta M/M_\odot$) by the models in the left panel as function of luminosity.
    }
    \label{pic:mloss_flash_l3a}
\end{figure*}

Standard models and evolutionary effects do not predict the increased mass loss needed to describe the location of the 2Ge HB stars in the CMD. To show this we compare the evolution trough the helium flash of a series of stellar tracks with different initial helium, in Figure \ref{pic:mloss_flash_l3a}.

The models shown here have [Fe/H]$=\,$--2.44 and --1.44, and are obtained with $\rm \eta_R=0.3$. All tracks reach the corresponding RGB tip in 12.0\,Gyr. Since each model refers to different helium abundance, we have smaller masses for higher helium. In particular  for [Fe/H]$=\,$--2.44 with  Y$=\,$0.25,0.28,0.35,0.40 we have $\rm M= 0.806,\,0.765,\,0.674,\,0.613,\, M_{\odot}$, respectively; for [Fe/H]$=\,$--1.44 
and the same Y's we have $\rm M= 0.828,\,0.786,\,0.692,\,0.628\, M_{\odot}$. The considerations we make in this section are valid for our entire database.

We plot in the left panel of Figure \ref{pic:mloss_flash_l3a} the fraction of the model luminosity coming from the 3 alpha reaction chains ($\rm log(L_{3\alpha}/L_\star)$), involved in the helium flash, as a function of the total luminosity of the star ($\rm log(L_\star/L_\odot)$), a proxy of the evolutionary stage. 
 The right panel of Figure \ref{pic:mloss_flash_l3a}, on the other hand, shows the total mass lost ($\rm \Delta M/M_\odot$) as function of the total luminosity for the models.
  
From the figure we get a crucial clue:  the helium flash starts ($\log \rm L_{3\alpha}/L_*>-2.5$) at progressively lower luminosity for increasing Y.  This implies that, within a single family of models describing a Type I cluster, the mass loss rate 
can not vary too much without any additional inputs.
If we consider that the models complete their evolution at the same time, and that helium rich models have slightly shorter RGB phases, the total mass lost should be decreasing as Y increases (note that the increase in the mass lost between the two families of models is due to their different metallicity and not their helium abundance, see Figure\,\ref{pic:mloss_models_comparison}).
These considerations strengthen the need to include additional physical inputs to the standard models to obtain results in agreement with the observations. 

Again, we notice that an independent alternative explanation of the data might be possible. The physics of mass loss in the framework of the combination of pulsation and magneto-acoustic processes \citep[see e.g.][]{mcdonald_2007}, at the time of this writing, is not modelled in 1-D stellar evolution codes, so there is some space left to argue that the enhanced helium abundance of the 2Ge stars might concur to increase their surface activity in the RGB phase, ending in an increase in the instantaneous mass loss rate, so that the 2Ge stars naturally lose more mass during their evolution. The correlation with helium shown in Figure \ref{pic:dmue_dymax} would then be the natural outcome of such interactions.

%%%%%%%%%%%%%%%%%%%%%%%%%%%%%%%%%%%%%%%%%%%%%%%%%%

% Don't change these lines
\bsp    % typesetting comment
\label{lastpage}
\end{document}